\documentclass[11pt]{article}
\pdfoutput=1

\usepackage{jheppub} 

\usepackage{amsmath,amssymb,epsfig,amsfonts}
\usepackage{graphicx,subfigure}
\usepackage{epsf}
\usepackage{makeidx}
\usepackage[all]{xy}
\usepackage{slashed}
\usepackage{verbatim} 
\usepackage{float}
\restylefloat{table}
\usepackage[all]{xy}
\usepackage{pdflscape}
\usepackage{array}
\usepackage{youngtab}
\usepackage{multirow}
\usepackage{color}
\usepackage{ulem}
\usepackage{cleveref}
\usepackage{wrapfig}
\usepackage{mathrsfs}
\usepackage{amssymb}
\usepackage{stmaryrd}
\usepackage{setspace}
\usepackage{ctable}
\usepackage{bbm}

\usepackage{tensor}

\makeatletter

\newcommand{\be}{\begin{equation}}
\newcommand{\ee}{\end{equation}}
\newcommand{\ba}{\begin{aligned}}
\newcommand{\ea}{\end{aligned}}

\newcommand{\aaa}{{a}}
\newcommand{\bbb}{{b}}
\newcommand{\ccc}{{c}}
\newcommand{\ddd}{{d}}
\newcommand{\llll}{{l}}
\newcommand{\kkkk}{{k}}

\newcommand{\nn}{\nonumber}

\newcommand{\dd}{\mathrm{d}}
\newcommand{\me}{\mathrm{e}}
\newcommand{\ii}{\mathrm{i}}

\newcommand{\vol}{\mathrm{vol}}

\newlength{\sswidth}

\newcommand{\lb}{\left(}
\newcommand{\rb}{\right)}

\def\cG{\mathcal{G}}
\def\cF{\mathcal{F}}
\def\cH{\mathcal{H}}
\def\bs{\boldsymbol}

\newcommand{\cN}{\mathcal{N}}

\newcommand{\bea}{\begin{eqnarray}}
\newcommand{\eea}{\end{eqnarray}}

\def\unit{{1\kern-.65ex {\rm l}}}
\def\1{{1\kern-.65ex {\rm l}}}

\newcount\hour \newcount\minute
\hour=\time \divide \hour by 60
\minute=\time
\def\now{%
\ifnum \hour<13
  \ifnum \hour=0 \advance \hour by 12 \number\hour:\else \number\hour:\fi%
     \ifnum \minute<10 0\fi%
     \number\minute%
\ A.M.%
\else \advance \hour by -12 \number\hour:%
  \ifnum \minute<10 0\fi%
  \number\minute%
  \ P.M.%
\fi%
}

\makeatother

\begin{document}

\baselineskip=18pt  
\numberwithin{equation}{section}  
\allowdisplaybreaks  


\thispagestyle{empty}

\baselineskip=15pt

\begin{titlepage}
\begin{flushright}
\parbox[t]{1.8in}{\begin{flushright} ~ \\
~ \end{flushright}}
\end{flushright}

\begin{center}

\vspace*{ 1.2cm}

\begin{spacing}{2.5}
\bf{{{\fontsize{26}{1}\selectfont  Twisted $\boldsymbol{\mathcal N=1}$ SCFTs and their AdS$_{\boldsymbol{3}}$ duals}}}
\end{spacing}

\vskip 1.2cm

\renewcommand{\thefootnote}{}
\begin{center}
 {Christopher Couzens, Huibert het Lam and Kilian Mayer

 \footnotetext{c.a.couzens@uu.nl~ \ ~ h.hetlam@uu.nl~ \ ~ k.mayer@uu.nl~}}
\end{center}
\vskip .2cm

{
Institute for Theoretical Physics and \\
Center for Extreme Matter and Emergent Phenomena,\\

}

\vspace*{.2cm}

\end{center}

 \renewcommand{\thefootnote}{\arabic{footnote}}
 
\begin{center} {\bf Abstract } \end{center}

\noindent We study compactifications of an infinite family of four-dimensional $\mathcal N=1$ SCFTs on a Riemann surface in the presence of arbitrary background fluxes for global symmetries. The four-dimensional parent theories have holographic Sasaki--Einstein duals in type IIB string theory. We compute central charges and R-charges of baryonic operators in the resulting two-dimensional $\mathcal N=(0,2)$ theories in three distinct ways: from the field theory side utilizing c-extremization, its recently discovered geometric dual formulation, and holographically using new AdS$_3$ duals of two-dimensional field theories.

\end{titlepage}


\tableofcontents
\newpage


\section{Introduction}

Since the advent of AdS/CFT in \cite{Maldacena:1997re} many examples of dual pairs have been conjectured. The first (and most studied) was AdS$_5\times$S$^5$ with five-form flux in type IIB and is proposed to be dual to four-dimensional $\mathcal{N}=4$ supersymmetric Yang--Mills \cite{Maldacena:1997re}. The five-sphere can be replaced by an arbitrary five-dimensional Sasaki--Einstein manifold $Y_5$ and gives rise to AdS$_5\times Y_5$ spaces dual to 4d $\mathcal{N}=1$ quiver gauge theories. In general, it is hard to find explicit metrics for these spaces and this makes it difficult to obtain evidence for new dualities.\footnote{One can make progress in understanding the gravity theory if one knows the topological data of the solution \cite{Martelli:2005tp,Martelli:2006yb}.} Despite the difficulties, the infinite family of explicit Sasaki--Einstein metrics $Y_5=Y^{p,q}$ \cite{Gauntlett:2004yd} and its generalization $Y_5=L^{a,b,c}$ \cite{Cvetic:2005ft} have been discovered. These manifolds are \emph{toric} and specified by coprime integers $p$ and $q,$ and $a$, $b$ and $c$ respectively. The duality can be understood via its construction using D3-branes probing the Calabi--Yau cone singularity with base $Y_5$. The 4d theory is the worldvolume theory on these D3-branes whilst AdS$_5\times Y_5$ is the near horizon geometry of the type IIB D3-brane solution. 

Of course, one can also study the correspondence in lower dimensions. One way is to take the aforementioned four-dimensional theories and compactify them on a (smooth) Riemann surface $\Sigma_g$ with genus $g$ performing a topological twist to preserve supersymmetry. Additionally, background magnetic fluxes for the continuous flavour symmetries in the 4d theory can be turned on. Since these fluxes are quantized, every 4d theory leads to a discrete family of 2d theories. In the IR one expects these theories to flow to a 2d superconformal field theory (SCFT), however proving the existence of this flow is in general challenging. Assuming the existence of the SCFT one can compute its central charges and R-charges from UV data using the technique of c-extremization \cite{Benini:2012cz,Benini:2013cda} which is based on anomaly considerations. In this scenario we are not considering a conventional RG flow from a 2d UV theory to the 2d IR theory but instead we consider an \emph{RG flow across dimensions}. One can obtain evidence for the existence of the 2d SCFT using holography by finding an explicit supergravity solution with the same central charges and R-charges. Employing the D3-brane picture, one expects the solution to correspond to the near horizon geometry of a D3-brane compactified on the Riemann surface. This solution will be of the form AdS$_3\times Y_7$ in which $Y_7$ is a fibration of the Sasaki--Einstein manifold $Y_5$ over the Riemann surface, with self-dual five-form flux sourced by the D3-branes. 

The above has been successfully studied in the case that the 4d parent theory is dual to AdS$_5\times Y_5$ for $Y_5=\text{S}^5$ \cite{Bershadsky:1995vm,Bershadsky:1995qy,Maldacena:2000mw,Benini:2012cz,Benini:2013cda}, i.e.~$\mathcal{N}=4$ super Yang--Mills, and for $Y_5=Y^{p,q}$ \cite{Benini:2015bwz}, see also \cite{Amariti:2017cyd,Amariti:2017iuz}. For the latter case of $Y_5=Y^{p,q}$, local supergravity solutions have been found in \cite{Benini:2015bwz} but their global regularity has not been studied so far, see also \cite{Donos:2008ug,Gauntlett:2006af}. However, even if explicit solutions are not known, one can still employ the recently found geometric dual of c-extremization \cite{Couzens:2018wnk,Gauntlett:2018dpc,Gauntlett:2019pqg,Hosseini:2019use}\footnote{The closely related geometric dual of $\mathcal{I}$-extremization \cite{Benini:2015eyy,Hosseini_2016} was also put forward in \cite{Couzens:2018wnk} and further studied in \cite{Gauntlett:2019roi,Hosseini:2019ddy,Kim:2019umc}.}. For this technique one has to assume the existence of a solution and can compute its central charges and R-charges using only topological data characterizing the internal manifold. This has been successfully applied to certain geometries including a subset of $Y_5=Y^{p,q}$ \cite{Gauntlett:2018dpc}. The dual of c-extremization from the perspective of gauged supergravity has been found in \cite{Karndumri:2013iqa} and further investigated in \cite{Karndumri:2013dca}.

In this paper we extend the above and study compactifications of the 4d $L^{a,b,c}$ quiver gauge theories \cite{Benvenuti:2005ja,Franco:2005sm,Butti:2005sw} on a Riemann surface $\Sigma_g$. Note that these theories include the previously studied $Y^{p,q}$ theories \cite{Benvenuti:2004dy} as a special limit. We twist the 4d $L^{a,b,c}$ theories such that the resulting 2d theories preserve $\mathcal{N}=(0,2)$ supersymmetry. On the field theory side 
we compute their central charges and R-charges using c-extremization. On the geometry side we compute these same charges, first by employing geometric c-extremization which has been developed for toric $Y_5$ in \cite{Gauntlett:2018dpc}. However, as mentioned above, to use this technique we must assume the existence of the explicit solution. We therefore construct and study solutions dual to 2d $\mathcal{N}=(0,2)$ SCFTs and match the central charges and R-charges to the expressions obtained via c-extremization. To complete 
the existing literature, in appendix \ref{sec return ypq} we also study the global regularity of the local AdS$_3 \times Y_7$ solutions, where $Y^{p,q}\hookrightarrow Y_7 \rightarrow \Sigma_g$, as given in \cite{Benini:2015bwz} and match 
their central charges and R-charges to the field theory.

The most important part of this paper contains the construction of new solutions of the form AdS$_3\times Y_7$ where $L^{a,b,c}\hookrightarrow Y_7 \rightarrow \Sigma_g$. Such solutions are contained in the classification of supersymmetric AdS$_3$ solutions of type IIB with only five-form flux \cite{Kim:2005ez}. The ten-dimensional solutions are completely determined by a six-dimensional K\"ahler metric solving a \emph{master equation}. We make a sufficiently general ansatz for this 6d metric which is of cohomogeneity two and depends on three functions. The ansatz is motivated by the work of \cite{Martelli:2005wy} in which the $L^{a,b,c}$ solutions of \cite{Cvetic:2005ft} were recovered using a K\"ahler orthotoric metric ansatz \cite{Apostolov:2001} in four dimensions. Their success motivates us to choose a fibration of the orthotoric metric over a Riemann surface for our ansatz. 
We extend these local metrics to globally well-defined metrics by studying their regularity without specifying the functions entering the ansatz. The resulting conditions from the regularity analysis place constraints on the roots of the functions in the ansatz and the first derivative of these functions evaluated there. Subsequently, we perform flux quantization, and compute the central charges and R-charges of baryonic operators while leaving the functions in the metric generic. Amazingly, the integrals one needs to compute can be integrated explicitly, even without solving the master equation. The resulting expressions only depend on the roots and first derivative of these functions. As a final step one can then determine the functions by solving the master equation. Plugging the result into the general expressions we obtain the various charges. The solution we find for the three functions in the ansatz solving the master equation do not contain enough free parameters to account for all the 2d SCFTs. However, given these general expressions for the regularity conditions, the central charges and R-charges, it is simple to match them to the field theory once the most general solution has been found. The internal manifold of the solutions we find are examples of \emph{GK geometries} \cite{Gauntlett:2007ts}. These include previously known solutions in the literature of this form \cite{Gauntlett:2006ns,Gauntlett:2006qw,Gauntlett:2007ts,Donos:2008ug,Kim:2012ek,Benini:2013cda,Benini:2015bwz} as special limits. 
For completeness note that other type IIB AdS$_3$ solutions have been found and classified in \cite{Couzens:2017way,Couzens:2017nnr,Eberhardt:2017uup,Lam:2018jln,Macpherson:2018mif,Passias:2019rga,Couzens:2019iog}. In (massive) type IIA AdS$_3$ solutions (and some type IIB solutions obtained via dualities) can be found in \cite{Lozano:2015bra,Macpherson:2018mif,Dibitetto:2018ftj,Lozano:2019emq,Lozano:2019jza,Lozano:2019zvg,Lozano:2019ywa}. 

Despite the fact that the central charges and R-charges we find from the solution formally match the charges we compute from the twisted compactifications of the $L^{a,b,c}$ quiver gauge theories we 
have only found holographic dual pairs for the higher genus Riemann surfaces $\Sigma_{g>1}$. The gravity solutions we find for S$^2$ and T$^2$ Riemann surfaces are characterized by integers $a<0$ and $b,$ $c>0$, whilst the 4d field theory and therefore also the resulting 2d theories are characterized by 
integers $a,$ $b,$ $c>0$. When $a>0$ (S$^2$ and T$^2$ cases) at least one of the R-charges in the field theory result becomes negative, which is in contradiction with the chiral origin of the 2d fields. 
It is therefore not clear what the 4d theory compactified on $\text{S}^2, \text{T}^2$ flows to in the IR. For the solutions containing a surface with $g>1$ we find gravity solutions with both $a>0$ and $a<0$.  The former are the solutions which are holographically dual to $L^{a, b, c}$ quivers compactified on $\Sigma_{g>1}$. In the latter case where $a<0$ the central charges and R-charges from field theory and the corresponding solution again formally agree in spite of the different domains for the integers. Identifying the 2d SCFTs dual to the AdS$_3$ solutions with $a<0$ constructed in this paper remains an open problem. This same phenomenon has been 
noted previously in \cite{Couzens:2017nnr,Couzens:2018wnk,Gauntlett:2018dpc} for solutions with internal space T$^2\times Y^{p,q}$, T$^2\times L^{a,b,c}$ and a specific fibration of $Y^{p,q}$ over 
S$^2$. In the $Y^{p,q}$ cases the field theory is characterized by integers $p>q>0$, however the naive dual supergravity solutions, when known are instead characterized by integers $q>p>0$. In appendix \ref{sec return ypq} we extend the results of the $Y^{p,q}$ case by studying more general fibrations over a Riemann surface $\Sigma_g$ and performing the regularity analysis. We find that the solutions for Riemann surfaces with $g=0,$ $1$ are also only globally 
regular for $q>p>0$. However, for $g>1$ this is not the case and solutions are valid for $p>q>0$, therefore in the $g>1$ case, and only this case, we have found evidence that the solutions and field theory are dual to each other.

The plan for the paper is as follows. We first study the field theory side in section \ref{field theory part} where we review the technique of c-extremization and use it to compute the central charge and R-charges of the 2d $\mathcal{N}=(0,2)$ SCFTs. In section \ref{geodual section} we review geometric c-extremization and use it to compute the central charges and R-charges from the geometric side assuming the existence of the solutions. Subsequently in section \ref{geometry} we turn our attention to finding these explicit solutions. We first study the regularity of our ansatz and compute expressions for the central charge, R-charges and flux quantitation in terms of this general ansatz. Afterwards we solve the master equation and the conditions that follow from the regularity analysis to find globally regular solutions dual to 2d SCFTs. We use the general expressions to compute the central charge and R-charges and match with the field theory results. We conclude in section \ref{sec: discussion}. Some technical material is relegated to appendices \ref{App:ansatz-analysis} and \ref{app: eff coordinate systems}. In appendix \ref{sec return ypq} we study the $Y^{p,q}$ case.

\newpage

\section{Twisted $\mathcal N=1$ field theories and c-extremization} \label{field theory part}

In this section we study the central charges and R-charges of baryonic operators of 2d $\mathcal N=(0,2)$ SCFTs arising from twisted compactifications of certain 4d $\mathcal N=1$ SCFTs on a compact genus $g$ Riemann surface $\Sigma_g$ from field theory. The 4d parent theories are realized in string theory by $N$ D3-branes probing certain Calabi--Yau threefold singularities. The class of non-compact Calabi--Yau threefolds we consider here are metric cones over the five-dimensional Sasaki--Einstein manifolds $L^{a,b,c}$ discovered in \cite{Cvetic:2005ft}. This infinite family of metrics is labelled by a triplet of integers $a, b, c$ that can be chosen such that
\begin{equation}\label{conditions triples field th}
0<a\leq d\leq c \leq b\,,\quad \text{gcd}(a,c)=\text{gcd}(a,d)=\text{gcd}(b,c)=\text{gcd}(b,d)=1\,,
\end{equation}
where $d=a+b-c$. The 4d $\mathcal N=1$ field theories arising from this setup are quiver gauge theories \cite{Benvenuti:2005ja,Franco:2005sm,Butti:2005sw}. These field theories generically have two flavour symmetries $U(1)_{F_1}\times U(1)_{F_2}$, a baryonic $U(1)_B$ symmetry and in addition a $U(1)_R$ R-symmetry. We are interested in twisted compactifications of these 4d theories on a Riemann surface $\Sigma_g$ with the most general flux configuration of the aforementioned global $U(1)$ symmetries preserving supersymmetry switched on. The IR fixed point (if it exists) is typically a strongly coupled 2d SCFT which may not admit a Lagrangian description. Assuming the compactified theory flows to a 2d $\mathcal N=(0,2)$ SCFT, we shall compute the central charges and R-charges of baryonic operators of this 2d theory. In order to make progress we will utilize the principle of c-extremization put forward in \cite{Benini:2012cz}. This enables the computation of the central charges of 2d SCFTs using UV data.

We begin by briefly reviewing c-extremization and the 4d $\cN=1$ $L^{a,b,c}$ SCFTs in order to keep this paper as self-contained as possible. Readers familiar with both may safely skip to section \ref{c-extr examples} in which we apply c-extremization to twisted compactifications of the 4d $L^{a,b,c}$ theories on a Riemann surface $\Sigma_g$ with constant curvature, turning on fluxes of global symmetries consistent with preserving supersymmetry. The trial central charge will be constructed and extremized with respect to the mixing parameters. While performing these computations in full generality is possible, we will only present the results of certain special cases which will be compared to their geometric counterparts later, since the results quickly become unwieldy. We have attached a Mathematica package to this paper in which the reader can find the results for the general case if they so wish. 

\subsection{Topologically twisted 4d field theories and c-extremization}

In order to determine the (right-moving) central charge of the 2d SCFTs we employ the principle of c-extremization \cite{Benini:2012cz}. Similar to the more familiar a-maximization in 4d \cite{Intriligator:2003jj}, one constructs a trial central charge $c_r^{\rm trial} (\varepsilon_I)$, a function depending on parameters $\varepsilon_I$ which parametrize the mixing of the UV R-symmetry with the remaining global symmetries. Given a 2d $\cN=(0,2)$ UV theory with R-symmetry $U(1)_R$ and global symmetries $U(1)_I$ we construct the trial central charge as follows. First define the trial R-symmetry via the generator
\begin{equation}
T_R^{\text{trial}}=T_R+\sum_{I\neq R} \varepsilon_I T_I \, .\label{Rtrial}
\end{equation}
The 't Hooft anomalies of the 2d theory are encoded in the anomaly four-form polynomial $I_4$ of the 2d theory. The trial central charge is defined via 
\begin{equation}
c_r^{\text{trial}}(\varepsilon_I)=3 k_{RR}^{\text{trial}}\, ,
\end{equation}
where $k_{RR}^{\text{trial}}$ is the quadratic 't Hooft anomaly coefficient of the trial R-symmetry in the anomaly polynomial
\begin{equation}
I_4 \supset  \frac{1}{2} k_{RR}^{\text{trial}} \, c_1(F_R^{\text{trial}})\wedge c_1(F_R^{\rm trial})\, .
\end{equation}
Here $F_R^{\text{trial}}$ is the background gauge field of the trial R-symmetry. The exact superconformal R-symmetry of the 2d $(0,2)$ SCFT is then obtained at the extremum of the trial central charge
\begin{equation}
c_R=c_r^{\rm trial}(\varepsilon_I^*) \,  ,\qquad   \frac{\partial c_r^{\rm trial}(\varepsilon_J)}{\partial \varepsilon_I}\Big|_{\varepsilon_I=\varepsilon_I^*}=0\, .
\end{equation}
This extremization condition follows from the fact that the superconformal R-symmetry at the IR fixed point has no mixed 't Hooft anomalies with any global symmetry, i.e.~$k_{R I}=0$. The trial central charge evaluated at the extremum $c_R=c_r^{\rm trial}(\varepsilon^*_I)$ is then the right-moving central charge of the SCFT. Furthermore the R-charges of operators can be computed by inserting $\varepsilon^*_I$ into the trial R-symmetry \eqref{Rtrial}. Let us note that the left-moving central charge can be computed from the result for the right-moving central charge and the gravitational anomaly of the 2d SCFT. The difference of the left- and right-moving central charge is determined by the gravitational anomaly coefficient
\begin{equation}
I_4 \supset -\frac{c_L-c_R}{24}p_1(TM_2)
\end{equation}
in the anomaly polynomial. This contribution however, scales as $\mathcal O(N^0)$ in the theories under consideration and since we are only interested in the leading $\mathcal O(N^2)$ contribution in the large $N$ limit in this paper, we shall not comment further on the subleading corrections.

The 4d $\mathcal N=1$ field theories which we wish to twist on a Riemann surface $\Sigma_g$ are quiver gauge theories with $SU(N)$ nodes and bifundamental matter linking the nodes. The gauge theory data necessary for our purposes are the bifundamental matter multiplets and gauginos together with their multiplicities and charges under the global $U(1)_{F_1}\times U(1)_{F_2}\times U(1)_B \times U(1)_R$ symmetry. The data of the six types of chiral bifundamentals included in the gauge theory, which are conventionally denoted by $Z,$ $Y$, $U_{1,2}$ and $ V_{1, 2}$, and the vector multiplets, is summarized in table \ref{quiverdata}.\footnote{The R-charge column is the (fiducial) R-charge $r$ of the fermionic component of the multiplet, the R-charge of the boson is $r+1$.}
\begin{table}[h]
\begin{centering}
\def\arraystretch{1.0}
\begin{tabular}{lccccc}
\specialrule{.07em}{0.07em}{0em}
Field & multiplicity & $U(1)_{F_1}$ & $U(1)_{F_2}$ & $U(1)_B$ & $U(1)_R$ \tabularnewline 
\specialrule{.07em}{0.07em}{0em}
$Z$ &  $N^2 a$ & $0$ & $k$ & $b$ & $-1$\\
$U_2$ & $N^2 c$ & $-1$ & $-k-l$ & $-d$ & $0$\\
$Y$ & $N^2 b$ & $1$ & $0$ & $a$ & $-1$\\
$U_1$ & $N^2 d$ & $0$ & $l$ & $-c$ & $0$\\
$V_1$ & $N^2(b-c)$ & $-1$ & $-l$ & $c-a$ & $0$\\
$V_2$ & $N^2 (c-a)$ & $0$ & $k+l$ & $b-c$ & $0$\\
$\lambda$     & $(N^2-1)(a+b)$& $0$ & $0$    & $0$   & $1$
\end{tabular}
\par\end{centering}
\caption{\label{quiverdata} Bifundamental matter and vector multiplets in the $L^{a,b,c}$ quiver and their charges. In this paper we often refer to the first four multiplets with $(X_{1},X_{2},X_{3},X_{4})=(Z,U_{2},Y,U_{1})$.}
\end{table}
The integers $k,$ $l$ are solutions of $c k+b l =1$ which exist by B\'ezout's identity. Note that our fiducial R-symmetry is not the R-charge assignment of the 4d theory at the 4d IR fixed point. Instead we choose a convenient set of R-charges such that the R-charge of the superpotential is two as usual. 

Upon wrapping the 4d theory on a genus $g$ Riemann surface $\Sigma_g$ we perform a partial topological twist to preserve $\mathcal N=(0,2)$ supersymmetry in 2d. We study the most general twists involving flux for the R-symmetry, flavour, and baryonic symmetries on the Riemann surface. The total background gauge field configuration on the Riemann surface is  
\begin{equation}
\mathscr F_{\rm{\tiny{flux}}}=\frac{\kappa}{2} \mathscr R_g+f_1 \mathscr{F}_{1}+f_2 \mathscr{F}_{2}+B \mathscr F_B\, ,\label{background}
\end{equation}
where $\mathscr F_{A}= T_A \text{dvol}(\Sigma_g)$ is the field strength associated to the symmetry $U(1)_A$ with its generator $T_A$. Here $\kappa$ is the (constant) normalized curvature of the Riemann surface with Ricci form $\mathscr R_g$. The volume form on the Riemann surface is normalized such that 
\begin{equation}
\text{vol}(\Sigma_g)=\int_{\Sigma_g}\text{dvol}(\Sigma_g)=2\pi \eta_g\equiv
\begin{cases}
4\pi |g-1|\, , \qquad g \neq 1\\
4\pi\, ,\qquad \qquad ~~\,   g=1
\end{cases}
\! \! \! \! \!.
\end{equation}

In order to construct the trial central charge for the compactified 4d theories that we will study in this paper, it is useful to work with the anomaly six-form polynomial of the parent 4d $\mathcal N=1$ theory. The anomaly polynomial for an $\mathcal N=1$ theory with global symmetries $U(1)_I$ and associated background field strengths $\mathcal F_I$ takes the form
\begin{equation}
I_6=\frac{1}{3!}k^{IJK} c_1(\mathcal F_I)\wedge c_1(\mathcal F_J) \wedge c_1(\mathcal F_K)-\frac{1}{24}k^I c_1(\mathcal F_I) \wedge p_1(TM_4)\, ,
\end{equation}
where $p_1(TM_4)$ is the first Pontryagin class of the four-manifold $M_4$ on which the 4d theory is defined and $c_{1}(\mathcal F_I)$ are the first Chern classes associated with the global symmetries. The coefficients in the anomaly polynomial depend on the chiral spectrum of the 4d theory via the cubic and linear 't Hooft anomalies
\begin{equation}
k^{IJK}=\sum_{{\text{Weyl fermions }} \,\psi} Q^I_\psi Q^J_\psi Q^K_\psi \, ,\qquad\qquad  k^I=\sum_{{\text{Weyl fermions }} \,\psi} Q^I_\psi\, ,
\end{equation}
where $Q^I_\psi$ is the charge of the fermion $\psi$ under $U(1)_I$. We can then compute the anomaly polynomial $I_4$ of the 2d $\mathcal{N}=(0, 2)$ theory by integrating the 4d one over the Riemann surface $\Sigma_g$. The dimensional reduction of the anomaly polynomial can be performed by including both the background flux and the mixing of the 2d background gauge fields in the decomposition of the 4d background gauge fields. The ansatz for the mixing of the UV R-symmetry with the remaining flavour symmetry can be parametrized by the generator
\begin{equation}
T_{R}^{\rm trial}=T_R+\varepsilon_1 T_{F_1}+\varepsilon_2 T_{F_2}+\varepsilon_B T_B\, ,
\end{equation}
 where $\varepsilon_i,$ $\varepsilon_B$ parametrize the mixing. The field strengths of $U(1)_{F_1} \times U(1)_{F_2}\times U(1)_B\times U(1)_R$ including fluxes and mixing parameters are then identified as
\begin{align}
\mathcal F_R&=F_R^{\rm trial}+\frac{\kappa}{2} \mathscr R_g\,, \nn\\
\mathcal F_{i}&= F_{i}+\varepsilon_{i} F_{R}^{\text{trial}}+f_i \mathscr{F}_{F_i}\,,\label{bundledecomp}\\
\mathcal F_B&=F_B+\varepsilon_B F_R^{\text{trial}}+B \mathscr{F}_B\nonumber \, ,
\end{align}
where now $F_{i},$ $F_B$ and $F_{R}^{\text{trial}}$ are the 2d field strengths of the global symmetries and $F_R^{\text{trial}}$ is the field strength of the 2d trial R-symmetry. The trial central charge is then given by the quadratic 't Hooft anomaly associated with the trial R-symmetry, i.e.~$c_r^{\rm trial}(\varepsilon_i, \varepsilon_B)=3 k_{R R}^{\rm trial}$. The 't Hooft anomaly of the trial R-symmetry can now easily be computed by plugging the bundle decomposition \eqref{bundledecomp} into the anomaly polynomial $I_6$ of the 4d field theory and integrating the latter over the Riemann surface $\Sigma_g$, thereby obtaining the anomaly polynomial $I_4$ of the 2d $(0, 2)$ theory. Reading off the coefficient
\begin{equation}
I_4 = \int_{\Sigma_g}I_6 \supset \frac{1}{2} k_{RR}^{\text{trial}} \, c_1(F_R^{\text{trial}})\wedge c_1(F_R^{\rm trial})\,,
\end{equation}
one finds
\begin{equation}
c_r^{\rm trial}=-3 \eta_g \sum_{\text{fermions}\, \psi}\!\!\!\text{mult}(\psi) T_\psi^{\rm flux} \big(Q_{R, \psi}^{\rm trial} \big)^2\, .\label{ctrial}
\end{equation}
The sum runs over all 4d fermions in the spectrum, mult$(\psi)$ is the multiplicity of $\psi$, $T_\psi^{\rm flux}$ is the charge of the fermion under the background gauge field \eqref{background}, and $Q_{R, \psi}^{\rm trial}$ is the charge under the trial R-symmetry. Extremizing the trial central charge $c_r^{\rm trial}(\varepsilon_i, \varepsilon_B)$ with respect to $\varepsilon_I=(\varepsilon_i, \varepsilon_B)$ and evaluating it at the extremum gives the value of the right-moving central charge at the (putative) IR fixed point.

\subsection{Central charges and R-charges from c-extremization}\label{c-extr examples}

It is now straightforward to compute the trial central charge for the $L^{a, b, c}$ field theories compactified on $\Sigma_g$ with fluxes using the data given in table \ref{quiverdata} and the general formula for the trial central charge \eqref{ctrial}. One finds
\begin{align}
c_r^{\rm trial}(\varepsilon_I)&=-3 N^2 \eta_g \bigg[ (b-c)\big( B(c-a)-f_1-f_2 l \big)\nonumber\\
&+(c-a)\big(B (b-c)+f_2 (k+l) \big)\big((k+l)\varepsilon_2+(b-c)\varepsilon_B \big)^2+ d \big(f_2  l -B c \big)\big(l \varepsilon_2 - c\, \varepsilon_B\big)^2\nonumber\\
&-c \big(d\, B +f_1+f_2 (k+l) \big)\big(\varepsilon_1+ (k+l)\varepsilon_2+ d \, \varepsilon_B \big)^2+b (a \,\varepsilon_B +\varepsilon_1-1)^2\Big(a \, B+f_1 -\frac{\kappa}{2} \Big)\nonumber\\
&+a (b \, \varepsilon_B+k \varepsilon_2-1)^2 \Big(b \, B+f_2 k -\frac{\kappa}{2} \Big)+\frac{1}{2}(a+b)\kappa \bigg]\,\label{ctrialLabc}
\end{align}
for the most general twisted field theory and flux configuration on $\Sigma_g$. It is possible to extremize the trial central charge \eqref{ctrialLabc} in this most general setting and evaluate the latter at the corresponding extremum. One thus obtains the right-moving central charge and R-charges of baryonic operators in the 2d SCFT in full generality. However, since the full expressions are tedious, we refrain from giving the result for the most general case here and instead specialize to some interesting cases. The interested reader can find the most general result in a Mathematica file attached to this submission. 

\subsubsection{Field theory c-extremization for ${\Sigma_g=\mathrm{T}^2}$}\label{cextrT2}

We now present some explicit computations for the case when the 4d parent theories are compactified on a torus T$^2$. The formulas for the central charges and R-charges of the putative 2d SCFT, to which the 4d theory flows in the IR, are given and shall be compared to the geometric dual of c-extremization and supergravity in sections \ref{subsec:Specifying-to-} and \ref{subsec:matching} respectively.

\paragraph{Purely baryonic flux.} The first case we consider is the 4d $L^{a,b,c}$ field theory compactified on $T^2$ with purely baryonic flux i.e.~$f_1=f_2=0$. The values of the mixing parameters $\varepsilon_I$ extremizing the trial central charge \eqref{ctrialLabc} for this case can be easily computed and are given by
\begin{equation}
\varepsilon^*_1=\frac{a\big(a-b+ak (b-c)\big)}{(a-c)(c-b)}\, , \qquad \varepsilon^*_2=\frac{ab}{c-a}\, ,\qquad \varepsilon^*_B=\frac{a \big(k (c-b)-1 \big)}{(b-c)(c-a)}\, .
\end{equation}
With these values for the mixing parameters one finds the right-moving central charge
\begin{equation}
c_R=12 N^2 B \frac{abc\, (a+b-c)}{(a-c)(b-c)}\, ,\label{cT2bary}
\end{equation}
and the R-charges of the baryonic operators are given by
\begin{equation}
R[X_1]=R[X_3]=\frac{ab}{(a-c)(b-c)}\, ,\qquad R[X_2]=R[X_4]=-\frac{c d}{(a-c)(b-c)}\, .\label{cextrbaryonic}
\end{equation}
Note that \eqref{cextrbaryonic} are not well-defined R-charges for chiral multiplets as one of them necessarily is negative when $a,$ $b,$ $c$ satisfy \eqref{conditions triples field th}. This has been noticed in \cite{Couzens:2018wnk} where it was shown from a gravity computation that this is a generic feature of these compactifications on a T$^2$ with only baryonic flux. We can now examine for what domain of $a,$ $b,$ $c$ both the central charge and R-charges are positive. We wish to keep $0<d\leq c \leq b$ since the $Y^{p,q}$ limit is given by taking $a=p-q$, $b=p+q$ and $c=d=p$ for positive integers $p$ and $q$. Therefore the natural modification to make is to allow $a$ to take negative values.\footnote{Technically the $Y^{p,q}$ limit does not fix the order of $c$ and $d$, however as in the 4d theory we believe we can take $d\leq c$ without loss of generality.} Indeed with these ranges we can simultaneously find a positive central charge and positive R-charges.

\paragraph{Fluxes $\boldsymbol{B \propto f_2, \,f_1=0}$.} Consider now a twisted compactification of the $L^{a,b,c}$ quiver gauge theories without $U(1)_{F_1}$ flux, and baryonic flux proportional to the $U(1)_{F_2}$ flux. Concretely, we choose a flux configuration where $B=-\frac{k}{b} f_2, \, f_1=0$. Extremizing the trial central charge \eqref{ctrialLabc} for these fluxes with respect to the mixing parameters $\varepsilon_I$ gives the extremal value of the central charge
\begin{equation}
c_{R}=\tfrac{6ak f_2 N^2(-1+ak)\big[a+b-bck+c(-2+ck)\big]}{b(-1+ck)+c(2-ck)+a^{2}k^{2}\big[c(2-ck)+b(-1+ck)\big]+a\big[-1-2ck+c^{2}k^{2}+bk(1-ck)\big]}\, ,
\end{equation}
and the R-charges of baryonic operators are given by
\begin{eqnarray}
R[X_1] & = & -\tfrac{abk(-1+ak)}{b(-1+ck)+c(2-ck)+a^{2}k^{2}\big[c(2-ck)+b(-1+ck)\big]+a\big[-1-2ck+c^{2}k^{2}+bk(1-ck)\big]}\,,\nonumber \\
R[X_2] & = & \tfrac{(-1+ak)\big[a+b-bck-ac^{2}k^{2}+ack(1+bk)+c(-2+ck)\big]}{b(-1+ck)+c(2-ck)+a^{2}k^{2}\big[c(2-ck)+b(-1+ck)\big]+a\big[-1-2ck+c^{2}k^{2}+bk(1-ck)\big]}\,,\nonumber \\
R[X_3] & = & \tfrac{a^{2}k(1+bk-ck)(-1+ck)}{b(-1+ck)+c(2-ck)+a^{2}k^{2}\big[c(2-ck)+b(-1+ck)\big]+a\big[-1-2ck+c^{2}k^{2}+bk(1-ck)\big]}\,,\\
R[X_4] & = & \tfrac{a^{2}ck^{2}+c(2-ck)+b(-1+ck)-a(1+ck)}{b(-1+ck)+c(2-ck)+a^{2}k^{2}\big[c(2-ck)+b(-1+ck)\big]+a\big[-1-2ck+c^{2}k^{2}+bk(1-ck)\big]}\,.\nonumber 
\end{eqnarray}
We find that at least one of the R-charges is negative when $a,$ $b,$ $c$ satisfy \eqref{conditions triples field th}.

\subsubsection{Field theory c-extremization for ${\Sigma_{g\neq1}}$} \label{subsecfieldthsigman1}

We now move to twisted compactifications of the $L^{a,b,c}$ quiver gauge theories on a sphere and on higher genus Riemann surfaces. The parameter $\kappa \neq 0$ gives the distinction between $\Sigma_{g=0}=\text{S}^2$ and $\Sigma_{g >1}$.

\paragraph{Baryonic flux and fixed flavour flux.} For this field theory compactification we choose general baryonic flux given by $B$ and flavour fluxes with fixed values $f_1=-\frac{\kappa}{2} a k$ and $f_2=-\frac{\kappa}{2} b$. We note that there is nothing special in field theory for these values except that they give simple results. Extremizing the trial central charge we find
\begin{equation}
c_R=\frac{6 a b c N^2 (g-1)   (k - 
    2 B \kappa)^2 (1 + (a + b - c) (k - 2 B \kappa))}{
 1 + (k - 2 B \kappa) \big(a + 
     a (b - c) (k - 2 B \kappa) - (b - c) (-1 + 
        c (k - 2 B \kappa))\big)}\, .\label{s2s3centralfield}
\end{equation}
The R-charges at the conformal fixed point are given by
\begin{align}
R[X_1]&=\frac{b (k-2 B \kappa ) (a (k-2 B \kappa )+1)}{(k-2 B \kappa ) \big(a (b-c) (k-2 B \kappa )+a-(b-c) (c (k-2 B \kappa
   )-1)\big)+1}\,,\nonumber\\
R[X_2]&=  \frac{2-(a+b-c) (k-2 B \kappa ) (c (k-2 B \kappa )-1)}{(k-2 B \kappa ) \big(a (b-c) (k-2 B \kappa )+a-(b-c) (c (k-2 B
   \kappa )-1)\big)+1}\,, \nonumber\\
   R[X_3]&=\frac{a (k-2 B \kappa ) (b (k-2 B \kappa )+1)}{(k-2 B \kappa ) \big(a (b-c) (k-2 B \kappa )+a-(b-c) (c (k-2 B \kappa
   )-1)\big)+1}\,,\label{s2s3Rchargesfield}\\
 R[X_4]&=\frac{c (2 B \kappa -k) ((a+b-c) (k-2 B \kappa )+1)}{(k-2 B \kappa ) \big(a (b-c) (k-2 B \kappa )+a-(b-c) (c (k-2 B \kappa
   )-1)\big)+1}\,\nonumber .
\end{align}
Here we also find that at least one of the R-charges is negative when $a,$ $b,$ $c$ satisfy \eqref{conditions triples field th}. However, the central charge and R-charges take positive values when $a$ is negative and 
${0<d\leq c \leq b}$.

\section{Geometric dual of c-extremization}\label{geodual section}

In this section we utilize the geometric dual of c-extremization \cite{Couzens:2018wnk,Gauntlett:2018dpc}
to compute the central charges and R-charges of the $L^{a,b,c}$ quiver
theories reduced on a Riemann surface $\Sigma_{g}$ with constant curvature. We apply geometric c-extremization to geometries of the form $\text{AdS}_{3}\times Y_{7},$
where $Y_{7}$ is a fibration of $L^{a,b,c}$ over $\Sigma_{g}$. Note
that the previously studied $Y^{p,q}$
fibrations over $\Sigma_{g}$ considered in \cite{Gauntlett:2018dpc} can be obtained from $L^{a,b,c}$ by taking a certain limit, see appendix \ref{sec return ypq}. We first describe the constraints on the geometry of the holographic duals of the 2d SCFTs imposed by supersymmetry in section \ref{Holographic duals}. We then briefly review the geometric c-extremization procedure in section \ref{subsec:Geometric-dual-of}. Again, the reader already familiar with these concepts may skip to section \ref{subsec:Specifying-to-} where we compute
the central charges and R-charges applying this formalism. We subsequently compare the results obtained from the geometric dual to the field theory results of section \ref{field theory part} and find perfect agreement. The matching that we find is a priori a formal matching since we have not proven the existence of the supergravity solution yet, this will be the subject of section \ref{geometry}, where we construct explicit solutions.

\subsection{Holographic AdS$_3$ duals with five-form flux} \label{Holographic duals}

In this paper we are interested in the 2d SCFTs arising from wrapping D3-branes on a constant curvature Riemann surface. The holographic duals of these SCFTs fall within the classification of \cite{Kim:2005ez}. There, the most general supersymmetric AdS$_3$ solutions of type IIB supergravity with only five-form flux are shown to take the form
\begin{align}
\dd s^{2} &= L^{2}\me^{-\frac{w}{2}}\Big( \dd s^{2}_{\text{AdS}_{3}}+\frac{1}{4}( \dd z+P)^{2} +\me^{w} \dd s^{2}(\mathcal{M}_{6})\Big)\,, \label{metric susy sol}\\
F_{5}&=(1+*)\dd \vol_{\text{AdS}_{3}}\wedge \Big(-2 J+ \frac{1}{2} \dd \left[ \me^{-w}(\dd z +P)\right]\Big)\,, \label{five-form}\\
\me^{w}&= \frac{1}{8} R\,. \label{warp}
\end{align}
The metric on $\text{AdS}{}_{3}$ with unit radius is denoted by $\text{d}s_{\text{AdS}_{3}}^{2}$
and its corresponding volume form by $\text{dvol}_{\text{AdS}_{3}}$. Furthermore, $\mathcal{M}_{6}$ is a transverse K\"ahler metric with K\"ahler form $J$ and Ricci scalar $R$, satisfying the \textit{master equation}
\be\label{Mastereq}
\square R= \frac{1}{2} R^{2} - R_{ij}R^{ij}\,.
\ee
The one-form $P$ is the Ricci-form potential of the K\"ahler metric satisfying $\dd P= \rho$ with $\rho$ the Ricci-form of $\mathcal{M}_{6}$. The scalar $w$ is determined in terms of $R$. Finally, $L$ is an overall dimensionful length scale. For a well-defined solution one must impose the flux quantization condition 
\begin{equation}\label{flux quantisation}
\frac{1}{(2\pi\ell_{s})^{4}g_{s}}\int_{\Sigma_{A}}F_{5}=N_{A}\in\mathbb{Z}\,,
\end{equation}
where $\ell_{s}$ is the dimensionful string length, $g_{s}$ is the
constant string coupling and $\Sigma_{A}$ form an integral basis
for the free part of $H_{5}(Y_{7},\mathbb{Z}).$ Note that the solution \eqref{metric susy sol} has a unit norm Killing vector
$\xi=2\partial_{z}$, called the \emph{R-symmetry vector}, which is dual to the R-symmetry of the SCFT and defines a foliation $\mathcal{F}_{\xi}$
of $Y_{7}$. Finally, the real cone over $Y_{7},$ i.e. $C(Y_{7})\equiv\mathbb{R}_{>0}\times Y_{7}$
with metric
\begin{equation}
\text{d}s^{2}=\text{d}r^{2}+r^{2}\text{d}s_{7}^{2}\,,
\end{equation}
is complex but not K\"ahler. There is a nowhere-zero, closed and holomorphic $(4,0)$-form
$\Psi,$ which has R-charge two, i.e.
\begin{equation}
\mathcal{L}_{\xi}\Psi=2i\Psi\,,\label{R-charge Psi}
\end{equation}
hence $\xi$ is a holomorphic vector field.

\subsection{Geometric dual of c-extremization: general formalism \label{subsec:Geometric-dual-of}}

Taking the SCFT as described in section \ref{field theory part} we will assume the existence of a dual solution satisfying the supersymmetry conditions and the master equation \eqref{Mastereq}. Let us review how geometric c-extremization \cite{Couzens:2018wnk,Gauntlett:2018dpc} determines the central charges and R-charges of the dual SCFT. To apply this technique we take the solution off-shell which is done by only imposing
the BPS-equations. We thus consider solutions of the form \eqref{metric susy sol}-\eqref{warp} satisfying \eqref{flux quantisation}. The solutions can then be put on-shell by additionally
applying the five-form equation of motion \eqref{Mastereq}. The metric cone over the manifold $Y_7$, which we denote by $C(Y_7)$, is a complex cone with holomorphic volume form $\Psi$ and holomorphic $U(1)^{s}$
action generated by real holomorphic vector fields $\partial_{\varphi_{i}},$ $i=1,...,s.$ The R-symmetry vector $\xi$ is holomorphic and can be expanded as
\begin{equation}
\xi=\sum_{i=1}^{s}b_{i}\partial_{\varphi_{i}}\,.
\end{equation}
We can choose a basis $\lbrace \partial_{\varphi_i} \rbrace_{i=1}^s$ such that $\Psi$ is only charged under $\partial_{\varphi_{1}}$, which by
(\ref{R-charge Psi}) implies that $b_{1}=2$. Once the vector field $\xi$
and thus foliation $\mathcal{F}_{\xi}$ is fixed, we have to choose
a transverse K\"ahler metric with K\"ahler class $[J]\in H_{B}^{1,1}(\mathcal{F}_{\xi}).$
As remarked earlier, we must impose flux quantization on the off-shell solution over all compact five-cycles. The relevant part of the five-form flux is the part solely on the internal space and may be written as
\be
F_5|_{\text{internal}}=\frac{1}{4} \Big(4 \eta_{\text{R}} \wedge \dd \eta_{\text{R}} \wedge J +\frac{1}{2} * \dd R\Big)\,,
\ee
where $\eta_R=\tfrac{1}{2}(\text{d}z+P)$. With this expression the flux quantization conditions \eqref{flux quantisation} are equivalent to
\begin{equation}
\int_{\Sigma_{A}}\eta_R\wedge\rho\wedge J=\frac{2(2\pi\ell_{s})^{4}g_{s}}{L^{4}}N_{A}\,,\label{flux quantization}
\end{equation}
where $\Sigma_{A}$ are representative five-cycles on $Y_{7}$. This is only consistent in homology when 
\begin{equation}
\int_{Y_{7}}\eta_R \wedge\rho^{2}\wedge J=0\label{constraint eq}
\end{equation}
is imposed in addition and $H^{2}(Y_{7},\mathbb{R})\cong H_{B}^{2}(\mathcal{F}_{\xi})/[\rho],$
which holds for the fibered geometries $Y_{7}$ we consider here.
The constraint equation (\ref{constraint eq}) is equivalent to
integrating the master equation (\ref{Mastereq}) over $Y_{7}.$ The {\it off-shell central charge} is defined by
\begin{equation}
\mathscr{Z}\equiv\frac{3L^{8}}{(2\pi)^{6}g_{s}^{2}\ell_{s}^{8}}S_{\text{SUSY}}=\frac{1}{2}\frac{3L^{8}}{(2\pi)^{6}g_{s}^{2}\ell_{s}^{8}}\int_{Y_{7}}\eta_R \wedge\rho\wedge J^{2}\,.\label{trial central charge}
\end{equation}
The flux quantization condition
(\ref{flux quantization}), constraint equation (\ref{constraint eq}) and trial central charge (\ref{trial central charge})
only depend on the geometry via the choice of $\xi$ and the K\"ahler
class $[J].$ The dual of c-extremization is given by extremizing
(\ref{trial central charge}) over these choices. The central charge of the dual SCFT is then
given by
\begin{equation}
c_{\text{sugra}}=\mathscr{Z}|_{\text{on-shell}}\,.
\end{equation}

So far we have given a general description of geometric c-extremization, but now we will specialize this procedure to geometries
that are fibrations of the form
\begin{equation}
Y_{5}\hookrightarrow Y_{7}\rightarrow\Sigma_{g}\,,
\end{equation}
where the fibers $Y_{5}$ are toric, i.e. they have a $U(1)^{3}$
isometry group generated by $\partial_{\varphi_{i}},$ $i=1,2,3,$
with $\varphi_{i}$ having period $2\pi.$ Furthermore, we assume that
the R-symmetry $\xi$ is tangent to $Y_{5}$ such that we can write
\begin{equation}
\xi=\sum_{i=1}^{3}b_{i}\partial_{\varphi_{i}\,.}
\end{equation}
Note that $\xi$ again defines a foliation $\mathcal{F}_{\xi}$ on
$Y_{5}$ and we use the same notation for $\eta_R$ and its restriction
to a fiber. We denote the transverse K\"ahler form for $\mathcal{F}_{\xi}$
by $\omega.$ Since $C(Y_{7})$ is complex, the cone $C(Y_{5})$ is
also complex with fixed complex structure. Moment map coordinates
can be defined by
\begin{equation}
y_{i}\equiv\tfrac{1}{2}r^{2}\partial_{\varphi_{i}}\lrcorner \,\eta_R\,,\qquad i=1,2,3\,,
\end{equation}
which are standard coordinates on $\mathbb{R}^{3}.$ The image of
the moment map is the convex polyhedral cone
\begin{equation}
\mathcal{C}=\{\vec{y}\in\mathbb{R}^{3}\,|\,(\vec{y},\vec{v}_{a})\geq0\,,\quad a=1,...,f\}\,,
\end{equation}
where $\vec{v}\in\mathbb{Z}^{3}$ are the inward pointing primitive
normals to the facets, which we take to be ordered in an anti-clockwise
direction. Furthermore, $f$ is the number of facets and $(\cdot,\cdot)$
is the Euclidean inner product. We assume that the cone $C(Y_{5})$ admits a global holomorphic $(3,0)$-form such that a basis exists in
which we can write $\vec{v}_{a}=(1,\vec{w}_{a}),$ for $\vec{w}_{a}\in\mathbb{Z}^{2}.$
In this basis the holomorphic $(3,0)$-form has unit charge under
$\partial_{\varphi_{1}}$ and is uncharged under $\partial_{\varphi_{2}}$
and $\partial_{\varphi_{3}}$, therefore, $b_{1}=2.$
The K\"ahler class of the transverse metric of $Y_{5}$ $[\omega]\in H_{B}^{2}(\mathcal{F}_{\xi})$
is specified by parameters $\lambda_{a},$ $a=1,...,f$ of which only
$f-2$ are independent \cite{Abreu}. The K\"ahler class of the Riemann surface is
determined by a parameter $A.$ The fibration is specified by the
first Chern numbers $\vec{n}=(n_{1},n_{2},n_{3})\in\mathbb{Z}^{3}$
that are associated to the three $U(1)$ bundles over the Riemann
surface. Using the transition functions of these bundles the manifold
$Y_{5}$ is then fibered over $\Sigma_{g}.$ Requiring the existence of
the global $(4,0)$-form $\Psi$ on $C(Y_{7})$ results in the
condition 
\begin{equation}
n_{1}=2-2g\,,
\end{equation}
which is the geometric equivalent of the field theory twist with the fiducial R-symmetry. Flux quantization \eqref{flux quantization}, the
master equation \eqref{constraint eq} and trial central charge \eqref{trial central charge}
can all be expressed using the so-called \textit{master volume}
\begin{equation}
\mathcal{V}(\vec{b};\{\lambda_{a}\})=\frac{(2\pi)^{3}}{2}\sum_{a=1}^{f}\lambda_{a}\frac{\lambda_{a-1}(\vec{v}_{a},\vec{v}_{a+1},\vec{b})-\lambda_{a}(\vec{v}_{a-1},\vec{v}_{a+1},\vec{b})+\lambda_{a+1}(\vec{v}_{a-1},\vec{v}_{a},\vec{b})}{(\vec{v}_{a-1},\vec{v}_{a},\vec{b})(\vec{v}_{a},\vec{v}_{a+1},\vec{b})}\,\label{volume function}
\end{equation}
and its derivatives, where $(\vec v,\vec w,\vec u) \equiv \text{det} (\vec v,\vec w,\vec u)$ denotes the determinant. The constraint equation
\eqref{constraint eq} is now equal to 
\begin{equation}
A\sum_{a,b=1}^{f}\frac{\partial^{2}\mathcal{V}}{\partial\lambda_{a}\partial\lambda_{b}}-2\pi n_{1}\sum_{a=1}^{f}\frac{\partial\mathcal{V}}{\partial\lambda_{a}}+2\pi b_{1}\sum_{a=1}^{f}\sum_{i=1}^{3}n_{i}\frac{\partial^{2}\mathcal{V}}{\partial\lambda_{a}\partial b_{i}}=0\,,\label{constraint equation}
\end{equation}
and the flux quantization \eqref{flux quantization} becomes
\begin{eqnarray}
\frac{2(2\pi\ell_{s})^{4}g_{s}}{L^{4}}N & = & -\sum_{a=1}^{f}\frac{\partial\mathcal{V}}{\partial\lambda_{a}}\,,\label{flux n}\\
\frac{2(2\pi\ell_{s})^{4}g_{s}}{L^{4}}M_{a} & = & \frac{A}{2\pi}\sum_{b=1}^{f}\frac{\partial^{2}\mathcal{V}}{\partial\lambda_{a}\partial\lambda_{b}}+b_{1}\sum_{i=1}^{3}n_{i}\frac{\partial^{2}\mathcal{V}}{\partial\lambda_{a}\partial b_{i}}\,.\label{fluxes}
\end{eqnarray}
The flux $N$ comes from taking the fiber $Y_{5}$ as five-cycle and
the fluxes $M_{a},$ $a=1,...,f$ correspond to torus-invariant three-manifolds
$S_{a}\subset Y_{5}$ fibered over $\Sigma_{g}.$ The latter satisfy
the three relations \cite{Gauntlett:2018dpc}
\begin{equation}
\sum_{a=1}^{f}v_{a}^{i}M_{a}=-n_{i}N\,,\qquad i=1,2,3\,,\label{equations fluxes}
\end{equation}
in agreement with the number of independent five-cycles $\text{dim}\,H_{5}(Y_{5},\mathbb{R})=f-3.$
The trial central charge \eqref{trial central charge} is given in terms
of $\mathcal{V}$ by
\begin{equation}
\mathscr{Z}=\frac{3L^{8}}{(2\pi)^{6}g_{s}^{2}\ell_{s}^{8}}\left(-A\sum_{a=1}^{f}\frac{\partial\mathcal{V}}{\partial\lambda_{a}}-2\pi b_{1}\sum_{i=1}^{3}n_{i}\frac{\partial\mathcal{V}}{\partial b_{i}}\right)\label{central charge}
\end{equation}
and depends after setting $b_{1}=2$ on two parameters since we
have imposed $f$ relations on $f+2$ parameters. The value of $\mathscr{Z}$
at its extremal point is equal to $c_{\text{sugra}}$. The R-charges
$R_{a}=R[S_{a}]$ of the baryonic operators dual to D3-branes
wrapping the supersymmetric three-manifolds $S_{a}\subset Y_{7}$
on a fixed point of $\Sigma_{g}$ are given by \cite{Couzens:2018wnk}
\begin{equation}
R_a=R[S_a]= \frac{L^4}{(2\pi)^3 g_s \ell_s^4}\int_{S_a}\text{e}^{-w} \text{dvol}(S_a)=\frac{L^4}{(2\pi)^3 g_s \ell_s^4}\int_{S_a} \eta_R \wedge \omega\, ,\label{Rcharges general}
\end{equation}
where
\begin{equation}\label{omega expr}
\omega=\dd \xi\wedge \big(\dd \phi+\eta \dd \psi \big)+\dd \eta \wedge \big(\dd \phi+\xi \dd \psi \big)\, .
\end{equation}
This may be rewritten in terms of the master volume as
\begin{equation}
R_{a}=-\frac{L^{4}}{(2\pi\ell_{s})^{4}g_{s}}\frac{\partial\mathcal{V}}{\partial\lambda_{a}}\,.\label{R-charge}
\end{equation}
Only $f-3$ of these are independent because of the relations \cite{Gauntlett:2018dpc}
\begin{equation}
\sum_{a=1}^{f}\vec{v}_{a}R_{a}=\vec{b}N\,.\label{values b2}
\end{equation}
Let us make a final remark that the central charges
and R-charges can be expressed in field theory data using the transformation
\begin{equation}
M_{a}=-{\eta}_g t[X_{a}]N\,,\label{relation field theory geometry}
\end{equation}
which was conjectured in \cite{Gauntlett:2018dpc}. In this relation, $t[X_{a}]$ are the charges of the bosonic fields $(X_{1},X_{2},X_{3},X_{4})=(Z,U_{2},Y,U_{1})$
of the quiver gauge theory with respect to the background gauge field \eqref{background}. In the remainder of this section we use this transformation to compare c-extremization results with their geometric dual and provide further evidence for this identification of parameters.

\subsection{Specifying to ${{Y_{5}=L^{a,b,c}}}$ \label{subsec:Specifying-to-}}

We now take $Y_{5}=L^{a,b,c}$ and compute the central charges and R-charges
using the formalism reviewed in the previous section. We show how these
results agree with the results of c-extremization on the field theory
side. The Sasaki--Einstein metrics $L^{a,b,c}$ were constructed in \cite{Cvetic:2005ft} and the inward pointing normal vectors are
given by
\begin{equation}
\vec{v}_{1}=(1,-al,c)\,,\qquad\vec{v}_{2}=(1,0,0)\,,\qquad\vec{v}_{3}=(1,1,0)\,,\qquad\vec{v}_{4}=(1,ak,b)\,.\label{normal vectors}
\end{equation}
For the computation we use the same steps and substitutions as in
\cite{Gauntlett:2018dpc} for the $Y^{p,q}$ case; in particular we separate the cases with genus $g\neq1$ and the case $g=1$. Using the normal vectors (\ref{normal vectors}) we have an explicit
expression for the volume function (\ref{volume function}) and thus for the constraint equation (\ref{constraint equation}), the fluxes (\ref{flux n}) and (\ref{fluxes}), the off-shell central charge (\ref{central charge}) and the R-charges (\ref{R-charge}), where we impose $b_{1}=2$ after taking the derivatives. In the following we give explicit expressions for the central charge and R-charges for values
of the parameters in which they simplify sufficiently and compare the results to the field theory computations shown in the previous sections. However, the interested reader can find the most general solutions to this extremization problem in a Mathematica notebook attached to the submission, and can compare to the field theory results to their hearts' content.

\subsubsection{Geometric dual of c-extremization for ${\Sigma_g=\mathrm{T}^2}$}\label{sec geom dual torus}

For the genus one case we define rescaled fluxes by
\begin{equation}
M_{a}\equiv m_{a}N\,,\label{rescaling}
\end{equation}
and compute the dictionary translating between geometric and field theory parameters using \eqref{relation field theory geometry}:
\begin{equation}
B=-\frac{m_{1}+kn_{3}}{2b}\,,\qquad f_{1}=\frac{n_{2}}{2}\,,\qquad f_{2}=\frac{n_{3}}{2}\,.\label{T2trafo}
\end{equation}
The latter relations will be useful once we wish to compare the results from this section to the ones from section \ref{cextrT2}.

\paragraph{$\bs{n_2=n_3=0}$: baryonic flux only.}We first solve the constraint equation (\ref{constraint equation}), the equation for $N$ (\ref{flux n})
and the equation for $M_{1}$ in (\ref{fluxes}) in terms of $A,$ $b_{3}$
and $\lambda_{4}.$ After solving the remaining equations in (\ref{fluxes}) for $m_{2},$ $m_{3}$ and $m_{4}$ we find
\begin{equation}
m_{2}=-\frac{(a+b-c)m_{1}}{b}\,,\qquad m_{3}=\frac{am_{1}}{b}\,,\qquad m_{4}=-\frac{cm_{1}}{b}\,.
\end{equation}
As a check one can easily see that (\ref{equations fluxes}) is satisfied. The off-shell central charge (\ref{central charge}) is then given
by
\begin{equation}
\mathscr{Z}=\frac{6c(a+b-c)\big[2a+b_{2}(b-c)\big]\big[-cb_{2}+a(-2+b_{2}+2ck)\big]m_{1}N^{2}}{a\big[a+b-bck+c(-2+ck)\big]^{2}}\,,
\end{equation}
which can be shown to match with the trial central charge even off-shell. Extremizing with respect to $b_{2}$ we find the central charge
\begin{equation}
c_{R}^{\text{geo}}=\frac{6ac(a+b-c)m_{1}N^{2}}{(a-c)(c-b)}\,.
\end{equation}
Using that for $n_2=n_3=0$ \eqref{T2trafo} boils down to $m_1=-2 b B, \, f_1=f_2=0$, we find perfect agreement with the field theory result \eqref{cT2bary}. The R-symmetry vector is given by $\vec{b}=(2,b_{2}^*,b_{3}^*)$
where
\begin{eqnarray}
b_{2}^* & = & \frac{a\big(a-c^{2}k+b(-1+ck)\big)}{(a-c)(c-b)}\,,\nonumber \\
b_{3}^{*} & = & -\frac{bc}{a-c}\,,
\end{eqnarray}
are the values of the R-symmetry vector components at the extremum of the trial central charge. For the R-charges of baryonic operators one then finds that they can be written as
\begin{eqnarray}\label{r-charges n2=n3=0}
R_1 &=& \frac{ab N}{(a-c)(b-c)}+b N \gamma \, , \qquad
R_2 = \frac{(a+b-c)cN}{(a-c)(c-b)}-(a+b-c) N \gamma \, ,\nn \\
R_3 &=& \frac{abN}{(a-c)(b-c)}+a N \gamma \, , \qquad
R_4 = \frac{(a+b-c)cN}{(a-c)(c-b)}-c N \gamma\, , 
\end{eqnarray}
where $\gamma$ is given by
\begin{equation}
\gamma=\frac{c }{(b-c)(c-a)}+\frac{L^4\big[b^2(\lambda_2-\lambda_1)+b\big( 2a \lambda_2+ c(\lambda_1-3\lambda_2) \big) +(a-c)\big(a(\lambda_2-\lambda_3)-2c \lambda_2 \big)\big]}{2 \pi g_s \ell_s^4 N a b c (a+b-c)}\,.
\end{equation}
As in \cite{Gauntlett:2018dpc} we can interpret $\gamma$ as parametrizing an undetermined transverse K\"ahler class which is not fixed by c-extremization. Note that one has to set $\gamma=0$ to identify the result \eqref{r-charges n2=n3=0} with the field theory c-extremization result \eqref{cextrbaryonic} upon identifying $R_a=R[X_a] N$. Setting $\gamma=0$ is also implied by taking the $n_2,$ $n_3\rightarrow 0$ limit of the R-charges in the next paragraph. See \cite{Gauntlett:2018dpc} for a more extensive discussion of the parameter $\gamma$ in the context of c-extremization for $Y^{p,q}$.

\paragraph{$\bs{n_2}$ and $\bs{n_3}$ not both zero.} In this case the constraint equation (\ref{constraint equation}) does not just depend on $b_2$ and $b_3$ but also on $\lambda_i$. Therefore we can follow a slightly different procedure than in the previous section and first solve for $A,$ $\lambda_{3}$ and $\lambda_{4}$ using the constraint equation (\ref{constraint equation}), the equation for $N$ (\ref{flux n}) and the equation for $M_{1}$ in (\ref{fluxes}). Subsequently we solve the remaining equations in (\ref{fluxes}) for $m_{2},$ $m_{3}$ and $m_{4}$. This results in 
\begin{eqnarray}
m_{2}&=&\frac{cm_{1}+b(n_{2}-m_{1})+n_{3}-a(m_{1}+kn_{3})}{b}\,, \nonumber \\
m_{3}&=&\frac{am_{1}-bn_{2}+akn_{3}}{b}\,, \\
m_{4}&=&-\frac{cm_{1}+n_{3}}{b}\,. \nonumber
\end{eqnarray}
Following this, one can extremize the off-shell central charge with respect to $b_{2}$ and $b_{3}.$ The resulting central charge and R-charges
are equal to the field theory result using the transformation (\ref{T2trafo}). For the sake of readability we only show the expressions for the central charge and R-charges in a special case again, namely we take $m_1=n_2=0$. The R-symmetry vector is given by $\vec b=(2, b_2^{*}, b_3^{*})$ with
\begin{eqnarray}
b_{2}^{*} & = & \tfrac{ak\big(a^{2}k+c(2-ck)+b(-1+ck)+a\big[-3+2ck-c^{2}k^{2}+bk(-1+ck)\big]\big)}{b(-1+ck)+c(2-ck)+a^{2}k^{2}\big[c(2-ck)+b(-1+ck)\big]+a\big[-1-2ck+c^{2}k^{2}+bk(1-ck)\big]}\,,\nonumber \\
b_{3}^{*} & = & \tfrac{b\big[-a+c(2-ck)+b(-1+ck)\big]}{b(-1+ck)+c(2-ck)+a^{2}k^{2}\big[c(2-ck)+b(-1+ck)\big]+a\big[-1-2ck+c^{2}k^{2}+bk(1-ck)\big]}\,.
\end{eqnarray}
The extremal value of the off-shell central charge is then given by
\begin{equation}
c_{R}^{\text{geo}}=\tfrac{3ak(-1+ak)\big[a+b-bck+c(-2+ck)\big]n_{3}N^{2}}{b(-1+ck)+c(2-ck)+a^{2}k^{2}\big[c(2-ck)+b(-1+ck)\big]+a\big[-1-2ck+c^{2}k^{2}+bk(1-ck)\big]}\, .
\end{equation}
Using the translation from geometric to field theory parameters \eqref{T2trafo} adapted to this case, we find $B=-\frac{k}{b} f_2$ with $f_2=\frac{n_3}{2}$ and $f_1=0$. This is precisely the class of twisted compactifications studied in section \ref{cextrT2} from the field theory perspective and again, we find perfect agreement between the field theory and geometric c-extremization results. Finally, the R-charges evaluated at the extremum are given by
\begin{eqnarray}
R_1& = & -\tfrac{abk(-1+ak)N}{b(-1+ck)+c(2-ck)+a^{2}k^{2}\big[c(2-ck)+b(-1+ck)\big]+a\big[-1-2ck+c^{2}k^{2}+bk(1-ck)\big]}\,,\nonumber \\
R_2& = & \tfrac{(-1+ak)\big[a+b-bck-ac^{2}k^{2}+ack(1+bk)+c(-2+ck)\big]N}{b(-1+ck)+c(2-ck)+a^{2}k^{2}\big[c(2-ck)+b(-1+ck)\big]+a\big[-1-2ck+c^{2}k^{2}+bk(1-ck)\big]}\,,\nonumber \\
R_3 & = & \tfrac{a^{2}k(1+bk-ck)(-1+ck) N}{b(-1+ck)+c(2-ck)+a^{2}k^{2}\big[c(2-ck)+b(-1+ck)\big]+a\big[-1-2ck+c^{2}k^{2}+bk(1-ck)\big]}\,, \\
R_4 & = & \tfrac{a^{2}ck^{2}+c(2-ck)+b(-1+ck)-a(1+ck)N}{b(-1+ck)+c(2-ck)+a^{2}k^{2}\big[c(2-ck)+b(-1+ck)\big]+a\big[-1-2ck+c^{2}k^{2}+bk(1-ck)\big]}\,,\nonumber
\end{eqnarray}
which are once more found to be in agreement with the field theory computation upon identifying $R_a=R[X_a]N$.

\subsubsection{Geometric dual of c-extremization for ${\Sigma_{g \neq 1}}$}

In this case we rescale the fluxes $M_{a}$ and flavour flux parameters $n_{2}$
and $n_{3}$ according to
\begin{equation}
M_{a}\equiv m_{a}(g-1)N\,,\qquad n_{2}\equiv s_{2}(g-1)\,,\qquad n_{3}\equiv s_{3}(g-1)\,.
\end{equation}
We again solve for $A,$ $\lambda_{3, 4}$ and $m_{2,3,4}$ using (\ref{constraint equation})--(\ref{fluxes}). We find
\begin{eqnarray}
m_{2}&=&\frac{cm_{1}+b(2-m_{1}+s_{2})+s_{3}-a(m_{1}+ks_{3})}{b}\,, \nonumber \\
m_{3}&=&\frac{a(m_{1}+ks_{3})-bs_{2}}{b}\,, \\
m_{4}&=&-\frac{cm_{1}+s_{3}}{b}\,. \nonumber
\end{eqnarray}
With the expressions for $A,$ $\lambda_{3, 4},$ $m_{2,3,4},$ the off-shell central charge (\ref{central charge})
becomes a function of $g,$ $N,$ $m_{1},$ $s_{2, 3},$ $b_{1, 2}$. The dependence on $\lambda_{1}$ and $\lambda_{2}$ drops out, since only two of the $\lambda_{i}$ are independent.
We then extremize the trial central charge with respect to $b_{2}$
and $b_{3}$ giving expressions for the central charge
(\ref{central charge}) and R-charges (\ref{R-charge}) for the most general flux configuration. Using the charge assignments in table \ref{quiverdata} and \eqref{relation field theory geometry} we can solve for $B$ and $f_{1, 2}$ in terms
of $m_{1}$ and $s_{2, 3}$. We obtain the equalities
\begin{equation}
B=\kappa\frac{m_{1}+ks_{3}+bk}{2b}\,,\qquad f_{1}=-\kappa\frac{s_{2}+ak}{2}\,,\qquad f_{2}=-\kappa\frac{b+s_{3}}{2}\label{Sigmagtrafo}
\end{equation}
for the dictionary between geometric and field theory data for the case $g\neq1$. Substituting these relations into the central charge and R-charges of the field theory, we find that the results of c-extremization and its geometric dual agree.

\paragraph*{$\bs{s_2=s_3=0}$.} We now focus on the case where the parameters $s_{2,3}$ are both taken to be zero. The R-symmetry vector $\vec{b}=(2,b_{2}^{*},b_{3}^{*})$, is
\begin{eqnarray}
b_{2}^{*} & = & \frac{a\big[a-c^{2}k+b(-1+ck)\big]m_{1}^{2}}{b^{2}(m_{1}-1)-b(a-c)(m_{1}-1)m_{1}+(a-c)cm_{1}^{2}}\,,\nonumber \\
b_{3}^{*} & = & \frac{b(b-c)cm_{1}^{2}}{b^{2}(m_{1}-1)-b(a-c)(m_{1}-1)m_{1}+(a-c)cm_{1}^{2}}\,.
\end{eqnarray}
The central charge is given by
\begin{eqnarray}
c_{R}^{\text{geo}} & = & \frac{6ac(g-1)m_{1}^{2}\big[b(m_{1}-1)+(a-c)m_{1}\big]N^{2}}{b^{2}(m_{1}-1)-b(a-c)(m_{1}-1)m_{1}+(a-c)cm_{1}^{2}}\,,\label{s2s3central}
\end{eqnarray}
and the R-charges are found to be
\begin{eqnarray}
R_{1}& = & \frac{bm_{1}(b-am_{1})N}{b^{2}(m_{1}-1)-b(a-c)(m_{1}-1)m_{1}+(a-c)cm_{1}^{2}}\,,\nonumber \\
R_{2} & = & \frac{\big[b^{2}(m_{1}-2)+b\big(a+c(m_{1}-1)\big)m_{1}+(a-c)cm_{1}^{2}\big]N}{b^{2}(m_{1}-1)-b(a-c)(m_{1}-1)m_{1}+(a-c)cm_{1}^{2}}\,,\nonumber \\
R_{3}& = & \frac{-ab(m_{1}-1)m_{1}N}{b^{2}(m_{1}-1)-b(a-c)(m_{1}-1)m_{1}+(a-c)cm_{1}^{2}}\,,\label{s2s3Rcharges}\\
R_{4}& = & \frac{-cm_{1}\big[b(1-m_{1})+(c-a)m_{1}\big]N}{b^{2}(m_{1}-1)-b(a-c)(m_{1}-1)m_{1}+(a-c)cm_{1}^{2}}\,.\nonumber 
\end{eqnarray}
From \eqref{Sigmagtrafo} we see that on the field theory side this corresponds to both non-vanishing baryonic flux $B=\kappa \frac{m_1+b k}{2b}$ and flavour fluxes $f_1=-\frac{\kappa}{2} a k$, $f_2=-\frac{\kappa}{2}b$. Plugging these relations into the expressions for the central charge \eqref{s2s3central} and R-charges \eqref{s2s3Rcharges} we recover the field theory results \eqref{s2s3centralfield} and \eqref{s2s3Rchargesfield} as expected.


\section{Explicit supergravity solutions}\label{geometry}

Having computed central charges and R-charges using c-extremization and its geometric dual, in this section we will turn our attention to computing these charges with explicit supergravity solutions. In section \ref{Holographic duals} we saw that these solutions are characterized by solutions to the master equation \eqref{Mastereq}. Our goal here is to find solutions that can be extended to be globally regular and whose central charges and R-charges can be matched to the results of the previous sections. In this endeavor we utilize an ansatz of cohomogeneity two which depends on four unknown functions. We argue that for our purposes we can constrain the specific form of two of the four functions without loss of generality and then solve the master equation \eqref{Mastereq} for the remaining two functions. Regularity of the solutions imposes additional conditions on these functions and as usual we impose flux quantization for the solutions to be well-defined string theory backgrounds. It transpires that it is possible to perform the integrals for the expressions for the fluxes, central charge and R-charges of these solutions without detailed knowledge of these two functions. This means we are not constrained to perform the analysis independently for all the different cases, but can obtain final expressions that simply require one to input the final solution. 
Finally using the explicit expressions of these functions, we compute the charges of the dual field theories and match them to expressions derived via c-extremization.

We begin by motivating the ansatz we use in section \ref{ansatz}. In section \ref{regularity} we then turn to examining the regularity of these metrics which imposes additional constraints on the solutions. Following this we impose flux quantization in section \ref{general expressions} and then in section \ref{sec: gen formula central charges} we derive general expressions for the central charges and R-charges. We solve the master equation \eqref{Mastereq} in section \ref{solving master eq} and subsequently in section \ref{subsec:matching} we solve the regularity conditions, compute the central charges and R-charges and match the expressions of the previous sections.

\subsection{Ansatz}\label{ansatz}

We have seen in section \ref{Holographic duals} that the geometry is completely determined by the choice of K\"ahler metric $\dd s^{2}(\mathcal{M}_{6})$ satisfying \eqref{Mastereq}. To proceed we must determine a suitable ansatz for the K\"ahler metric. We wish to describe the $L^{a,b,c}$ quiver theories compactified on a Riemann surface and therefore we demand that $Y_{7}$ is a fibration over a Riemann surface $Y_{5} \hookrightarrow Y_{7}\rightarrow \Sigma_{g}$, with $Y_{5}$ admitting a $U(1)^{3}$ isometry. One of the $U(1)$'s in $Y_{5}$ will be identified with the R-symmetry vector, leaving two $U(1)$'s in the ansatz for the K\"ahler base metric $\dd s^{2}(\mathcal{M}_{6})$.
For the K\"ahler metric we shall modify the orthotoric K\"ahler metric \cite{Apostolov:2001} used in \cite{Martelli:2005wy} to recover the $L^{a,b,c}$ metrics \cite{Cvetic:2005ft} to include a general fibration over a Riemann surface. The ansatz we use is
\begin{align}
\dd s^{2}(\mathcal M_6)= &~\mathcal{H}(\eta,\xi)\dd s^{2}(\Sigma_{g})+ \frac{\eta-\xi}{\mathcal{F(\xi)}}\dd \xi^{2}+\frac{\mathcal{F}(\xi)}{\eta-\xi}\Big(\dd \phi+\partial_{\xi}\mathcal{L}(\eta,\xi)\dd \psi + \partial_{\xi}\mathcal{H}(\eta,\xi) \mathcal{A}_{g}\Big)^{2}\nonumber\\
&+\frac{\eta-\xi}{\mathcal{G}(\eta)} \dd \eta^{2}+\frac{\mathcal{G}(\eta)}{\eta-\xi}\Big( \dd \phi+ \partial_{\eta}\mathcal{L}(\eta,\xi)\dd \psi+\partial_{\eta}\mathcal{H}(\eta,\xi)\mathcal{A}_{g}\Big)^{2}\label{ansatz}\,.
\end{align}
The metric $\dd s^{2}(\Sigma_{g})$ is the metric on a genus $g$ Riemann surface of constant curvature normalized to be $0,$ $\pm1$ depending on the genus.\footnote{For a Riemann surface with our normalization, the Ricci-form satisfies $\rho_{g}= R J_{g}=\kappa J_{g}$ where the last equality follows from our choice of normalization of the metric. For $g\neq1$ the Riemann-Roch theorem implies that the volume of the Riemann surface satisfies $\vol(\Sigma_{g})=4 \pi |g-1|$.}
The one-form $\mathcal{A}_{g}$ is the potential for the K\"ahler form of the Riemann surface and satisfies $\dd \mathcal{A}_{g}=J_{g}$. Explicit expressions are given by
\begin{equation}\label{tab:riemann surfaces metrics and}
\dd s^{2}(\Sigma_g)= \frac{1}{1-\kappa x^{2}} \dd x^{2} +(1- \kappa x^{2}) \dd y^{2} \, ,\qquad  \mathcal{A}_{g}=x \dd y\,.
\end{equation}
In the remainder of this section we shall analyze the ansatz that we have chosen. First note that the function $\mathcal{H}$ should be positive. When $\mathcal{L}=\eta \xi$ and $\mathcal{H}$ is constant, the metric \eqref{ansatz} is the direct product of the orthotoric metric and a Riemann surface. Lastly, observe that the ansatz \eqref{ansatz} admits some `gauge' transformations which leaves the internal manifold invariant. One can perform an overall rescaling of the K\"ahler metric by a constant parameter $\gamma^{2}$ whilst concurrently implementing the coordinate transformations
\be \label{gauge transf}
(\eta, \xi,\phi,\psi) \rightarrow (\alpha \eta +\beta, \alpha \xi + \beta, \frac{\phi}{\alpha \gamma^{2}},  \psi)~
\ee
and redefining
\begin{eqnarray}
\hat{\mathcal{F}}(\xi) &=& \frac{\mathcal{F}(\alpha \xi + \beta)}{\alpha^{3}\gamma^{2}} \, ,  \qquad \hat{\mathcal{H}}(\eta,\xi)= \gamma^{2}\mathcal{H}(\alpha \eta +\beta, \alpha \xi + \beta)\,, \nn \\
\hat{\mathcal{G}}(\eta) &=& \frac{\mathcal{G}(\alpha \eta + \beta)}{\alpha^{3}\gamma^{2}} \, , \qquad \, \, \hat{\mathcal{L}}(\eta,\xi)=\gamma^{2} \mathcal{L}(\alpha \eta +\beta, \alpha \xi + \beta)\,,
\end{eqnarray}
and preserve the form of the metric on the internal space. This symmetry will be used later to simplify the function $\mathcal{H}$ such that the master equation \eqref{Mastereq} becomes more tractable.

By construction the metric \eqref{ansatz} admits a closed two-form given by
\begin{equation}\label{kahler form}
J=\mathcal{H}J_{g}+\dd \xi \wedge\big(\dd\phi+\eta \dd \psi+\partial_\xi (\mathcal{H}) \mathcal{A}_g \big)+\dd \eta \wedge \big(\dd\phi+\xi \dd \psi+\partial_\eta (\mathcal{H}) \mathcal{A}_g\big)\, .
\end{equation}
However, it is not in general complex for arbitrary $\mathcal{H}$ and $\mathcal{L}$. We impose that the metric is complex in appendix \ref{App:ansatz-analysis} which is equivalent to a set of second order non-linear partial differential equations given by \eqref{Compcond}. We are unable to solve these conditions in general but if we additionally require that the universal twist solution\footnote{This is the standard topological twist solution when wrapping on $\Sigma_{g>1}$ and is obtained by taking a direct product of the orthotoric metric and the Riemann surface.} can be obtained from our ansatz, i.e. by fixing $\mathcal{L}$, we find that the most general solution is given by 
\be \label{solution H}
\mathcal{H}(\eta,\xi)= \alpha_{0}+\alpha_{1}(\eta + \xi) + \alpha_{2} \eta \xi \, ,\qquad \mathcal{L}(\eta,\xi)=\eta \xi\,,
\ee
with $\mathcal{F}$ and $\mathcal{G}$ still undetermined. For the time being we shall refrain from using the explicit solution for $\mathcal{H}$ in order to keep expressions as compact as possible.

Using the expression for $\mathcal{L}$ the Ricci scalar of the metric is easily computed to be
\be
R=\frac{2 \kappa(\xi-\eta)+ \mathcal{H}\big(\mathcal{F}''(\xi)+\mathcal{G}''(\eta)\big)+2\big(\mathcal{F}'(\xi)\partial_{\xi}\mathcal{H}+\mathcal{G}'(\eta) \partial_{\eta}\mathcal{H}\big)+ 3 \big(\mathcal{F}(\xi)\partial_{\xi}^{2} \mathcal{H}+\mathcal{G}(\eta)\partial_{\eta}^{2} \mathcal{H}\big)}{(\xi-\eta)\mathcal{H}}
\ee
whilst the Ricci-form connection is
\begin{align}\label{ricci-form connection}
P=\,&P_g -\frac{\partial_{\xi} (\mathcal{F}\mathcal{H})}{2(\eta-\xi) \mathcal{H}} \big(\dd \phi +\eta\dd \psi+\partial_{\xi} (\mathcal{H}) \mathcal{A}_{g}\big)-\frac{\partial_{\eta} (\mathcal{G} \mathcal{H})}{2(\eta-\xi) \mathcal{H}} \big(\dd \phi +\xi\dd \psi+\partial_{\eta} (\mathcal{H}) \mathcal{A}_{g}\big)\, .\nonumber\\
\end{align}
Here $P_{g}$ is the Ricci-form connection on the Riemann surface. The full ten-dimensional solution follows from using \eqref{metric susy sol}-\eqref{warp} and the explicit values of $\mathcal{F}$ and $\mathcal{G}$ given later.


\subsection{Regularity analysis} \label{regularity}

Before proceeding with determining the explicit solutions we shall present a preliminary analysis of the global regularity of the full seven-dimensional metric. By \eqref{warp} we are allowed to restrict to the seven-dimensional solution instead of the full ten-dimensional one if we also require that the Ricci scalar is strictly positive. As the remaining space is a fibration of the form $Y_{5}\hookrightarrow Y_{7}\rightarrow \Sigma_{g}$ we may first check the global regularity of $Y_{5}$ before checking the fibration over $\Sigma_{g}$. The metric on $Y_{5}$ is given by\pagebreak[4]
{\begin{align}
\dd s^{2}(Y_{5})=&\frac{1}{4}\left(\dd z -\frac{1}{2(\eta-\xi)}\bigg[\bigg(\mathcal{F}'(\xi)+ \frac{\mathcal{F}\partial_{\xi} \mathcal{H}}{\mathcal{H}}\bigg)(\dd \phi + \eta \dd \psi)+\bigg(\mathcal{G}'(\eta)+ \frac{\mathcal{G}(\eta)\partial_{\eta} \mathcal{H}}{\mathcal{H}}\bigg)(\dd \phi + \xi \dd \psi)\bigg]\right)^{2}\nonumber\\
&+\frac{R}{8}\left[\frac{\eta-\xi}{\mathcal{F(\xi)}}\dd \xi^{2}+\frac{\mathcal{F}(\xi)}{\eta-\xi}\big(\dd \phi+\eta\dd \psi \big)^{2}\right.+\left.\frac{\eta-\xi}{\mathcal{G}(\eta)} \dd \eta^{2}+\frac{\mathcal{G}(\eta)}{\eta-\xi}\big( \dd \phi+ \xi\dd \psi \big)^{2}\right]\,.
\end{align}}
To make the metric compact we must determine suitable ranges for the $\eta$ and $\xi$ coordinates, which are fixed by finding two roots of the functions $\mathcal{G}$ and $\mathcal{F}$ respectively. Moreover we require that $\eta-\xi$ is non-vanishing everywhere.\footnote{It may be possible to obtain a regular solution if the two ranges overlap at an endpoint. This however, will lead to a manifold with different topology to those where this is not the case. We shall therefore ignore this possibility and take $\eta\neq \xi$ everywhere.} 
 Without loss of generality we may take $\eta> \xi$ everywhere and consequently we require that both $\mathcal{F}$ and $\mathcal{G}$ are positive in their respective domains, i.e. they admit a local maximum between the two roots. The other case is simply a relabeling $\eta\leftrightarrow \xi$ and $\mathcal{G}\leftrightarrow -\mathcal{F}$. Moreover since we must impose that the function $\mathcal{H}$ is strictly positive we take without loss of generality that $0<\xi<\eta$, the other choice can be recovered by the coordinate transformation $(\eta ,\xi)\rightarrow (-\eta,-\xi)$.  We denote the roots by $\xi_{\pm}$ and $\eta_{\pm}$ for $\mathcal{F}$ and $\mathcal{G}$ respectively with the subscript denoting the larger (smaller) of the two roots. 
 
Consider now the metric at one of the roots. At each root the metric has a two-dimensional surface which degenerates. In the language of toric geometry this is equivalent to one of the $U(1)$'s shrinking on the facet of the polytope. This $U(1)$ corresponds to a specific Killing vector. Like in the classic example of a round S$^2$, requiring that the surface degenerates smoothly and locally looks like flat space imposes constraints on the period of the aforementioned Killing vector. The by now well established method for performing the regularity analysis, is to normalize the surface gravity 
\be
\kappa^{2}_{\rm grav} =\frac{\partial_{\mu}|V|^{2}\partial^{\mu}|V|^{2}}{4 |V|^{2}}\,,
\ee
of the degenerating Killing vector $V$ to be unity on the degeneration surface, see \cite{Cvetic:2005ft}. The metric on $Y_{5}$ has a $U(1)^3$ isometry group. The three commuting Killing vectors are $\partial_{z},~\partial_{\psi},~\partial_{\phi}$ and linear combinations of these three Killing vectors degenerate at the four degeneration surfaces. The four such Killing vectors located at the roots $\eta_{\pm}$ and $\xi_{\pm}$ are
\begin{align}
k_{\eta_\pm}&=\partial_{z}-\frac{2}{\mathcal{G}'_\pm}(\partial_{\psi} - \eta_{\pm} \partial_{\phi})\,,\nonumber\\
l_{\xi_\pm}&=\partial_{z}+\frac{2}{\mathcal{F}'_\pm}(\partial_{\psi}-\xi_{\pm}\partial_{\phi})\,,\label{Killing}
\end{align}
and correspond to translation generators with period $2 \pi$.
Here and in the following we denote the evaluation of the functions $\cF,$ $\cG$ at their respective roots with a $\pm$ subscript, e.g.~$\cF'_\pm\equiv \cF'(\xi_\pm)$. Clearly as we have four Killing vectors in a three-dimensional vector space there is a linear relation between them
\be
\ccc k_{\eta_-}+ \ddd k_{\eta_+} -\aaa l_{\xi_-}-\bbb l_{\xi_+}=0\,,
\ee
for relatively prime integers $\aaa,\bbb,\ccc,\ddd$. To avoid conical singularities on the intersection of two degeneration surfaces we must demand that $\aaa, \bbb$ are each coprime to each of $\ccc,\ddd$. The linear relation implies the three conditions
\begin{align}\label{regularity conditions}
\aaa+\bbb-\ccc-\ddd&=0\,,\nonumber\\
\frac{\aaa}{\mathcal{F}'_-}+\frac{\bbb}{\mathcal{F}'_+}+\frac{\ccc}{\mathcal{G}'_-}+\frac{\ddd}{\mathcal{G}'_+}&=0\,,\\
\frac{\aaa \xi_{-}}{\mathcal{F}'_-}+\frac{\bbb \xi_{+}}{\mathcal{F}'_+}+\frac{\ccc \eta_{-}}{\mathcal{G}'_-}+\frac{\ddd \eta_{+}}{\mathcal{G}'_+}&=0\,.\nonumber
\end{align}
The Killing vectors $\partial_{z},\partial_{\psi},\partial_{\phi}$ do not generate an effective torus action. Such a basis, $\partial_{\psi_{i}}$ with $2\pi$ period, is given by
\be \label{toric data}
\begin{pmatrix}
k_{\eta_+}\\
k_{\eta_-}\\
l_{\xi_+}\\
l_{\xi_-}
\end{pmatrix}
=
\begin{pmatrix}
1 & 0 & 0\\
1 &  \aaa \kkkk & \bbb\\
1 & -\aaa \llll & \ccc\\
1 & 1 & 0
\end{pmatrix}
\begin{pmatrix}
\partial_{\psi_{1}}\\
\partial_{\psi_{2}}\\
\partial_{\psi_{3}}
\end{pmatrix}
\,.
\ee
Here, by using B\'ezout's identity, we have introduced the integers $\kkkk,\llll$ satisfying
\be \label{bezout}
\bbb \llll +\ccc \kkkk=1\,.
\ee
The manifolds $Y_5$ are characterized by the three integers $a$, $b$, $c$. The toric data of the family of manifolds $Y_5$ studied in this section, which can be read off from \eqref{toric data}, coincides with the toric data of the $L^{a,b,c}$ Sasaki--Einstein metrics. Therefore we will, with slight abuse of notation, also refer in the following to the manifolds fibered over the Riemann surfaces as $Y_5=L^{a,b,c}$. Of course, at the moment we have not been able to impose any inequalities on the integers $\aaa,\bbb,\ccc,\ddd$ let alone state that they are positive, this can only be imposed with an explicit solution.


\subsection{Flux quantization} \label{general expressions}

In order for the class of supersymmetric solutions of type IIB supergravity to lead to well-defined string theory backgrounds we need to impose flux quantization. The only non-vanishing flux in the class of solutions we consider is the self-dual five-form flux \eqref{five-form}. Therefore we need to impose flux quantization over all five-cycles in the compact geometry $Y_7$. There are two classes of five-cycles to consider. The first is given by the fiber $Y_5$ for a fixed point on the Riemann surface. Since we consider string theory setups with D3-branes wrapped over a Riemann surface $\Sigma_g$ and probing a Calabi--Yau singularity, we can identify the number of flux quanta through $Y_5$ with the total number of D3-branes $N$. The second class consists of five-cycles $\Sigma_a$ that are given by three-cycles $S_a\subset Y_5$ fibered over the Riemann surface. In going to one of the degeneration surfaces considered above one necessarily finds the three-cycle $S_{a}$.
In order to impose flux quantization on the supergravity background we need to compute integrals of the five-form over these five-cycles. Once all supersymmetry conditions, regularity conditions and equations of motion are solved, this is of course a straightforward task. However, as stated before, we keep the discussion here as general as possible and will not assume any particular form of the functions $\mathcal F(\xi)$, $\mathcal G(\eta)$ and $\mathcal H(\eta, \xi)$ solving the master equation \eqref{Mastereq}. Remarkably, all the integrals can be performed without detailed knowledge of these functions. To perform the integrals we will only assume that the functions are solutions to the constraints \eqref{Compcond} and $\mathcal{L}=\eta \xi$.

\paragraph{Integration over $\boldsymbol{Y_5}$.} We start by integrating the five-form field strength over the five-cycle $Y_5$. The flux \eqref{flux quantisation} can be easily inferred from the general expression \eqref{five-form} for supersymmetric solutions. Defining
\begin{equation}
N=\frac{1}{(2\pi \ell_s)^4 g_s}\int_{Y_5}F_5 
\end{equation}
one then finds
\begin{align}
N=\frac{L^4}{(2\pi \ell_s)^4 g_s}\int_{Y_5} {\rm{d}}(z, \eta, \xi, \phi, \psi) \, \frac{1}{8 \mathcal H^2} \Big[ &\mathcal H^2 \big(\mathcal F''+\mathcal G''\big)-\mathcal F( \partial_\xi \mathcal H)^2-\mathcal G( \partial_\eta \mathcal H)^2\label{Nintegr1}\\
+&\mathcal H \big(\mathcal F' \partial_\xi \mathcal H+\mathcal F \partial^2_\xi \mathcal H+\mathcal G' \partial_\eta \mathcal H+\mathcal G \partial^2_\eta \mathcal H \big)\Big]\, ,\nonumber
\end{align}
where we used the shorthand notation $\dd (\omega_1, \dots, \omega_n)=\bigwedge_{i=1}^n \dd \omega_i$. The integrand in \eqref{Nintegr1} can be rewritten as a total derivative
\begin{align}
N&=\frac{L^4}{(2\pi \ell_s)^4 g_s}\int_{Y_5}{\rm{d}}(z, \eta, \xi, \phi, \psi) \, \frac{1}{8}\bigg[ \partial_{\xi}\partial_\eta \big(\eta \mathcal F'+\xi \mathcal G' \big)+\partial_\xi \Big(\mathcal F \frac{\partial_\xi \mathcal H}{\mathcal H} \Big)+\partial_\eta \Big(\mathcal G \frac{\partial_\eta \mathcal H}{\mathcal H} \Big)\bigg]\, ,\nonumber\\
&~\label{Nintegr2}
\end{align}
which we can now integrate explicitly. The last last two terms in \eqref{Nintegr2} vanish upon performing the $\eta$ and $\xi$ integration respectively, since $\mathcal F(\xi_{\pm})=\mathcal G(\eta_{\pm})=0$. For the first term in \eqref{Nintegr2} the $\eta,$ $\xi$ integrations are now trivial. To perform the angular integration we use three coordinates generating an effective action for the three-torus. These follow from the relation \eqref{toric data} which implies the coordinate transformation
\begin{align}
z&=\psi_{1} \, ,\nonumber\\
\psi&=-\frac{2}{\mathcal{G}'_+} \psi_{1} + 2 \lb \frac{1}{\mathcal{F}'_-}+\frac{1}{\mathcal{G}'_+}\rb \psi_{2} +2 \lb \frac{\kkkk}{\mathcal{F}'_+}-\frac{\llll}{\mathcal{G}'_-}+\frac{\llll+\kkkk}{\mathcal{G}'_+}\rb \psi_{3}\,,\label{effcoords}\\
\phi&= \frac{2 \eta_{+}}{\mathcal{G}'_+} \psi_{1} - 2 \lb \frac{\xi_{-}}{\mathcal{F}'_-}+\frac{\eta_{+}}{\mathcal{G}'_+}\rb \psi_{2} -2 \lb \frac{\kkkk \xi_{+}}{\mathcal{F}'_+}-\frac{\llll \eta_{-}}{\mathcal{G}'_-}+\frac{(\llll+\kkkk)\eta_{+}}{\mathcal{G}'_+}\rb \psi_{3}\,,\nonumber
\end{align}
with $\psi_{i}$ all having period $2\pi$.
The angular part is then equal to
\begin{equation}
\int \dd(z,\phi,\psi)=A_{J} \int \dd(\psi_1,\psi_2,\psi_3)=(2\pi)^3 A_{J}\,,
\end{equation}
where, using the regularity conditions \eqref{regularity conditions} and the relation \eqref{bezout} for $l$, we can write
\begin{align}
A_{J}=&\frac{4}{(a+b-c)c(\eta_+-\eta_-)(\cF'_- \cF'_+)^2} \Big(b(\eta_--\xi_+)(\xi_+-\eta_+)(\cF'_{-})^{2}+\big[-a(\eta_+-\xi_-)(\eta_--\xi_+) \nn \\
&+b(\eta_--\xi_-)(\eta_+-\xi_+)-c(\eta_--\eta_+)(\xi_--\xi_+)\big]\cF'_- \cF'_+-a(\eta_--\xi_-)(\xi_--\eta_+)(\cF'_+)^2\Big) \, .
\end{align}
Combining the angular part with the part coming from the integration over $\xi$ and $\eta$ results in
\begin{equation}
N=\frac{L^4}{16 \pi g_s \ell_s^4}A_J \big((\eta_- -\eta_+)(\cF'_--\cF'_+)+(\xi_--\xi_+)(\cG_--\cG_+)\big)\, .\label{Nfinal}
\end{equation}
Note that \eqref{Nfinal} is independent of the function $\cH$ and is therefore independent of the flavour twisting. 

\paragraph{Integration over $\boldsymbol{\Sigma_a}$.}
We now turn to the five-cycles $\Sigma_a$ which are fibrations of three-cycles $S_a \subset Y_5$ over the Riemann surface. These three-cycles are the degeneration surfaces of the Killing vectors \eqref{Killing} and we denote them by $(S_{\xi_+},S_{\eta_+},S_{\xi_-},S_{\eta_-})=(S_1,S_2,S_3,S_4)$. Thus we have four additional five-cycles $\Sigma_a \in H_5(Y_7, \mathbb{Z})$. However, as can be seen from \eqref{equations fluxes}, these five-cycles are in fact not independent. Due to these relations there is only one independent flux quantum number in this class. Nevertheless, we proceed in the following by performing the integrals of the five-form field strength over all five-cycles $\Sigma_{\xi_\pm}$ and $\Sigma_{\eta_\pm}$. 

We first perform the integral over $\Sigma_{\eta_+}$, for which we have to integrate over $\xi$, the Riemann surface and two angular coordinates. From \eqref{toric data} we see that the coordinate $\psi_1$ in \eqref{effcoords} corresponds to the degenerating Killing vector $k_{\eta_+}$. Therefore, $\psi_2$ and $\psi_3$ will give an effective action of the two-torus that is obtained by contracting the one-cycle of the degenerating Killing vector of the toric $T^{3}$. 
It turns out that the integral of the field strength can be written as
\begin{align}
M_{\eta_+}=&\frac{L^4}{(2\pi \ell_s)^4 g_s} \frac{\cG'_+ A_{J} }{16}\int_{\Sigma_{\eta_+}} \!\!\!J_g \wedge \text{d}(\xi, \psi_2, \psi_3) \Bigg[ \partial_\xi \big(2\kappa \xi \big)-\partial_\xi \bigg(\cH(\eta_+, \xi)\frac{\cF'+\cG'_+}{\eta_+-\xi} \bigg)\label{Metap1}\\
&-\partial_\xi \bigg(\partial_\xi \cH(\eta_+, \xi)\frac{\cF}{\eta_+-\xi} \bigg)-\frac{2 \cF \partial^2_\xi \cH(\eta_+, \xi)-\cG'_+ \big[\partial_\xi \cH(\eta_+, \xi)-\partial_\eta \cH(\eta_+, \xi) \big]}{\eta_+-\xi}\Bigg]\, .\nonumber
\end{align}
The last piece in \eqref{Metap1} seems to spoil the possibility to integrate this equation completely, as it is not manifestly written as a total derivative. We can now utilize the first equation in \eqref{Compcond} and evaluate it at $\eta=\eta_+$ from which we find that $\partial^2_\xi \cH(\eta_+, \xi)=0$. The third equation of \eqref{Compcond} evaluated at $\eta=\eta_+$ allows us to trade the difference of the $\xi$ and $\eta$ derivatives with a total derivative $\partial_\xi \partial_\eta \cH(\eta_+, \xi)$. We have therefore shown that, independent of the details of the functions in the metric ansatz, we can write the integrand once again as a total derivative and we can integrate it explicitly. We obtain
\begin{align}
M_{\eta_+}=M_2=\frac{L^4}{(2\pi)^2 g_s \,\ell_s^4}&\frac{\cG'_+ A_{J} }{16} \text{vol}(\Sigma_g)\Big(2\kappa (\xi_+-\xi_-)+\frac{(\cF'_-+\cG'_+)\cH(\eta_+, \xi_-)}{\eta_+-\xi_-}\label{Metap}\\
&-\frac{(\cF'_++\cG'_+)\cH(\eta_+, \xi_+)}{\eta_+-\xi_+}+\cG'_+\big(\partial_\eta \cH(\eta_+, \xi_+)-\partial_\eta \cH(\eta_+, \xi_-)\big) \Big)\, .\nonumber
\end{align} 
Similarly we can integrate the five-form over the remaining three five-cycles $\Sigma_{\eta_-}$ and $\Sigma_{\xi_\pm}$. Instead of giving the full details about the integration we only give the final results for the remaining fluxes. Note that for each integration one has to find a coordinate system as in \eqref{effcoords} to perform the integration, such that $\partial_{\psi_1}$ is equal to the Killing vector that degenerates at the relevant degeneration surface. In appendix \ref{app: eff coordinate systems} we provide the coordinate systems that we use for the different degeneration surfaces. The value of the flux $M_{\eta_-}$ threading through the five-cycle defined by the degeneration surface $\eta=\eta_-$ is given by
\begin{align}
M_{\eta_-}=M_4=-\frac{L^4}{(2\pi)^2 g_s\, \ell_s^4}&\frac{\cG'_- A_{J} }{16} \text{vol}(\Sigma_g)\Big(2 \kappa  (\xi _+-\xi_-)+\frac{(\cF'_-+\cG'_-) \cH (\eta _-,\xi _-)}{\eta_--\xi _-}\label{Metam}\\
&-\frac{(\cF'_++\cG'_-) \cH (\eta_-,\xi _+)}{\eta _--\xi _+}+\cG'_- \big(\partial_\eta\cH (\eta _-,\xi _+)-\partial_\eta\cH (\eta _-,\xi _-)\big) \Big)\, .\nonumber
\end{align} 
The fluxes associated with the $\xi_+$ and $\xi_-$ degeneration surfaces are given by
\begin{align}
M_{\xi_+}=M_1=\frac{L^4}{(2\pi)^2 g_s\, \ell_s^4}&\frac{\cF'_+ A_{J} }{16} \text{vol}(\Sigma_g)\Big(2 \kappa (\eta _+-\eta _-)+\frac{(\cF'_++\cG'_-) \cH (\eta _-,\xi _+)}{\eta _--\xi _+}\label{Mxip}\\
&-\frac{(\cF'_++\cG'_+) \cH (\eta_+,\xi _+)}{\eta _+-\xi _+}+\cF'_+ (\partial_\xi \cH (\eta _-,\xi _+)-\partial_\xi \cH (\eta _+,\xi _+)\big) \Big)\, \nonumber
\end{align} 
and 
\begin{align}
M_{\xi_-}=M_3=-\frac{L^4}{(2\pi)^2 g_s\, \ell_s^4}&\frac{\cF'_- A_{J} }{16} \text{vol}(\Sigma_g)\Big(2 \kappa (\eta _+-\eta _-)+\frac{(\cF'_-+\cG'_-) \cH (\eta _-,\xi _-)}{\eta _--\xi _-}\label{Mxim} \\
&-\frac{(\cF'_-+\cG'_+) \cH (\eta_+,\xi _-)}{\eta _+-\xi _-}+\cF'_- \big(\partial_\xi \cH (\eta _-,\xi _-)-\partial_\xi \cH (\eta _+,\xi _-)\big) \Big)\, \nonumber
\end{align}
respectively.

\subsection{General formula for central charge and R-charges}\label{sec: gen formula central charges}
We now continue the general discussion of the AdS$_3$ solutions of interest with the computation of the central charge and R-charges of the dual 2d SCFT. To leading order in the large $N$ limit the central charge can be computed by evaluating \eqref{trial central charge} for the solution. Again we assume no additional form of the functions in the metric ansatz. Using the explicit expression \eqref{kahler form} for $J$, and the expression \eqref{ricci-form connection} for $P$ to compute $\eta_R$ and $\rho$, one obtains the central charge
\begin{align}
c_{\text{sugra}}=\frac{3 L^8}{64 \pi ^6 g_s^2 \ell_s^8}\frac{1}{4}\int_{Y_7}&\dd (z, \eta, \xi, \phi, \psi)\wedge J_g \Big( \cH(\eta ,\xi ) \left(\cF''(\xi )+\cG''(\eta )\right)+2 \cF'(\xi ) \partial_\xi\cH(\eta ,\xi )\nonumber\\
+&2 \kappa  (\xi- \eta)+3 \cF(\xi )\partial^2_\xi \cH(\eta ,\xi )+2 \cG'(\eta )\partial_\eta \cH(\eta ,\xi )+3 \cG(\eta ) \partial_\eta^2\cH(\eta
   ,\xi ) \Big)\, .\nonumber\\
   & \label{cchargebare}
\end{align}
Remarkably, it is again possible to rewrite \eqref{cchargebare} as a total derivative,
\begin{align}
c_{\text{sugra}}=&\frac{3 L^8}{64 \pi ^6 g_s^2 \ell_s^8}\frac{1}{4}\int_{Y_7} \dd (z, \eta, \xi, \phi, \psi)\wedge J_g\, \partial_\eta \partial_\xi \bigg[\frac{1}{2} \Big(\eta  \cF'(\xi ) \left(2 \cH(\eta ,\xi )-\eta  \partial_\eta\cH(\eta ,\xi )\right)\nonumber\\
&+\eta  \cF(\xi ) \left(2 \partial_\xi\cH(\eta ,\xi )-\eta  \partial_\eta \partial_\xi\cH(\eta ,\xi )\right)+\xi 
   \cG'(\eta ) \left(2 \cH(\eta ,\xi )-\xi  \partial_\xi\cH(\eta ,\xi )\right)\\
   &+\xi  \cG(\eta ) \left(2 \partial_\eta\cH(\eta ,\xi )-\xi  \partial_\eta \partial_\xi\cH(\eta ,\xi )\right)\Big)+\kappa  \left(\eta 
   \xi ^2-\eta ^2 \xi \right)\bigg]\, ,\nonumber
\end{align}
up to terms which vanish upon evaluation on the boundaries of the $(\eta, \xi)$-integration. Integrating the latter equation is now straightforward and leads to the lengthy but fully general expression for the central charge
\begin{align} 
c_{\text{sugra}}=&\frac{3 L^8}{(2\pi)^3 g_s^2 \ell_s^8}\frac{A_J}{8}\text{vol}(\Sigma_g) \bigg( 2 \left(\eta _- \cF'_-+\xi _- \cG'_-\right)\cH\left(\eta _-,\xi _-\right)-2 \left(\eta _- \cF'_++\xi _+
   \cG'_-\right)\cH\left(\eta _-,\xi _+\right) \nonumber\\[0.3cm]
&-2 \left(\eta _+ \cF'_-+\xi _- \cG'_+\right)\cH\left(\eta _+,\xi _-\right)+2 \left(\eta _+ \cF'_++\xi _+ \cG'_+\right) \cH\left(\eta _+,\xi _+\right)-\eta _-^2 \cF'_-\partial_\eta \cH\left(\eta
   _-,\xi _-\right)\label{centralgeneral}\nonumber \\[0.5cm]
   &+\eta _-^2 \cF'_+ \partial_\eta\cH\left(\eta _-,\xi _+\right)+\eta _+^2 \cF'_- \partial_\eta\cH\left(\eta _+,\xi _-\right)-\eta _+^2
   \cF'_+ \partial_\eta\cH\left(\eta _+,\xi _+\right)-\xi _-^2 \cG'_- \partial_\xi\cH\left(\eta _-,\xi
   _-\right)\nonumber\\[0.5cm]
   &+\xi _+^2 \cG'_- \partial_\xi\cH\left(\eta _-,\xi _+\right)+\xi _-^2 \cG'_+ \partial_\xi\cH\left(\eta _+,\xi _-\right)-\xi _+^2 \cG'_+
   \partial_\xi\cH\left(\eta _+,\xi _+\right)\\[0.3cm]
   &-2 \kappa \left(\eta _--\eta _+\right)  \left(\xi _--\xi _+\right) \left(\eta
   _-+\eta _+-\xi _--\xi _+\right) \bigg)\, .\nonumber
\end{align}
The central charge of the dual 2d SCFT is thus expressed in terms of the functions $\cF,$ $\cG,$ $\cH$ and their derivatives evaluated at the roots. Let us now compute the R-charges of the baryonic operators which are given by the integrals \eqref{Rcharges general} evaluated on the three-cycles $S_{\xi_\pm}$ and $S_{\eta_\pm}$. For these integrals we again use the coordinate system relevant for the particular degeneration surface. As opposed to the flux integrals and the central charge, where rewriting the integrands as total derivatives was non-trivial, the integrals for the R-charges are considerably easier to perform. For all four three-cycles $S_a$ the corresponding integrands are simply constant. The R-charges are given by:
\pagebreak[4]
\begin{align}
R_{\eta_+}=R_2&=-\frac{1}{\cF'_+ \cG'_- }\frac{L^4 }{2 \pi  a {g_s} \ell_s^4}\left(\xi _+-\xi _-\right)  \left[\left(\eta _+-\eta _-\right) \cF'_++\left(\eta _+-\xi _+\right) \cG'_-+\left(\xi _+-\eta
   _-\right) \cG'_+\right]\, ,\nonumber\\
R_{\eta_-}=R_4&=\frac{1}{\cF'_+ \cG'_+}\,\frac{L^4 }{2 \pi  a g_s \ell_s^4 }\left(\xi _+-\xi _-\right) \left[\left(\eta _+-\eta _-\right) \cF'_++\left(\eta _+-\xi _+\right) \cG'_-+\left(\xi _+-\eta
   _-\right) \cG'_+\right]\, ,\label{Rchargesgravresult} \nn \\
R_{\xi_+}=R_1&=-\frac{1}{\cG'_- \cG'_+}\frac{L^4}{2 \pi  a {g_s} \ell_s^4}\left(\eta _+-\eta _-\right) \left[\left(\eta _+-\eta _-\right) \cF'_++\left(\eta _+-\xi _+\right) \cG'_-+\left(\xi
   _+-\eta _-\right) \cG'_+\right]\, ,\nn \\
   R_{\xi_-}=R_3&=\frac{\cF'_-}{\cF'_+ \cG'_- \cG'_+}\frac{L^4 }{2 \pi  a {g_s} \ell_s^4}\left(\eta _+-\eta _-\right)  \left[\left(\eta _+-\eta _-\right) \cF'_++\left(\eta _+-\xi _+\right) \cG'_-+\left(\xi _+-\eta
   _-\right) \cG'_+\right]\, .
\end{align}
For writing the R-charges in this form we used the regularity conditions \eqref{regularity conditions}. If we solve \eqref{Nfinal} for the string coupling it is not hard to show that the first component of the sum \eqref{values b2} is indeed satisfied.\footnote{Recall that $b_1=2$ is fixed.} From the remaining two components of \eqref{values b2} we can determine the (off-shell) R-symmetry vector $\vec b$.

\subsection{Solving the master equation}\label{solving master eq}

In this section we solve the master equation (\ref{Mastereq}) in
order to determine explicit $\text{AdS}_{3}$ solutions. We will use
ans\"atze for the functions $\mathcal{L},$ $\mathcal{H},$ $\mathcal{F}$
and $\mathcal{G}.$ As previously stated we take $\mathcal{L}=\eta\xi$ in order to recover the universal twist solution \cite{Gauntlett:2006qw} and this fixes
$\mathcal{H}$ to be
\begin{equation}
\mathcal{H}=\alpha_{0}+\alpha_{1}(\eta+\xi)+\alpha_{2}\eta\xi\,.
\end{equation}
However, some of the parameters are redundant and can be removed by using the symmetries \eqref{gauge transf}
of the metric \eqref{ansatz}. In writing the simplest form of $\mathcal{H}$ we have to
distinguish between different cases. Firstly, when $\alpha_{2}\neq0$ we use
the symmetry \eqref{gauge transf} to set $\alpha_{1}=0$ and $\alpha_{2}=1$ by taking
\begin{equation}
\alpha=\frac{1}{\sqrt{|\alpha_{2}|}}\,,\qquad \beta=-\frac{\alpha_{1}}{\alpha_{2}}\,,\qquad \gamma =1\, .
\end{equation}
Therefore, we can take
\begin{equation}
\mathcal{H}=\alpha_{0}+\eta\xi\,.\label{a3 is not 0}
\end{equation}
When $\alpha_{2}=0$ and $\alpha_{1}\neq0$ we can use similar transformations
to write $\mathcal{H}$ in the form 
\begin{equation}
\mathcal{H}=\eta+\xi\,.\label{a3=00003D0}
\end{equation}
The last case is when $\alpha_{1}=\alpha_{2}=0,$ for which we have 
\begin{equation}
\mathcal{H}=\alpha_{0}\,.
\end{equation}
Note that the metric \eqref{ansatz} in this last case reduces to a direct product of the
Riemann surface and the orthotoric metric $\text{d}s^{2}(\mathcal{M}_{4})$,
\begin{equation}
\text{d}s^{2}=\text{d}s^{2}(\Sigma_{g})+\text{d}s^{2}(\mathcal{M}_{4})\,,
\end{equation}
which corresponds to the case of no twisting with the flavour symmetries.
Since this case has been considered in the literature for $\mathrm{T}^2$ in \cite{Couzens:2018wnk} and $\Sigma_{g>1}$ in \cite{Gauntlett:2006qw} (the S$^2$ case does not exist), we will not consider it further in this paper.

For the functions $\mathcal{F}$ and $\mathcal{G}$ we take the ans\"atze
\begin{equation}
\mathcal{F}(\xi)=\sum_{i=-5}^{5}\tilde f_{i}\xi^{i}\,,\qquad \qquad \mathcal{G}(\eta)=\sum_{i=-5}^{5}\tilde g_{i}\eta^{i}\,.
\end{equation}
Using the expression for $\mathcal{H}$ above, we solve the master equation
to determine the coefficients $\tilde f_{i},$ $\tilde g_{i}$ and the
remaining coefficients in $\mathcal{H}.$ The case (\ref{a3=00003D0}) with $\alpha_{2}=0$ turns out to never give a sensible globally defined compact solution and therefore we restrict ourselves solely to the
case $\alpha_{2}\neq0$ for which $\mathcal{H}$ is given by (\ref{a3 is not 0}).
To solve the master equation we use the expression  \eqref{tab:riemann surfaces metrics and} for $\mathcal{A}_{g}$ and the metric $\text{d}s^{2}(\Sigma_{g})$ to treat all Riemann surfaces universally. Solving the master
equation with $\mathcal{H}$ given by (\ref{a3 is not 0}) and our polynomial ans\"atze for $\mathcal{F}$ and $\mathcal{G}$ implies $\alpha_{0}=0$ and after a suitable redefinition of parameters the functions $\mathcal{F}$ and $\mathcal{G}$ are given by
\begin{equation}
\mathcal{F}(\xi)=\frac{-A(\xi+C)^{2}+\kappa\xi^{2}+D\xi^{3}}{\xi}\,,\qquad\mathcal{G}(\eta)=\frac{-B(\eta+C)^{2}-\kappa\eta^{2}-D\eta^{3}}{\eta}\,,\label{functions f and g}
\end{equation}
where $\kappa=1,$ $0,$ $-1$ for $\Sigma_{g}=\text{S}^{2},$
$\mathrm{T}^{2},$ $\Sigma_{g>1}$ respectively. The functions $\mathcal{F},$
$\mathcal{G}$ and $\mathcal{H}$ are thus determined by four parameters
$A,$ $B,$ $C$ and $D$.\footnote{Note that here the coefficients $A$ and $B$ do not correspond to the variables in the previous sections where they correspond to respectively the K\"ahler class of the Riemann surface and the baryonic flux.} Recall that we have not used up all of the gauge symmetry and we may use the remainder to set $D=\pm 1$.  These functions need to satisfy the
regularity conditions \eqref{regularity conditions} which allows us to solve for two of the remaining three
parameters in terms of the other one and $a,$ $b$ and $c.$ The
final solution thus depends on four parameters. Consider now the central charge \eqref{centralgeneral}. The coefficient $L^8/(g_s^2 \ell_s^8)$ can be expressed in terms of $N^2$ using \eqref{Nfinal}. The central charge, and also the R-charges \eqref{Rchargesgravresult}, then depend on five parameters instead of the seven
parameters $a,$ $b,$ $c,$ $m_{1},$ $N,$ $n_{2}$ and $n_{3}$
that we find on the c-extremization side. Ergo, we have not found
the most general class of solutions. We have motivated that the ansatz for $\mathcal{H}$ is the most general that we could take, which implies that the ans\"atze for $\mathcal{F}$ and
$\mathcal{G}$ should be modified to account for the extra parameters. Note
that the functions (\ref{functions f and g}) have three roots which need to be real. Their domains, given by $[\xi_{-},\xi_{+}]$ and
$[\eta_{-},\eta_{+}]$ respectively, should be such that $0<\xi_{-}<\xi_{+}<\eta_{-}<\eta_{+}$, as assumed without loss of generality in the regularity analysis.
Lastly, the functions should be positive on their respective domains given the above choice.
These conditions translate into restrictions on $A,$ $B,$ and $C$, which we examine in more detail in the next subsection.

\subsection{Solving the regularity equations and matching to the field theory}\label{subsec:matching}

So far we have derived the local form of solutions satisfying \eqref{Mastereq}, in this section we extend these local solutions to globally well-defined ones by imposing the regularity conditions \eqref{regularity conditions}. With the global solutions in hand we can compute the central charges and R-charges of the solutions and compare to the field theory results, finding perfect agreement. We perform this matching numerically for a large set of solutions, a few of which we present for exposition.

To solve the regularity conditions we use that the functions $\mathcal{F}$ and $\mathcal{G}$ (\ref{functions f and g}) can be rewritten in terms of their roots as 
\begin{equation}
\mathcal{F}(\xi)=\frac{D(\xi-\xi_{-})(\xi-\xi_{+})(\xi-\xi_{*})}{\xi}\,,\qquad\mathcal{G}(\eta)=-\frac{D(\eta-\eta_{-})(\eta-\eta_{+})(\eta-\eta_{*})}{\eta}\,.\label{torus fibration in terms of roots}
\end{equation}
Here we have denoted the third root of $\mathcal{F}(\xi)$ and $\mathcal{G}(\eta)$ by $\xi_{*}$ and $\eta_{*}$ respectively.
The derivatives of the functions evaluated at $\xi_{\pm},$ $\eta_{\pm}$ are given by
\begin{eqnarray}
\mathcal{F}'_- & = & \frac{D(\xi_--\xi_{+})(\xi_--\xi_{*})}{\xi_{-}}\,,\qquad  \mathcal{G}'_- = -\frac{D(\eta_--\eta_{+})(\eta_--\eta_{*})}{\eta_{-}} \,,\nonumber \\
\mathcal{F}'_+&=&\frac{D(\xi_+-\xi_{-})(\xi_+-\xi_{*})}{\xi_{+}}\,,\qquad\mathcal{G}'_+=-\frac{D(\eta_+-\eta_{-})(\eta_+-\eta_{*})}{\eta_{+}}\,.\label{derivatives}
\end{eqnarray}
With these expressions the fluxes \eqref{Metap}--\eqref{Mxim}, central charge \eqref{centralgeneral} and R-charges \eqref{Rchargesgravresult} can
be expressed in terms of the six roots and $D$\footnote{Recall, this may be normalized to $\pm1$.}. However, these parameters are not independent. Equating (\ref{functions f and g}) and (\ref{torus fibration in terms of roots}) results in the conditions
\begin{eqnarray}
\xi_{-}\xi_{+}\xi_{*} & = & \frac{AC^{2}}{D}\,,\qquad \, \xi_{-}\xi_{+}+\xi_{-}\xi_{*}+\xi_{+}\xi_{*} =-\frac{2AC}{D}\,,\quad \; \, \xi_{-}+\xi_{+}+\xi_{*}=\frac{A-\kappa}{D}\,,\label{conditions equation functions}\\
\eta_{-}\eta_{+}\eta_{*} & = & -\frac{BC^{2}}{D}\,,\quad\eta_{-}\eta_{+}+\eta_{-}\eta_{*}+\eta_{+}\eta_{*}=\frac{2BC}{D}\,,\qquad\eta_{-}+\eta_{+}+\eta_{*}=-\frac{B+\kappa}{D}\,.\nonumber 
\end{eqnarray}
Furthermore, the substitution of (\ref{derivatives}) into the regularity
conditions \eqref{regularity conditions} results in two further conditions
\begin{eqnarray}\label{reg condtions 2}
a\xi_{-}(\xi_{+}-\xi_{*})(\xi_{-}-\eta_{-})(\eta_{*}-\eta_{+}) & = & (\xi_{-}-\xi_{*})\big[b\xi_{+}(\xi_{+}-\eta_{-})(\eta_{*}-\eta_{+})-d\eta_{+}(\xi_{-}-\xi_{+})(\xi_{+}-\xi_{*})\big]\,,\nonumber \\
a\xi_{-}(\xi_{+}-\xi_{*})(\xi_{-}-\eta_{+})(\eta_{*}-\eta_{-}) & = & (\xi_{-}-\xi_{*})\big[b\xi_{+}(\xi_{+}-\eta_{+})(\eta_{*}-\eta_{-})-c\eta_{-}(\xi_{-}-\xi_{+})(\xi_{+}-\xi_{*})\big]\,. \nonumber \\
&&
\end{eqnarray}
We thus have eight equations for ten parameters which we will solve for $A,$
$B$ and the roots. We first study $\kappa=0$ since it is the easiest case before turning our attention to the $\kappa\neq0$ cases.


\subsubsection{Torus fibrations} \label{torus fibrations}

\begin{figure}[b]
\begin{centering}
\includegraphics[scale=0.5]{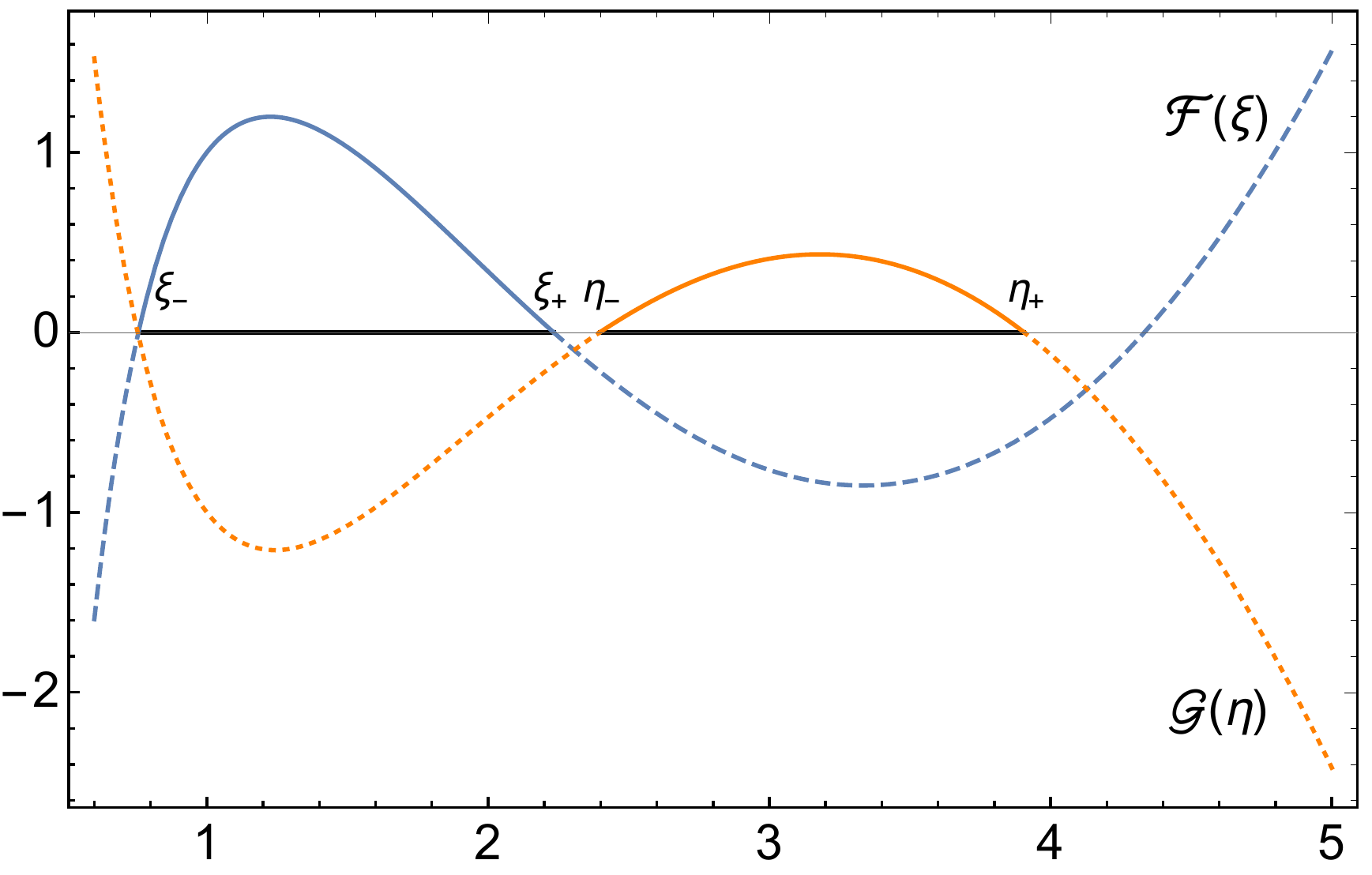}
\par\end{centering}
\caption{\label{fig:Plot-of-}Plot of $\mathcal{F}(\xi)$ and $\mathcal{G}(\eta)$ for $a=-1,$
$b=19,$ $c=16,$ $C=-1$ and $D=1$. The domain of $\mathcal{F}$ is given
by $[\xi_{-},\xi_{+}]$ and the domain of $\mathcal{G}$ by $[\eta_{-},\eta_{+}]$,
this is drawn using the solid line. Outside of their
domains $\mathcal{F}$ is given by the dashed blue line and $\mathcal{G}$ by the dotted
orange line.}

\end{figure}%

There are two choices that one can make, i.e.~one can take $D=\pm1$. It transpires that the $D=-1$ solution is incompatible with our choice of $0<\xi<\eta$ and therefore we take $D=1$ throughout this section. Since $D=1$ we have that the auxiliary root $\xi_{*}$ satisfies $\xi_{*}>\xi_+>\xi_{-}>0$ and therefore \eqref{conditions equation functions} implies firstly that $A>0$ and secondly that $C<0$. Moreover the positivity of the discriminant of $\xi \mathcal{F}(\xi)$ and $\eta \mathcal{G}(\eta)$ implies
\be
A>-\frac{27}{4}C>0\, , \qquad B<\frac{27}{4} C <0\,.
\ee
It then follows from \eqref{conditions equation functions} that $\eta_{*}>0$. To aid the reader we have plotted a representative example of the functions $\mathcal{F}$ and $\mathcal{G}$ and their roots in figure \ref{fig:Plot-of-}.

It is convenient to perform the following redefinition of the auxiliary roots:
\be\label{roots COC}
\eta_{*}=-\frac{C}{4}(v^{2}+3)\, ,\qquad \xi_{*}= -\frac{3C}{4} \Big( \frac{3}{u^{2}}+1\Big)\, .
\ee
After this change of coordinates one can solve \eqref{conditions equation functions} and find that the physical roots are located at
\be
\xi_{\pm}=-\frac{3C(3+ u^{2}) }{(u\mp 3)^{2}}\, ,\qquad \eta_{\pm}=-\frac{C(3+v^{2})}{(v\mp 1)^{2}}\, ,
\ee
whilst the constants $A$ and $B$ take the form
\be
A=-\frac{27 C (3+u^{2})^{3}}{4 u^{2}(u^{2}-9)^{2}}\, ,\qquad B= \frac{C (3+v^{2})^{3}}{4(v^{2}-1)^{2}}\, .
\ee
The domains of $u$ and $v$ are fixed by imposing $0<\xi_{-}<\xi_{+}<\eta_{-}<\eta_{+}$ and $\xi_{+}<\xi_{*}$, giving
\be \label{domainuv}
0<v<1\, ,\qquad 0< u< \frac{3(1-v)}{v+3}\, .
\ee
The only constraint on the constant $C$ is that it must be strictly negative, i.e. $C<0$. One can also see that these conditions are consistent with the Ricci scalar being strictly positive definite in the domain as is required by supersymmetry. 

One should now solve the regularity conditions to obtain expressions for $u$ and $v$ in terms of the integers $a,$ $b,$ $c$. Expressed in terms of $u$ and $v$, the conditions are given by 
\be\label{abtorus}
a=-b \frac{(1+u)((u+3)^{2}v^{2}-9(1-u)^{2})}{(1-u)((u-3)^{2}v^{2}-9(1+u)^{2})}\,  ,\qquad c= b \frac{u (3+v)(u (v-3)+3 (1+v))}{3 (1-u)(3(1-v)+u(3+v))}\, ,
\ee
where we have also solved them for the integers $a$ and $c$. The solution of these equations that expresses $u$ and $v$ as function of $a,$ $b$ and $c$ is a bit unwieldy. The variable $u$ is a root of the quintic
\begin{align}
Z[X]=&- b c(b-c)(X-1)\big(3+4 X +X^2\big)^2 +a^2 (1-X)^2 \big(8 b X^2 +c(3-X)^2(1+X)\big)+\nonumber\\
&a (1+X)\big[8 b^2 X^2 (1+X)-c^2 (3-4 X + X^2)^2 -2 b c \big(-9 +9 X -7 X^2 +7 X^3\big)\big]\, .
\end{align}
Given such a root fixes $v$ to be 
\begin{align}
v=&-\frac{1}{12 a b (a+b)(a+b-2 c)}\big[12 a b (a+b)^2 +3c (a+b-c)(13 a^2 -54 a b +13 b^2) \nonumber\\
&+(a^2-b^2)(16 a b -25 (a+b)c+25 c^2)u+(5a^2 -22 a b +5 b^2)(8 a b -7(a+b)c+7 c^2)u^2 \nonumber\\
&-(a^2-b^2)(32 ab -25 (a+b) c +25 c^2)u^3 -4(a-b)^2 (a+b-c) c u^4\big]\, .
\end{align}
One has to take the root $u$ such that $u,$ $v$ are in the domain \eqref{domainuv}.

The expressions \eqref{abtorus} can be used to determine the possible ranges of $a,b,c$ and $d$ compatible with globally well-defined solutions. Using the domain \eqref{domainuv} of $u$ and $v$ we find that regular solutions always have
\be\label{analysis torus ineq}
a<0<d\leq c \leq b\, ,\quad \mathrm{gcd}(a,c)=\mathrm{gcd}(a,d)=\mathrm{gcd}(b,c)=\mathrm{gcd}(b,d)=1\,,
\ee
where we have taken $b>0$ without loss of generality. In particular we see that $a<0$ whilst the other three integers are positive. Note that from the naive comparison with the Sasaki--Einstein $L^{a,b,c}$ metric one would have taken $a>0$ and \emph{not} $a<0$. This phenomenon has been noted in the literature previously for the Baryonic twist solutions of $L^{a,b,c}$ on the torus \cite{Couzens:2018wnk}, here we have proven that it also extends to the twist solutions presented here. 

We now turn to matching these solutions to the field theory side. We have done this numerically for explicit triples $a,$ $b,$ $c$ satisfying \eqref{analysis torus ineq}.\footnote{We have excluded values where $c=d$ since this is in fact the $Y^{p,q}$ limit and we make further comments about this in appendix \ref{sec return ypq}. In particular we show how to take this limit for the local solution.} More concretely we have matched the central charge $c_{\text{sugra }}$ and R-charges computed using the supergravity solution, i.e.~by evaluating \eqref{centralgeneral} and \eqref{Rchargesgravresult}, to the central charge $c^{\text{geo}}_R$ and R-charges that were obtained from geometric c-extremization (see section \ref{sec geom dual torus}). We have taken triples
$a,$ $b,$ $c$ with $-150<a<0,$ $b<150$ and satisfying
the conditions \eqref{analysis torus ineq}. This resulted in $48289$ possible triples to analyze. In order to
compare the two central charges we have to continue the domain of the result on the c-extremization side to allow for negative values of $a$. Furthermore, to compute $c^{\text{geo}}_R$ and the R-charges on the c-extremization side one needs the values of $m_1$, $n_2$ and $n_3$. These can be calculated from the solution via the fluxes \eqref{Metap}-\eqref{Mxim} and the relations \eqref{equations fluxes}, \eqref{rescaling}, i.e.
\begin{equation}\label{dictionary}
m_1=\tfrac{1}{N}M_{\xi_+}\,,\qquad n_2=-\tfrac{1}{N}(M_{\xi_-}+a k M_{\eta_-}-a l M_{\xi_+})\,,\qquad n_3=-\tfrac{1}{N}(bM_{\eta_-}+cM_{\xi_+})\,.
\end{equation} 
The factor $L^4/(g_s \ell_s^4)$ appearing in the fluxes is expressed in terms of $N$ via \eqref{Nfinal}. We found matching central charges and R-charges for all triples analyzed. For purposes of illustration explicit values for $m_{1},$ $n_{2},$ $n_{3},$ $c_{\text{sugra}}=c^{\text{geo}}_R$ and the R-charges for 3 triples with $a<0$
can be found in table \ref{tab:Values-of-}. 

When $a>0$ one can still solve the regularity conditions \eqref{abtorus} for $u$ and $v$. However, one does not find a solution in the domain \eqref{domainuv}. Even though the solution is not regular one can proceed and compute its central charges and R-charges. These values do still match with the field theory side, however it turns out that always either the central charge or one of the R-charges is negative. To illustrate this, we have also presented three triples with positive $a$ in table \ref{tab:Values-of-}. It is an opportune moment to remind the reader that the necessity of negative $a$ can also be shown at the field theory side, namely the cases considered in section \ref{c-extr examples} did not simultaneously have positive central charge and positive R-charges for $a>0$ for any choice of twist parameters.

We thus only obtain physically meaningful results when $a$ is negative. Note that in the field theory this analytic continuation is ill-defined if one considers the reduction from four dimensions since $a$ is related to the multiplicity of a particular field. Ignoring this subtlety we did find that the expressions for the central charge and R-charges computed at the field theory side formally agree with the expressions in this section. We emphasize that the supergravity solutions for negative $a$ constructed here have no pathologies and are singularity free, globally well-defined solutions. They are not regular for $a>0$ however. This result seems to suggest that one cannot construct a flow solution from AdS$_5$ to AdS$_3\times \text{T}^2$ in a consistent truncation of $L^{a,b,c}$ and it would be interesting to understand why this is the case. Moreover, identifying the field theory dual to these solutions remains an open problem. A similar discrepancy has previously been observed for direct products $Y_7=\text{T}^2 \times {Y}^{p,q},$\footnote{${Y}^{p,q}$ are again, in abuse of notation, the analogues of the Sasaki--Einstein manifolds $Y^{p,q}$.} $L^{a,b,c}$  and for a specific fibration of ${Y}^{p,q} \hookrightarrow Y_7 \to \text{S}^2$ \cite{Couzens:2017nnr,Couzens:2018wnk,Gauntlett:2018dpc}. The $Y^{p,q}$ field theory is characterized by integers $p>q>0$ whilst the solutions are characterized by integers $q>p>0$. In appendix \ref{sec return ypq} we extend the ${Y}^{p,q}$ case by studying more general fibrations over a Riemann surface $\Sigma_g$ and performing the required regularity analysis. We again find that the solutions with torus fibrations are only regular for $q>p>0$ in agreement with the results presented here.
\begin{table}[h]
\begin{centering}
\begin{tabular}{|c|c|c|c|c|c|c|c|c|c|}
\hline 
$(a,b,c)$ & $m_1$ & $n_2$ & $n_3$ & $c_R$ & $R_{\xi_-}$ & $R_{\xi_+}$ & $R_{\eta_-}$ & $R_{\eta_+}$\tabularnewline
\hline 
\hline 
$(-8, 116, 97)$ & -0.49 & 798.36 & 211.30 & -335.60 & 1.31 & 0.33 & 0.16 & 0.20  \tabularnewline
\hline 
$(-27, 91, 62)$ & -4.42 & 2681.15  & 411.61 & -196.26 & 1.13 & 0.84 & 0.01 & 0.02 \tabularnewline
\hline 
$(-53, 139, 79)$ & -3.36 & -7722.81 & 460.05 & -270.34 & 1.16 & 0.80 & 0.02 & 0.03 \tabularnewline
\hline 
$(7, 133, 111)$ & 0.26 & 55.57 & 185.60 & -387.21 & 1.32 & -0.26 & 0.49 & 0.44 \tabularnewline
\hline 
$(25, 139, 108)$ & 0.70 & -315.87 & 193.15 & -246.98 & 1.25 & -0.81 & 0.85 & 0.71 \tabularnewline
\hline 
$(50, 122, 119)$ & 12.04 & -2824.87 & 167.15 & 53355.54 & -2.37 & -14.06 & 13.69 & 4.75  \tabularnewline
\hline 
\end{tabular}
\par\end{centering}
\caption{\label{tab:Values-of-}Values of $m_{1},$ $n_{2},$ $n_{3}$ (divided by $C$), $c^{\text{geo}}_R=c_{\text{sugra}}$ (divided by $C N^2$) and the R-charges (divided by $N$) for certain triples
$a,$ $b$ and $c.$ The triples with negative $a$ correspond to globally well-defined solutions; note that $C<0$ and therefore the central charges are positive. The triples with positive $a$ always have at least one negative R-charge.}
\end{table}


\subsubsection{Sphere fibrations}

Let us now turn our attention to the solutions consisting of a fibration over a two-sphere. The analysis in this case and in the higher genus Riemann surfaces, that we tackle in the subsequent section, are even more cumbersome than the analysis presented in the previous section for the torus. The torus case was easier because we could find a coordinate transformation for the auxiliary roots which got rid of a square root in the physical roots. It also made the parameter $C$ drop out of the physical roots and the regularity conditions. For the $\kappa=\pm 1$ cases we have not found such a coordinate transformation and (therefore) also not found a solution that expresses $\xi_*$ and $\eta_*$ as function of $a,$ $b$, $c$ and $C$. Nevertheless we can still make progress and identify ranges for the parameters which give global solutions. We shall supplement the ranges of the parameters with some explicit numerical computations which show that the results obtained here agree with the results obtained from the field theory and geometric dual of c-extremization. Again, this matching is up to the domains of the integers $a$, $b$, $c$ characterizing the gravity solution and the field theory. The analysis is largely analogous to the previous case and we will thus be brief in places. In particular the matching of the dictionary in equation \eqref{dictionary} is the same. 

For the fibrations over a two-sphere it is possible to find solutions for both $D=\pm1$. It is also possible to rewrite the regularity conditions such that the integers $a$ and $c$ are expressed in terms of the auxiliary roots and $C$ as in \eqref{abtorus}. However, these expressions are exceedingly tedious and we therefore refrain from presenting them in this paper. Despite this it can be algebraically proven that the integers $a$, $b$, $c$ and $d$ are necessarily of the form \eqref{analysis torus ineq} in order to obtain globally regular solutions. This is the same as in the torus case studied above and the same discussion appearing there is equally applicable here. 

We now focus on the case that $D=1$ to provide a few more details. First we make the coordinate transformations
\begin{align}
\eta_{*}=-\frac{C(2+ v)}{4}\, , \qquad \xi_{*}=-C \Big( \frac{2}{u}+1\Big)\, . 
\end{align}
We now use the equations \eqref{conditions equation functions} to express the physical roots and parameters $A$ and $B$ as
\begin{align}
\eta_{\pm}=&\frac{2(2-C) v - C v^{2}\pm 2 \sqrt{4 v^2 +C^2 (v+2)^2(v-1)-C (v^3+6 v^2 +4 v -8)}}{(v-2)^2}\, ,\nonumber\\%
\xi_{\pm}=&\frac{1}{8}\Big( 4 u+u^2 -C(8+6 u+u^2)
\pm \sqrt{u\big[\big(C(2+u)-u\big)\big(C(16+10u+u^2)-(4+u)^2\big) \big]}\Big)\, , \nonumber\\%
A=&\frac{(u+2)^2 (u-C(u+2))}{4 u}\, ,\\
B=&\frac{(v+2)^2(C(v+2)-4)}{4(v-2)^2}\, .\nonumber
\end{align}
One of the domains, where the solution is globally regular, is given by
\be \label{domsphere}
0<v<1\, ,\qquad 0<u<v\, , \qquad \frac{v^{2}}{v^2+v-2}< C< \frac{u (u(v-2)+4 v)}{u^2(v-2)-(v-2)^2 +u (6 v-4)}\, .
\ee
We have solved the regularity conditions numerically using this domain for a large number of triples $a$, $b$, $c$ satisfying \eqref{analysis torus ineq}. For these triples we also compared the results for the central charges and R-charges obtained from the solutions to the field theory and geometric dual results, as we did for the torus case. We have presented the results for three triples in table \ref{S2 table}. For these triples we fixed $v=0.5$ and solved the regularity conditions for $u$ and $C$ within the domain \eqref{domsphere}.

\begin{table}[h]
\begin{centering}
\begin{tabular}{|c|c|c|c|c|c|c|c|c|}
\hline 
 $(a,b,c)$ & $m_1$ & $n_2$ & $n_3$ & $c_R$ & $R_{\xi_-}$ & $R_{\xi_+}$ & $R_{\eta_-}$ & $R_{\eta_+}$\tabularnewline
\hline 
\hline 
 $(-19, 29, 8)$ & -2.38 & 177.63 & -24.21 & 4.49 & 1.30 & 0.68 & 0.01 & 0.01 \tabularnewline
\hline 
  $(-8, 26, 15)$ & -0.71 & 40.41 & -17.38 & 9.09 & 1.52 & 0.38 & 0.05 & 0.06 \tabularnewline
\hline 
  $(-1, 25, 16)$ & -0.07 & 8.64 & -12.82 & 14.88 & 1.60 & 0.05 & 0.18 & 0.18 \tabularnewline
\hline 
\end{tabular}
\par\end{centering}
\caption{\label{S2 table}Values of $m_{1},$ $n_{2},$ $n_{3},$  $c^{\text{geo}}_R=c_{\text{sugra}}$ (divided by $N^2$), and the R-charges (divided by $N$) for certain triples
$a,$ $b,$ $c$ and $\kappa= 1$. Note that the central charges and R-charges are all positive.}
\end{table}


\subsubsection{Higher genus Riemann surface fibrations}

In this final section we will study the regularity of solutions for which the genus is $g>1$. This case is distinct from the two previous cases in that it is possible to find globally regular solutions for which the integers $a$, $b$, $c$ and $d$ are all positive as one would expect from the naive comparison with the $L^{a,b,c}$ Sasaki--Einstein solutions. Additionally, we find that there are also regular solutions for integers satisfying $a<0<d\leq c\leq b$. We shall therefore split this section in two parts to give examples of both cases. We furthermore find that there are no global solutions for $D=-1$; global solutions only exist for $D=1$.


\subsubsection*{Positive $\boldsymbol{a}$ solutions}

We first make the change of parameters
\be
\eta_{*}=- C v\, , \qquad \xi_{*}=-C u \, .
\ee
The physical roots and parameters $A$ and $B$ in terms of these variables are
\begin{align}
\eta_{\pm}=& \frac{1+(C-2) v -2 C v^2 \pm \sqrt{2C v (2 v^2-1)+(2v-1)^2+C^2 v^2 (4v-3)}}{2(v-1)^2}\, ,\nonumber\\
\xi_{\pm}=&\frac{1+(C-2)u-2 C u^2 \pm \sqrt{2 C u (2 u^2-1)+(2u-1)^2 + C^2 u^2 (4 u-3)}}{2(u-1)^2}\, ,\nonumber\\
A=&-\frac{u^2(C u+1)}{(u-1)^2}\, ,\\
B=& \frac{v^2(C v+1)}{(v-1)^2}\, .\nonumber
\end{align}
One of the domains where the solutions can be globally regular is given by
\begin{align}\label{domHab0}
\frac{2v}{2v-1} < u  < \frac{4 v-3}{4v-4}\, ,\qquad 
\frac{9}{8}&< v  <  \frac{1}{4}(1+\sqrt{13})\, ,\\
\frac{2 u v- u-v}{u^2+u v -2 u^2 v +v^2 -2 u v^2 +u^2 v^2}< &~C  < \frac{-1+4v-4v^2}{4v^2-3v}\, . \nonumber
\end{align}
For this domain the integers $a$, $b$, $c$, $d$ corresponding to global regular solutions are also all positive and satisfy the field theoretic conditions \eqref{conditions triples field th}, i.e.
\begin{equation}\label{conditions solapos}
0<a\leq d\leq c \leq b\,,\quad \text{gcd}(a,c)=\text{gcd}(a,d)=\text{gcd}(b,c)=\text{gcd}(b,d)=1\,.
\end{equation}
We can again compare the solutions to the field theory by numerically solving the regularity conditions and computing the central charge and R-charges. We have found matching values for all cases we checked. This is strong evidence for new dualities where the gravity side is given by the solutions we found and the field theories are twisted compactifications of the 4d $L^{a,b,c}$ quiver gauge theories. We have again given a few examples for explicit triples in table \ref{table H2 a>0}. Here we have fixed $v=1.13$ and solved the regularity conditions for $u$ and $C$ taking values according to \eqref{domHab0}.
\begin{table}[h]
\begin{centering}
\begin{tabular}{|c|c|c|c|c|c|c|c|c|}
\hline 
 $(a,b,c)$ & $m_1$ & $n_2$ & $n_3$ & $c_R$ & $R_{\xi_-}$ & $R_{\xi_+}$ & $R_{\eta_-}$ & $R_{\eta_+}$\tabularnewline
\hline 
\hline 
 $(1, 19, 11)$ & 0.03 & -6.14 & -15.77 & 53.48 & 0.88 & 0.29 & 0.40 & 0.42\tabularnewline
\hline 
 $(8,26,23)$ & 0.13 & 61.06 & -22.14 & 87.13 &  0.86 & 0.67 & 0.18 & 0.29 \tabularnewline
\hline 
 $(14,22,19)$ & 0.17 & -91.44 & -20.50 & 90.53 & 0.82 & 0.75 & 0.20 & 0.22 \tabularnewline
\hline 
\end{tabular}
\par\end{centering}
\caption{\label{table H2 a>0}Values of $m_{1},$ $n_{2}$ and $n_{3}$ (divided by $g-1$), $c^{\text{geo}}_R=c_{\text{sugra}}$ (divided by $(g-1)N^2$) and the R-charges (divided by $N$) for certain triples
$a,$ $b,$ $c$ with positive $a$ and $\kappa=- 1$. Observe that the central charges and R-charges are all positive.}
\end{table}

\subsubsection*{Negative $\boldsymbol{a}$ solutions}

To find globally regular solutions with $a<0$ one can make the parameter redefinitions
\be
\eta_{*}=- \frac{v}{C'}\, , \qquad \xi_{*}=-\frac{1}{C'u}\, , \qquad C=\frac{1}{C'}\, .
\ee
Using the conditions \eqref{conditions equation functions} we express the physical roots and $A$, $B$ as
\begin{align}
\eta_{\pm}=&\frac{C'+v -2 C v -2 v^2 \mp \sqrt{C'^2(2v-1)^2 +v^2 (4v-3)+ 2C' v(2 v^2-1)}}{2 C' (v-1)^2}\, , \nonumber\\
\xi_{\pm}=&\frac{-2 + u -2 C u+ C u^2\mp\sqrt{u (1+C'u)\big(4-3 u+C' (u-2)^2\big)}}{2 C'(u-1)^2}\, ,\nonumber\\
A=&- \frac{1+ C' u}{C' u (u-1)^2}\, , \\
B=&\frac{v^2(C'+v)}{C'(v-1)^2}\, . \nonumber
\end{align}
A range of parameters for which the solutions can be globally regular is given by
\be\label{domHan}
\frac{3}{4}<v<1\, ,\qquad 0< u<\frac{2(v-1)^2}{v(2v-1+\sqrt{4v-3})}\, ,\qquad -\frac{v(4v-3)}{(2v-1)^2}<C'<0\, .
\ee
For this domain we can show that the solutions of the regularity conditions necessarily have $a<0$ with the remaining integers all positive satisfying \eqref{analysis torus ineq}. We can indeed numerically solve the regularity conditions for explicit values of $a,$ $b$ and $c$ using these domains. The resulting solutions can again be compared to the field theory by computing their central charge and R-charges. Three examples can be found in table \ref{table H2 a<0}. For these triples we took $v=0.8$ and solved the regularity conditions for $u$ and $C'$ obeying the bounds \eqref{domHan}.

\begin{table}[h]
\begin{centering}
\begin{tabular}{|c|c|c|c|c|c|c|c|c|}
\hline 
 $(a,b,c)$ & $m_1$ & $n_2$ & $n_3$ & $c_R$ & $R_{\xi_-}$ & $R_{\xi_+}$ & $R_{\eta_-}$ & $R_{\eta_+}$\tabularnewline
\hline 
\hline 
 $(-19,29,8)$ & 14.38 & 1685.99 & -232.51 & 103.86 & 1.08 & 0.90 & 0.01 & 0.01 \tabularnewline
\hline 
 $(-8,26,15)$ &  3.93 & 355.01 & -161.14 & 175.14 & 1.18 & 0.71 & 0.04 & 0.07 \tabularnewline
\hline 
 $(-1,25,16)$ & 0.21 & 37.82 & -76.81 & 167.57 & 1.25 & 0.15 & 0.29 & 0.31 \tabularnewline
\hline 
\end{tabular}
\par\end{centering}
\caption{\label{table H2 a<0}Values of $m_{1},$ $n_{2}$ and $n_{3}$ (divided by $g-1$), $c^{\text{geo}}_R=c_{\text{sugra}}$ (divided by $(g-1)N^2$) and the R-charges (divided by $N$) for certain triples
$a,$ $b,$ $c$ with negative $a$ and $\kappa=- 1$. Observe that the central charges and R-charges are all positive.}
\end{table}

\section{Discussion}\label{sec: discussion}

In this paper we have studied the IR limits of compactifications of the 4d $L^{a,b,c}$ $\mathcal{N}=1$ quiver gauge theories on a Riemann surface $\Sigma_g$ with fluxes of global symmetries switched on. Assuming they flow to an $\mathcal{N}=(0,2)$ SCFT we have computed the central charges and R-charges of this IR fixed point in three different ways. Field theoretically we computed the latter by utilizing the technique of c-extremization. With knowledge of UV data it is then possible to calculate the central charge and R-charges by virtue of 't Hooft anomaly matching. The second way is the recently found geometric dual formulation of c-extremization. This technique only depends on topological data of the solution and assumes the existence of a supergravity solution. Lastly, we have constructed explicit solutions dual to 2d $\mathcal{N}=(0,2)$ SCFTs and computed their central charges and R-charges directly. We found an exact matching between the field theoretic and geometric values.

The solutions we have constructed are of the form AdS$_3\times Y_7$ where ${L}^{a,b,c}\hookrightarrow Y_7 \rightarrow \Sigma_g$. We have not found the most general solution of this form since it is lacking two parameters in comparison to the field theory. The ansatz we have used depends on four unknown functions $\mathcal{F}$, $\mathcal{G}$, $\mathcal{H}$ and $\mathcal{L}$ and by requiring the recovery of the universal twist solution the form of $\mathcal{H}$ and $\mathcal{L}$ is fixed. It is therefore expected, that a suitable generalization of the ans\"atze for the functions $\mathcal{F}$ and $\mathcal{G}$ accounts for the missing parameters. 

We have obtained generic expressions for the fluxes, central charge and R-charges without requiring a solution of the master equation. Recall that the central charge agrees off-shell for c-extremization and its geometric dual \cite{Gauntlett:2018dpc,Hosseini:2019use}. It would be interesting to see if one can also obtain a matching to the field theory with our off-shell expressions for the charges. 

Let us stress again that the matching of central charges and R-charges between field theory and the solutions is only a formal matching for the S$^2$, T$^2$ and $a<0$ case of $\Sigma_{g>1}$. While the field theory results are valid for $a>0$, globally regular solutions necessarily have $a<0$ in the S$^2$ and T$^2$ cases, whilst they can have both $a>0$ and $a<0$ for $g>1$ compactifications. For the field theories compactified on S$^2$ and T$^2$ with $a>0$ we find at least one negative R-charge, which is inconsistent with representation theory of the superconformal algebra. This is evidence that the compactified 4d $\cN=1$ theories on these spaces do not flow to an IR fixed point. It would be of great interest to study the RG flow trajectories of the compactified 4d theories from first principles, e.g~with field theoretical methods or holographic RG flow across dimensions as was done in \cite{Bobev:2017uzs}.

Nevertheless, even for the cases with $a<0$, we found well-defined supersymmetric AdS$_3$ solutions of type IIB supergravity, which according to the AdS/CFT correspondence should have 2d SCFT duals. The formulation of these SCFTs and if they admit an interpretation as descending via compactification from higher-dimensional SCFTs remains an interesting puzzle.

\paragraph*{Acknowledgements.}

It is a pleasure to thank Marcos Crichigno, Nava Gaddam, Jerome Gauntlett, Thomas Grimm, Dario Martelli, James Sparks and Stefan Vandoren for valuable discussions.

This work was supported in part by the D-ITP consortium, a program of the Netherlands Organization for Scientific Research (NWO) that is funded by the Dutch Ministry of Education,
Culture and Science (OCW), and by the NWO Graduate Programme.


\appendix


\section{Ansatz analysis details}\label{App:ansatz-analysis}

As explained in the main text, the metric ansatz always admits a closed non-degenerate two-form, $J$, which may be viewed as the almost complex structure of the metric. Despite this, the metric is not a complex manifold for arbitrary $\mathcal{H}$ and $\mathcal{L}$. The vanishing of the Nijenhuis tensor, and thus the integrability of the almost complex structure still needs to be imposed. This is most easily enforced by requiring that the exterior derivative of the holomorphic $(3,0)$ form $\Omega$, compatible with the chosen almost complex structure, satisfies
\be
\dd \Omega= \ii P \wedge \Omega
\ee
for some one-form $P$. The one-form $P$ is the canonical Ricci-form potential defined by the K\"ahler metric satisfying $\dd P= \rho$, with $\rho= \frac{1}{2}R_{ijkl} J^{kl} \dd x^{i} \wedge \dd x^{j}$ the Ricci-form of the K\"ahler metric.
After a few pages of computation one finds that the necessary and sufficient conditions for the integrability of the complex structure are the following five non-trivially coupled differential equations
\begin{align}
\mathcal{F}(\xi)^{2}\partial_{\xi}^{2} \mathcal{H}(\eta,\xi)&=\mathcal{G}(\eta)^{2} \partial_{\eta}^{2} \mathcal{H}(\eta,\xi)\,,~~\nonumber\\
\mathcal{F}(\xi)^{2} \partial_{\xi}^{2} \mathcal{L}(\eta,\xi)&=\mathcal{G}(\eta)^{2} \partial_{\eta}^{2} \mathcal{L}(\eta,\xi)\,,~~\nonumber\\
(\eta-\xi) \mathcal{G}(\eta) \partial_{\eta}^2 \mathcal{H}(\eta,\xi)&=-\mathcal{F}(\xi) \big[ (\partial_{\eta}-\partial_{\xi})\mathcal{H}(\eta,\xi)+(\eta-\xi) \partial_{\eta}\partial_{\xi} \mathcal{H}(\eta,\xi))\big]\,,~~\label{Compcond}\\
(\eta-\xi) \mathcal{G}(\eta) \partial_{\eta}^2 \mathcal{L}(\eta,\xi)&=-\mathcal{F}(\xi) \big[ (\partial_{\eta}-\partial_{\xi})\mathcal{L}(\eta,\xi)+(\eta-\xi) \partial_{\eta}\partial_{\xi} \mathcal{L}(\eta,\xi))\big]\,,~~\nonumber\\
\partial_{\eta}\partial_{\xi} \mathcal{H}(\eta,\xi)(\partial_{\eta}-\partial_{\xi})\mathcal{L}(\eta,\xi)&=\partial_{\eta}\partial_{\xi} \mathcal{L}(\eta,\xi)(\partial_{\eta}-\partial_{\xi})\mathcal{H}(\eta,\xi)\,.\nonumber
\end{align}
As a by-product of this analysis the Ricci-form potential is found to be
\begin{align}
P=&P_g -\frac{\partial_{\xi} \big( \mathcal{F}(\xi) \mathcal{H}(\eta,\xi)\big)}{2(\eta-\xi) \mathcal{H}(\eta,\xi)} \big(\dd \phi +\partial_{\xi} \mathcal{L}(\eta,\xi)\dd \psi+\partial_{\xi} \mathcal{H}(\eta,\xi) \mathcal{A}_{g}\big)\nonumber\\
&-\frac{\partial_{\eta} \big(\mathcal{G}(\eta) \mathcal{H}(\eta,\xi)\big)}{2(\eta-\xi) \mathcal{H}(\eta,\xi)} \big(\dd \phi +\partial_{\eta} \mathcal{L}(\eta,\xi)\dd \psi+\partial_{\eta} \mathcal{H}(\eta,\xi) \mathcal{A}_{g}\big)\nonumber\\
&- \frac{\mathcal{F}(\xi) \partial_{\xi}^{2} \mathcal{L}(\eta,\xi)+\mathcal{G}(\eta)\partial_{\eta}^{2} \mathcal{L}(\eta,\xi)}{\eta-\xi} \dd \psi- \frac{\mathcal{F}(\xi) \partial_{\xi}^{2} \mathcal{H}(\eta,\xi)+\mathcal{G}(\eta)\partial_{\eta}^{2} \mathcal{H}(\eta,\xi)}{\eta-\xi} \mathcal{A}_{g}\,,
\end{align}
where $P_{g}$ and $\mathcal{A}_{g}$ are the Ricci-form potential and K\"ahler form potential on the Riemann surface respectively.
We are unable to solve \eqref{Compcond} in full generality, but it is not too difficult to see that a family of solutions is given by
\be \label{HLsol}
\mathcal{H}(\eta,\xi)= \alpha_{0}+\alpha_{1}(\eta + \xi) + \alpha_{2} \eta \xi\,,\qquad \mathcal{L}(\eta,\xi)=\beta_{0}+\beta_{1}(\eta+\xi) +\beta_{2} \eta \xi\,.
\ee
The three constants in $\mathcal{L}$ are trivial and can be removed by coordinate transformations of the angular coordinates and in this manner one can set $\beta_{0}=\beta_{1}=0$ and $\beta_{2}=1$. The universal twist solution is obtained by taking $\mathcal{L}=\eta \xi$ and therefore it seems reasonable that this choice of solution is the correct one for our purposes. Once we have fixed this, $\mathcal{H}$ is fixed to take the form in \eqref{HLsol}. Namely, substituting $\mathcal{L}=\eta \xi$ into the last equation in \eqref{Compcond} implies $\mathcal{H}$ takes the form presented in \eqref{HLsol}.


\section{Effective coordinate systems}\label{app: eff coordinate systems}

In this appendix we state the effective coordinate systems which are
needed to perform the integrations over the angular parts. For an
explanation of how to construct the transformation \eqref{toric data}
see e.g. \cite{Butti:2005sw}. From this relation it is easy to derive
the corresponding coordinate transformation \eqref{effcoords}
which was needed for the computation of the flux $N$ and for integrations
over the degeneration surface corresponding to $k_{\eta_+}.$ We now state
the transformations needed for the integrations over the other three
degeneration surfaces.

\paragraph*{$\boldsymbol{\eta=\eta_{-}}$.}

The vectors generating the effective torus action are now related
to the Killing vectors via
\begin{equation}
\left(\begin{array}{c}
k_{\eta_-}\\
k_{\eta_+}\\
l_{\xi_-}\\
l_{\xi_+}
\end{array}\right)=\left(\begin{array}{ccc}
1 & 0 & 0\\
1 & bk & a\\
1 & bl & d\\
1 & 1 & 0
\end{array}\right)\left(\begin{array}{c}
\partial_{\psi_{1}}\\
\partial_{\psi_{2}}\\
\partial_{\psi_{3}}
\end{array}\right)\,,
\end{equation}
where $k,$ $l$ are solutions of $dk-al=1.$ The corresponding coordinate
transformation is then given by
\begin{eqnarray}
z & = & \psi_{1}\,,\nonumber \\
\psi & = & -\frac{2}{\mathcal{G}'_{-}}\psi_{1}+2\Big(\frac{1}{\mathcal{F}'_{+}}+\frac{1}{\mathcal{G}'_{-}}\Big)\psi_{2}+2\Big(\frac{k}{\mathcal{F}'_{-}}+\frac{l}{\mathcal{G}'_{+}}+\frac{k-l}{\mathcal{G}'_{-}}\Big)\psi_{3}\,,\\
\phi & = & \frac{2\eta_{-}}{\mathcal{G}'_{-}}\psi_{1}-2\Big(\frac{\xi_{+}}{\mathcal{F}'_{+}}+\frac{\eta_{-}}{\mathcal{G}'_{-}}\Big)\psi_{2}-2\Big(\frac{k\xi_{-}}{\mathcal{F}'_{-}}+\frac{l\eta_{+}}{\mathcal{G}'_{+}}+\frac{(k-l)\eta_{-}}{\mathcal{G}'_{-}}\Big)\psi_{3}\,.\nonumber 
\end{eqnarray}
Since $k_{\eta_-}=\partial_{\psi_{1}},$ at the $\eta_{-}$ degeneration surface the $T^{3}$ will degenerate to a $T^2$ which has as basis of the effective action the coordinates $\psi_{2}$ and $\psi_{3}$. 

\paragraph*{$\boldsymbol{\xi=\xi_{+}}$.}

We now have the relation

\begin{equation}
\left(\begin{array}{c}
l_{\xi_+}\\
l_{\xi_-}\\
k_{\eta_-}\\
k_{\eta_+}
\end{array}\right)=\left(\begin{array}{ccc}
1 & 0 & 0\\
1 & -dk & -c\\
1 & -dl & -a\\
1 & 1 & 0
\end{array}\right)\left(\begin{array}{c}
\partial_{\psi_{1}}\\
\partial_{\psi_{2}}\\
\partial_{\psi_{3}}
\end{array}\right)\,,
\end{equation}
where $k,$ $l$ solve the relation $cl-ak=1$. The coordinate transformation
that follows from this relation is 
\begin{eqnarray}
z & = & \psi_{1}\,,\nonumber \\
\psi & = & \frac{2}{\mathcal{F}'_{+}}\psi_{1}-2\Big(\frac{1}{\mathcal{F}'_{+}}+\frac{1}{\mathcal{G}'_{+}}\Big)\psi_{2}-2\Big(\frac{k}{\mathcal{G}'_{-}}+\frac{l}{\mathcal{F}'_{-}}+\frac{k-l}{\mathcal{F}'_{+}}\Big)\psi_{3}\,,\\
\phi & = & -\frac{2\xi_{+}}{\mathcal{F}'_{+}}\psi_{1}+2\Big(\frac{\xi_{+}}{\mathcal{F}'_{+}}+\frac{\eta_{+}}{\mathcal{G}'_{+}}\Big)\psi_{2}+2\Big(\frac{k\eta_{-}}{\mathcal{G}'_{-}}+\frac{l\xi_{-}}{\mathcal{F}'_{-}}+\frac{(k-l)\xi_{+}}{\mathcal{F}'_{+}}\Big)\psi_{3}\,,\nonumber 
\end{eqnarray}
where again $\psi_{2}$ and $\psi_{3}$ are the coordinates needed
for integration over the surface corresponding to $l_{\xi_+}.$

\paragraph*{$\boldsymbol{\xi=\xi_{-}}$.}

This time the relation between the Killing vectors and the vectors
generating an effective torus action is given by

\begin{equation}
\left(\begin{array}{c}
l_{\xi_-}\\
l_{\xi_+}\\
k_{\eta_-}\\
k_{\eta_+}
\end{array}\right)=\left(\begin{array}{ccc}
1 & 0 & 0\\
1 & dk & c\\
1 & -dl & b\\
1 & 1 & 0
\end{array}\right)\left(\begin{array}{c}
\partial_{\psi_{1}}\\
\partial_{\psi_{2}}\\
\partial_{\psi_{3}}
\end{array}\right)\,,
\end{equation}
in which $k,$ $l$ solve $cl+bk=1.$ The coordinate transformation
is then equal to
\begin{eqnarray}
z & = & \psi_{1}\,,\nonumber \\
\psi & = & \frac{2}{\mathcal{F}'_{-}}\psi_{1}-2\Big(\frac{1}{\mathcal{F}'_{-}}+\frac{1}{\mathcal{G}'_{+}}\Big)\psi_{2}-2\Big(\frac{k}{\mathcal{G}'_{-}}-\frac{l}{\mathcal{F}'_{+}}+\frac{k+l}{\mathcal{F}'_{-}}\Big)\psi_{3}\,,\\
\phi & = & -\frac{2\xi_{-}}{\mathcal{F}'_{-}}\psi_{1}+2\Big(\frac{\xi_{-}}{\mathcal{F}'_{-}}+\frac{\eta_{+}}{\mathcal{G}'_{+}}\Big)\psi_{2}+2\Big(\frac{k\eta_{-}}{\mathcal{G}'_{-}}-\frac{l\xi_{+}}{\mathcal{F}'_{+}}+\frac{(k+l)\eta_{-}}{\mathcal{F}'_{-}}\Big)\psi_{3}\,.\nonumber 
\end{eqnarray}
As in the previous cases $\psi_{2}$ and $\psi_{3}$ are the relevant
coordinates for the corresponding degenerate surface.


\section{The return of $Y^{p,q}$} \label{sec return ypq}

So far we have analyzed the case of $L^{a,b,c}$, where little is known in the literature. In contrast the case of $Y^{p,q}$ is far better understood but there are still certain solutions corresponding to different twists that are yet to be found. We shall extend the known solutions, giving a more general solution of the form $Y_{5}\hookrightarrow Y_7 \rightarrow \Sigma_{g}$ with $Y_{5}$ admitting isometry group $U(1)^2\times SU(2)$. Field theoretically this corresponds to twisting with only the baryonic, R-symmetry and the abelian flavour symmetry and not the non-abelian $SU(2)$ isometry. 


\subsection{Ansatz}

To begin we must give an ansatz for the K\"ahler base. We require a metric which is a fibration over a Riemann surface and has a $U(1)\times SU(2)$ isometry. The most general metric one can devise with this structure is
\be
\dd s^{2}= H(y) \dd s^{2}(\Sigma_{g})+\dd y^2 +F(y) (\dd \psi+\cos \theta \dd \phi+P(y) \mathcal{A}_{g})^{2} + G(y)(\dd \theta^2 +\sin^{2} \theta \dd \phi^2)
\ee
with $\mathcal{A}_{g}$ a one-form on the Riemann surface satisfying $\dd \mathcal{A}_{g}=J_{g}$ with $J_{g}$ the K\"ahler form on the Riemann surface. Imposing that the metric has a closed two-form preserving the $SU(2)$ isometry imposes
\be
F(y)=G'(y)^2\,,~~~ P(y)=-\frac{H'(y)}{G'(y)}\,.
\ee
Moreover, the condition that the metric is complex and the closed two-form is indeed the K\"ahler form requires
\be
\partial_{y} \frac{H'(y)}{G'(y)}=0\,.
\ee
Either we set the warp factor of the Riemann surface to vanish and therefore consider no flavour fibration, or we have
\be
H'(y)=c_1 G'(y)\,.
\ee
The K\"ahler metric is 
\be
\dd s^{2}=(c_{1}G(y)+4 c_{2}) \dd s^{2}(\Sigma_{g})+\dd y^{2} +G'(y)^{2} (\dd \psi+\cos \theta \dd \phi -c_{1} \mathcal{A}_{g})^{2}+G(y)(\dd \theta^{2}+\sin^{2}\theta\dd \phi^{2})\,.
\ee
We now introduce a new coordinate satisfying 
\be
4 x G(y(x))=1
\ee
and define the function 
\be
U(x)=\frac{G'(y(x))^{2}}{4x}\,.
\ee
The final form of the metric (after multiplying by an overall factor of 4) is
\be \label{ansats Ypq}
\dd s^{2}= \frac{c_{1}+c_{2}x}{x} \dd s^{2}(\Sigma_{g})+\frac{\dd x^{2}}{ x^{3} U(x)}+\frac{U(x)}{x}(\dd \psi +\cos \theta \dd \phi -c_{1} \mathcal{A}_{g})^2+\frac{1}{x}(\dd \theta^{2}+\sin^{2}\theta \dd \phi^{2})\,.
\ee
The parameter $c_2$ can be fixed to a non-zero value by a rescaling $x\rightarrow x/c_2$ and multiplying by an overall factor of $1/c_2$. We emphasize that \eqref{ansats Ypq} is the most general ansatz one can make admitting the required symmetries. Note that the parameter $c_{1}$ measures the twisting of the $U(1)$ over the Riemann surface and will therefore be related to the flavour twist parameter. 

The K\"ahler two-form and $(3,0)$-form preserving $SU(2)$ are
\begin{align}
J=&\frac{c_{1}+c_{2}x}{x} J_{g}+\frac{1}{x^2}\dd x \wedge (\dd \psi +\cos \theta \dd \phi - c_{1} \mathcal{A}_{g})+\frac{1}{x} \sin \theta \dd \theta \wedge \dd \phi\,,\\
\Omega=&\frac{\sqrt{U(x)(c_{1}+c_{2}x)}}{x^{\frac{3}{2}}}\Omega_{g}\wedge \Big( \frac{1}{x U(x)}\dd x + \ii (\dd \psi+\cos \theta \dd \phi-c_{1} \mathcal{A}_{g})\Big) \wedge (\dd \theta+\ii \sin \theta \dd \phi)\,.\nonumber
\end{align}
The Ricci scalar and Ricci-form potential are given by
\begin{align}
R&=2 x^{2} \partial_{x} f(x)-\frac{2 x(2 c_{1}+c_{2}x)}{c_{1}+c_{2}x}f(x)+\frac{2x(c_{1}+c_{2}x+\kappa)}{c_{1}+c_{2}x}\,,\nn \\
P&=P_{g}-\cos \theta \dd \phi +f(x) (\dd \psi +\cos \theta \dd \phi -c_{1} \mathcal{A}_{g})\,,
\end{align}
where
\be
~~f(x)=\frac{(3 c_{1}+2 c_{2} x)U(x)-x(c_{1}+c_{2}x)U'(x)}{2(c_{1}+c_{2}x)}\,.
\ee

\subsection{Regularity analysis}

Let us analyze the global regularity of the ten-dimensional metric,
which as in section \ref{geometry} can be done by first checking the regularity
of $Y_{5}$ . To make the metric \eqref{ansats Ypq} compact we must require that
$U$ takes its values in $[x_{-},x_{+}]$, where $U(x_{\pm})=0$ and
either $0<x_{-}<x_{+}$ or $x_{-}<x_{+}<0$. This follows from the fact that $x=0$ would lead to a singularity due to the explicit $\frac{1}{x}$ appearing in the metric.\footnote{It may be possible to obtain a regular solution with $x=0$ an endpoint, however the solution will not have the required topology.}
Furthermore, we need
that $U$ is positive on its domain. We have four surfaces at which
the metric degenerates specified by $x=x_{\pm}$ and $\theta=0,$
$\pi.$ As for $L^{a,b,c}$ we need to determine the Killing vectors
that vanish at the degeneration surfaces. By requiring the surface
gravity of the relevant Killing vector to be normalized to one, the
metric will extend smoothly onto the degeneration surface. For $Y^{p,q}$
we find that the Killing vectors located at the roots $x=x_{\pm}$
and $\theta=0,$ $\pi$ which satisfy these conditions are given by
\begin{eqnarray}
k_{x_\pm} & = & \partial_{z}+\frac{2}{x_{\pm}U'_{\pm}}\partial_{\psi}\,,\nonumber \\
l_{0} & = & \partial_{z}+\partial_{\phi}-\partial_{\psi}\,,\label{eq:killing vectors Ypq}\\
l_{\pi} & = & \partial_{z}-\partial_{\phi}-\partial_{\psi}\,.\nonumber 
\end{eqnarray}
As for $\mathcal{F}$ and $\mathcal{G}$ in the $L^{a,b,c}$ case, we denote the evaluation of $U$ at its roots by $U'_{\pm}\equiv U'(x_{\pm}).$
We again have a relation between the vectors (\ref{eq:killing vectors Ypq})
which is given by 
\begin{equation}
(p+q)k_{x_+}+(p-q)k_{x_-}-pl_{0}-pl_{\pi}=0\,,
\end{equation}
for relatively prime integers $p$ and $q.$ This implies the regularity
condition
\begin{equation}
\frac{p+q}{x_{+}U'_{+}}+\frac{q-p}{x_{-}U'_{-}}=-p\,.\label{regularity equation}
\end{equation}
So far we have not imposed any inequalities on $p$ and
$q,$ this will follow once we have obtained explicit solutions.

\subsection{General formulas for fluxes, central charge and R-charges}

Before solving the master equation \eqref{Mastereq} and obtaining explicit solutions
for the function $U,$ we derive formulas for the fluxes, central
charge and R-charges that are valid for arbitrary functions $U$ and
thus for arbitrary metrics of the form \eqref{ansats Ypq}. The integrations needed
to obtain these expressions are considerably easier than in the $L^{a,b,c}$
case. Therefore, we only state the effective system of coordinates
we have used for the integration and the resulting formulas.

\paragraph*{Effective coordinates systems.}

For the integration over the degeneration surface defined by $x=x_{+}$
we use the effective coordinates 
\begin{eqnarray}
z & = & \psi_{1}\,,\nonumber \\
\phi & = & \psi_{2}\,,\\
\psi & = & \frac{2}{x_{+}U'_{+}}\psi_{1}-\Big(1+\frac{2}{x_{+}U'_{+}}\Big)\psi_{2}+2\Big(\frac{\mu}{x_{-}U'_{-}}-\nu-\frac{\mu+2\nu}{x_{+}U'_{+}}\Big)\psi_{3}\,.\nonumber 
\end{eqnarray}
The integers $\mu$ and $\nu$ are a solution of B\'ezout's identity
$\mu p+\nu(p-q)=1.$ The relation of $\partial_{\psi_i}$ to the Killing vectors is given
by
\begin{equation}
\left(\begin{array}{c}
k_{x_+}\\
k_{x_-}\\
l_{\pi}\\
l_{0}
\end{array}\right)=\left(\begin{array}{ccc}
1 & 0 & 0\\
1 & 0 & p\\
1 & -1 & p-q\\
1 & 1 & 0
\end{array}\right)\left(\begin{array}{c}
\partial_{\psi_{1}}\\
\partial_{\psi_{2}}\\
\partial_{\psi_{3}}
\end{array}\right)\,.\label{jacobian Ypq}
\end{equation}
For the surface determined by $x=x_{-}$ we have the system

\begin{eqnarray}
z & = & \psi_{1}\,,\nonumber \\
\phi & = & -\psi_{2}+\mu\psi_{3}\,,\\
\psi & = & \frac{2}{x_{-}U'_{-}}\psi_{1}-\Big(1+\frac{2}{x_{-}U'_{-}}\Big)\psi_{2}+\Big(\frac{2\nu}{x_{+}U'_{+}}-\mu-\frac{2(\mu+\nu)}{x_{-}U'_{-}}\Big)\psi_{3}\,.\nonumber 
\end{eqnarray}
Here $\mu$ and $\nu$ solve $\mu(p+q)+\nu p=1.$ Then for the degeneration
surface determined by $\theta=0$ we used the coordinates
\begin{eqnarray}
z & = & \psi_{1}\,,\nonumber \\
\phi & = & \psi_{1}-\psi_{2}-(2\mu+\nu)\psi_{3}\,,\\
\psi & = & -\psi_{1}+\Big(1+\frac{2}{x_{+}U'_{+}}\Big)\psi_{2}+\nu\Big(1+\frac{2}{x_{-}U'_{-}}\Big)\psi_{3}\,,\nonumber 
\end{eqnarray}
where $\mu(p-q)+\nu p=1.$ Lastly, for the surface defined by $\theta=\pi$
the coordinates are given by
\begin{eqnarray}
z & = & \psi_{1}\,,\nonumber \\
\phi & = & -\psi_{1}+\psi_{2}+(2\mu+\nu)\psi_{3}\,,\\
\psi & = & -\psi_{1}+\Big(1+\frac{2}{x_{-}U'_{-}}\Big)\psi_{2}+\nu\Big(1+\frac{2}{x_{+}U'_{+}}\Big)\psi_{3}\,,\nonumber 
\end{eqnarray}
with $\mu(p+q)+\nu p=1.$ Note that to perform integrals over the
full angular part parametrized by $z,$ $\phi$ and $\psi,$ one can
use any of these coordinate systems and the result will agree. 

\paragraph*{Flux quantization.}

One can compute the flux $N$ corresponding to integrating the five-form
over $Y_{5}$ using \eqref{flux quantisation} and picking any of the effective
coordinate systems given in the previous paragraph to perform the
integration over $z,$ $\phi$ and $\psi.$ The other flux quantization conditions correspond
to integrating the five-form over the fibration of the different degeneration
surfaces over the Riemann surfaces. These fluxes $M_{x_{\pm}},$ $M_{\theta_{0}}$
and $M_{\theta_{\pi}}$, which correspond to the degeneration surfaces
defined by $x=x_{\pm},$ $\theta=0$ and $\theta=\pi$ respectively,
are again given by \eqref{flux quantisation}. The resulting
fluxes are then given by
\begin{eqnarray}
N & = & \frac{L^{4}}{4\pi g_{s}\ell_{s}^{4}}\frac{(2+x_{+}U'_{+})^{2}\big((p+q)(x_{-}-x_{+})+px_{-}x_{+}U'_{+}\big)}{(p-q)x_{-}x_{+}^{2}U'_{+}(p+q+px_{+}U'_{+}\big)}\,,\nonumber \\
M_{x_{+}} & = & \frac{L^{4}}{16\pi^{2}g_{s}\ell_{s}^{4}}\text{vol}(\Sigma_{g})\frac{2+x_{+}U'_{+}}{(p-q)x_{+}}\Big(2(c_{1}+c_{2}x_{+}+\kappa)+(2c_{1}+c_{2}x_{+})x_{+}U'_{+}\Big)\,,\nonumber \\
M_{x_{-}} & = & \frac{L^{4}}{16\pi^{2}g_{s}\ell_{s}^{4}}\text{vol}(\Sigma_{g})\frac{2+x_{+}U'_{+}}{x_{-}(p+q+px_{+}U'_{+})^{2}}\Big(2(p+q)(c_{1}+c_{2}x_{-}+\kappa) \nn\\
&&\qquad \qquad    +(2qc_{1}+(p+q)c_{2}x_{-}+2\kappa p)x_{+}U'_{+}\Big)\,\\
M_{\theta_{0}}=M_{\theta_{\pi}} & = & \frac{L^{4}}{16\pi^{2}g_{s}\ell_{s}^{4}}\text{vol}(\Sigma_{g})\frac{2+x_{+}U'_{+}}{(p-q)x_{-}x_{+}^{2}U'_{+}}\Big(2\kappa(x_{-}-x_{+})-(c_{1}+c_{2}x_{-})x_{-}x_{+}U'_{-}\nn \\
&&\qquad \qquad  +(c_{1}+c_{2}x_{+})x_{-}x_{+}U'_{+}\Big)\,.\nonumber 
\end{eqnarray}
Note that the equality of $M_{\theta_{0}}$ and $M_{\theta_{\pi}}$
results in a vanishing $n_{2}$ via the relation \eqref{equations fluxes} and the toric
data (\ref{jacobian Ypq}).

\paragraph*{Central charge and R-charges.}

One can also integrate the expression for the central charge \eqref{trial central charge} and
this results in
\begin{align}\label{central charge Ypq}
c_{\text{sugra}}=\frac{3L^{8}}{8\pi^{3}g_{s}^{2}\ell_{s}^{8}}\text{vol}(\Sigma_{g})\frac{2+x_{+}U'_{+}}{(p-q)x_{-}^{2}x_{+}^{3}U'_{+}}& \Big((c_{1}+2c_{2}x_{+}+\kappa)x_{-}^{2}+(c_{1}+c_{2}x_{+})x_{-}^{2}x_{+}U'_{+}\\& -(c_{1}+2c_{2}x_{-}+\kappa)x_{+}^{2}-(c_{1}+c_{2}x_{-})x_{-}x_{+}^{2}U'_{-}\Big)\,.\nn
\end{align}
For the R-charges \eqref{Rcharges general} we find
\begin{eqnarray}
R_{x_{+}} & = & \frac{L^{4}}{2\pi g_{s}\ell_{s}^{4}}\frac{2+x_{+}U'_{+}}{(p-q)x_{+}}\,,\nonumber \\
R_{x_{-}} & = & \frac{L^{4}}{2\pi g_{s}\ell_{s}^{4}}\frac{2+x_{+}U'_{+}}{(p+q+px_{+}U'_{+})x_{-}}\,,\label{R-charges Ypq}\\
R_{\theta_{0}}=R_{\theta_{\pi}} & = & \frac{L^{4}}{2\pi g_{s}\ell_{s}^{4}}\frac{(x_{-}-x_{+})(2+x_{+}U'_{+})}{(p-q)x_{-}x_{+}^{2}U'_{+}}\,.\nonumber 
\end{eqnarray}
With these expressions and the flux $N$, one can easily check that
the first equation in \eqref{values b2} is obeyed. Using the second and third relation
in \eqref{values b2} one computes the remaining two components of the R-symmetry
vector $\vec{b}.$

\subsection{Solving the master equation} \label{sec: ypq solution}

Inserting the ansatz \eqref{ansats Ypq} into the master equation \eqref{Mastereq} results in a fourth-order non-linear differential equation for the function $U$:
\begin{align}\label{joejoe}
&\left(c_1+c_2 x\right){}^2 \left[2 \left(c_1+c_2 x\right) \left(2 \kappa -x^2 \left(c_1+c_2 x+\kappa \right) U''(x)\right)+x^2 \left(3
   c_1^2+3 c_{1}c_{2} x-c_2^2 x^2\right) U'(x)^2\right.\nonumber\\
&  \left. +x \left(c_1+c_2 x\right) U'(x) \left(x^3 \left(c_1+c_2 x\right) U^{(3)}(x)+c_1 \left(4-3 x^2 U''(x)\right)+4 \kappa \right)\right]\nonumber\\
  & +\left(c_1+c_2
   x\right) U(x) \left[\left(c_1+c_2 x\right) \left\{x^2 \left(3 c_1^2+6 c_1 c_2 x+2 c_2^2 x^2\right) U''(x)\right.\right.\\
   &\left.\left.+\left(c_1+c_2 x\right) \left(x^3 \left(x \left(c_1+c_2
   x\right) U^{(4)}(x)+\left(c_1+4 c_2 x\right) U^{(3)}(x)\right)-4 \left(c_1+\kappa \right)\right)\right\}\right. \nonumber\\
   &\left.
   -2 c_1 x \left(3 c_1^2+6 c_1 c_2 x+4 c_2^2 x^2\right)
   U'(x)\right]+c_1 \left(3 c_1^3 +9 c_1^2 c_2 x+9 c_1 c_2^2 x^2+4 c_2^3 x^3\right) U(x)^2=0\,.\nonumber
\end{align}
At first sight this seems like a quite imposing differential equation. However it is possible to integrate this equation twice. One sees that it implies
\begin{align}
\frac{\dd}{\dd x} \left( \frac{U(x)(c_1+c_2x)}{x} \frac{\dd}{\dd x} R-\frac{4 f(x)(c_{1}+c_{2}x+\kappa)}{x}+\frac{4 \kappa}{x}+\frac{2f(x)^2(2 c_{1}+c_{2}x)}{x}\right)=0\,,
\end{align}
and therefore
\be\label{YpqKimint1}
A_1 x-4 \kappa=U(x)(c_1+c_2x) \frac{\dd}{\dd x} R-4 f(x)(c_{1}+c_{2}x+\kappa)+2f(x)^2(2 c_{1}+c_{2}x)\,.
\ee
Defining the function
\begin{align}
g(x)=&2x (c_1+c_2 x +\kappa)U(x) -x^3(c_1+c_2 x) U(x)U''(x)+\frac{x^3(c_1+c_2 x)}{2}U'(x)^2 \nonumber\\
&-\frac{c_1 x (3 c_1+4 c_2 x)}{2(c_1+c_2 x)} U(x)^2 +c_1 x^2 U(x) U'(x)\,,
\end{align}
equation \eqref{YpqKimint1} is equivalent to 
\be
A_1 x-4 \kappa=\frac{\dd}{\dd x}g(x)-\frac{3}{x} g(x) \, .
\ee
This simple linear ODE implies
\be
g(x)=2 \kappa x -A_{1}x^2 + x^3 A_{2}\,,
\ee
where $A_2$ is a second integration constant. We have therefore reduced the fourth order non-linear differential equation to the following second order one
\begin{align}
2 \kappa x -A_{1}x^2 + x^3 A_{2}&=2x (c_1+c_2 x +\kappa)U(x) -x^3(c_1+c_2 x) U(x)U''(x)+\frac{x^3(c_1+c_2 x)}{2}U'(x)^2 \nonumber\\
&-\frac{c_1 x (3 c_1+4 c_2 x)}{2(c_1+c_2 x)} U(x)^2 +c_1 x^2 U(x) U'(x)\,.
\end{align}
In principle one can plug into this equation an ansatz, e.g. 
\begin{equation}
U(x)=\sum_{i=-1}^{4}\tilde{u}_{i}x^{i}\,,
\ee
to obtain solutions. However, we take a slightly different approach, since solutions of
the form we are interested in have already appeared in \cite{Benini:2015bwz}. Indeed,
taking the solution (3.93) in \cite{Benini:2015bwz}\footnote{To write their solution in this form \eqref{ansats Ypq}, one has to set $a_{2}=c_{1},$
use $y=1/(\tfrac{4}{3}c_{2}x)$ and identify $U(x)=\tfrac{(4c_{2}x/3)^{3}Q(1/(4c_{2}x/3))}{c_{1}+c_{2}x}$.} we find that the function 
\begin{align}
U(x)=\frac{1}{c_{1}+c_{2}x}&\Big(\frac{1}{3}(1+c_{1}-4C_{1})+(1-4(a-1)c_{1}^{2})c_{2}x-\frac{4c_{1}}{3(c_{1}-1)}(a-1)(-5+c_{1}-4C_{1})c_{2}^{2}x^{2}\nn \\
&-\frac{4}{9(c_{1}-1)^{2}}(a-1)\big(7+c_{1}^{2}+8C_{1}-4c_{1}(1+C_{1})\big)c_{2}^{3}x^{3}\Big)\label{U bobev}
\end{align}
solves the master equation \eqref{joejoe} for $\kappa=1.$ Here\footnote{In \cite{Benini:2015bwz} the expression for $C_1$ has a plus sign in front of the square root. The function
$U$ with this plus sign also solves the master equation for $\kappa=1$,
however, after solving the regularity equation one finds that it does
not give a globally defined solution. }
\be
C_{1}=-\frac{1}{4}\big(1+c_{1}-2\sqrt{1-c_{1}+c_{1}^{2}}\big)\,.
\ee
We now want to generalize the function (\ref{U bobev}) to be a solution
of the master equation for all values of $\kappa$. It turns out that
this generalization is given by
\begin{align}
U(x)=&\frac{1}{c_{1}+c_{2}x}\Big(\frac{1}{3}(\kappa+c_{1}-4C_{1})+(1-4(a-1)c_{1}^{2})c_{2}x\nn \\
&-\frac{4c_{1}}{3(c_{1}-\kappa)}(a-1)(-5\kappa+c_{1}-4C_{1})c_{2}^{2}x^{2}-\frac{1}{3}(a-1)\frac{2\kappa-c_{1}+4C_{1}}{C_{1}}c_{2}^{3}x^{3}\Big)\,,\label{general solution}
\end{align}
where 
\be
C_{1}=-\frac{1}{4}\big(\kappa+c_{1}-2\sqrt{\kappa^{2}-\kappa c_{1}+c_{1}^{2}}\big)\,.\label{C1value}
\ee
In the limit $\kappa\rightarrow1$ this function reduces to (\ref{U bobev}).
In the limit $\kappa\rightarrow0$ we find that, when assuming that $c_{1}<0$,
the function (\ref{general solution}) reduces to 
\be
U(x)=\frac{12c_{1}+\big(9c_{2}-36(a-1)c_{1}^{2}c_{2}\big)x-48c_{1}(a-1)c_{2}^{2}x^{2}-16(a-1)c_{2}^{3}x^{3}}{9(c_{1}+c_{2}x)}\,.
\ee
The solution with this function is equal to (A.104) in \cite{Benini:2015bwz} when identifying
$a_{2}=c_{1}$ and $q_{0}=-\tfrac{4}{3}(a-1)c_{1}^{3},$ using $y=-c_{1}/(c_{2}x)$
and identifying\footnote{Actually we find their function $U$ with a minus sign, and we assume
$c_{1}<0$ while the authors of \cite{Benini:2015bwz} assume that $c_{1}>0.$ }
\be
U(x)=\frac{9}{64c_{1}^{3}}\frac{(-\tfrac{4}{3}c_{2}x)^{3}w(-\frac{c_{1}}{c_{2}x})}{c_{1}+c_{2}x}\,.
\ee
The
solutions depend on two parameters ($c_{2}$ can be scaled away)
of which one will be fixed by the regularity equation (\ref{regularity equation}).
In the field theory there are two parameters \cite{Benini:2015bwz,Gauntlett:2018dpc}, i.e. one
of the fluxes and one twist parameter. Therefore, the most general solution should have one parameter more than the one we give here. Since we have given expressions
for the fluxes, central charge and R-charges in terms of the general
function $U$ one can compute these quantities very easily once a more general solution is found.

The ansatz \eqref{ansats Ypq} with function $U$ given by \eqref{general solution} is a
limit of the class of $L^{a,b,c}$ solutions we have constructed in
section \ref{solving master eq}. Namely, starting from the ansatz \eqref{ansatz} with
functions $\mathcal{L}=\mathcal{H}=\eta\xi$ and $\mathcal{F},$ $\mathcal{G}$
given by \eqref{functions f and g}, we can first transform to the coordinates of \cite{Cvetic:2005ft} using \cite{Martelli:2005wy}
\begin{align}
\xi&=(\alpha-\beta)\sin^{2}\theta+\gamma\,,\nonumber \\ 
\eta&=\alpha-x+\gamma\,,\nonumber\\
\phi&=\frac{(\alpha-\beta+\gamma)\Psi}{2(\alpha-\beta)\beta}-\frac{\gamma\Phi}{2(\alpha-\beta)\alpha}\,,\label{transf 1}\\
\psi&=\frac{\Phi}{2(\alpha-\beta)\alpha}-\frac{\Psi}{2(\alpha-\beta)\beta}\,.\nonumber
\end{align}
The shift with $\gamma$ is different from \cite{Martelli:2005wy}, but needed here in order to take the $Y^{p,q}$ limit. After the transformation
\eqref{transf 1} the metric can be written as
\begin{eqnarray}
\text{d}s^{2} &  = & (\alpha-x+\gamma)\big((\alpha-\beta)\sin^{2}\theta+\gamma\big)\text{d}s^{2}(\Sigma_{g})+\frac{\rho^{2}}{4\Delta_{x}}\text{d}x^{2}+\frac{\rho^{2}}{\Delta_{\theta}}d\theta^{2}\nonumber\\
&&+\frac{\Delta_{x}}{\rho^{2}}\Big(\frac{1}{\alpha}\sin^{2}\theta\text{d}\Phi+\frac{1}{\beta}\cos^{2}\theta\text{d}\Psi+2\big((\alpha-\beta)\sin^{2}\theta+\gamma\big)\mathcal{A}_{g}\Big)^{2}\nonumber \\
 && +\frac{\Delta_{\theta}\sin^{2}\theta\cos^{2}\theta}{\rho^{2}}\Big(\frac{\alpha-x}{\alpha}\text{d}\Phi-\frac{\beta-x}{\beta}\text{d}\Psi+2(\alpha-\beta)(\alpha-x+\gamma)\mathcal{A}_{g}\Big)^{2},
\end{eqnarray}
where
\begin{eqnarray}
\Delta_{x} & = & \frac{1}{4}\mathcal{G}(\alpha-x+\gamma)\,,\nonumber \\
\Delta_{\theta} & = & \frac{1}{4(\alpha-\beta)^{2}\sin^{2}\theta\cos^{2}\theta}\mathcal{F}((\alpha-\beta)\sin^{2}\theta+\gamma)\,,\\
\rho^{2} & = & \alpha\cos^{2}\theta+\beta\sin^{2}\theta-x\,.\nonumber 
\end{eqnarray}
The $Y^{p,q}$ limit is then performed by sending $\beta\rightarrow\alpha$
and requiring that in this limit $\Delta_{\theta}=\alpha$ \cite{Cvetic:2005ft}.
This results in the metric
\begin{eqnarray}
\text{d}s^{2} & = & (\alpha-x+\gamma)\gamma\text{d}s^{2}(\Sigma_{g})+\frac{\alpha-x}{4\Delta_{x}}\text{d}x^{2}+\frac{\Delta_{x}}{(\alpha-x)\alpha^{2}}\big(\sin^{2}\theta\text{d}\Phi+\cos^{2}\theta\text{d}\Psi+2\gamma\alpha\mathcal{A}_{g}\big)^{2}\nonumber \\
 &  & +\frac{\alpha-x}{\alpha}\big(\text{d}\theta^{2}+\sin^{2}\theta\cos^{2}\theta(\text{d}\Phi-\text{d}\Psi)^{2}\big)\,.
\end{eqnarray}
We now transform
\begin{equation}
\theta\rightarrow\frac{1}{2}\theta\,,\qquad\Phi\rightarrow\frac{1}{2}(\Phi-\Psi)\,,\qquad\Psi\rightarrow-\frac{1}{2}(\Psi+\Phi)\,,\qquad x\rightarrow-\frac{4\alpha}{x}+\alpha\,,
\end{equation}
such that in the new coordinates the metric becomes
\begin{eqnarray}
\text{d}s^{2} & = & \frac{4\alpha\gamma+\gamma^{2}x}{x}\text{d}s^{2}(\Sigma_{g})+\frac{16\alpha^{3}}{x^{5}\Delta_{x}}\text{d}x^{2}+\frac{x\Delta_{x}}{16\alpha^{3}}\big(\text{d}\Psi+\cos\theta\text{d}\Phi-4\gamma\alpha\mathcal{A}_{g}\big)^{2}\nonumber \\
 &  & +\frac{1}{x}\big(\text{d}\theta^{2}+\sin^{2}(\theta)\text{d}\Phi^{2}\big)\label{metric ypq}
\end{eqnarray}
with
\begin{equation}
\Delta_{x}=\frac{1}{4}\mathcal{G}\Big(\frac{4\alpha}{x}+\gamma\Big)\,.
\end{equation}
The metric (\ref{metric ypq}) has the form of our ansatz \eqref{ansats Ypq}
when identifying $c_{1}=4\gamma\alpha,$ $c_{2}=\gamma^{2}$ and
\begin{equation}
U(x)=\frac{x^{2}\Delta_{x}}{16\alpha^{3}}=\frac{x^{2}c_{2}^{3/2}}{c_{1}^{3}}\mathcal{G}\Big(\frac{c_{1}}{\sqrt{c_{2}}x}+\sqrt{c_{2}}\Big)\,.\label{Ureduction}
\end{equation}
We still have to impose that $\Delta_{\theta}=\alpha$ in the limit
$\beta\to\alpha.$ This results in the conditions
\begin{eqnarray}
D\gamma^{3}-A(\gamma+C)^{2}+\kappa\gamma{}^{2} & = & 0\,,\nonumber \\
-2A(\gamma+C)+2\kappa\gamma+3D\gamma^{2} & = & 4\gamma\alpha(\alpha-\beta)\,,\\
-A+\kappa+3D\gamma & = & -4\gamma\alpha\,,\nonumber 
\end{eqnarray}
which are solved by
\begin{eqnarray}
A & = & 4C_{1}\,,\nonumber \\
C & = & \frac{\sqrt{c_{2}}}{6(c_{1}-\kappa)}\big(-5c_{1}-4C_{1}+\kappa\big)\,,\\
D & = & \frac{-c_{1}-\kappa+4C_{1}}{3\sqrt{c_{2}}}\,.\nonumber 
\end{eqnarray}
In the latter expressions we have taken the limit $\beta\to\alpha.$
Here $C_{1}$ is given by \eqref{C1value}. The function $U$ \eqref{Ureduction}
is equal to \eqref{general solution} when reparametrizing
\[
B=4(-c_{1}^{3}+ac_{1}^{3}-C_{1})\,.
\]

\subsection{Solving the regularity equation and matching to the field theory}
We now intend to solve the regularity condition (\ref{regularity equation})
in order to see whether the solutions \eqref{ansats Ypq} with function (\ref{general solution})
are indeed globally well-defined. We do this as in section \ref{subsec:matching}. First
of all, we write the function $U$ as
\begin{equation}
U(x)=\frac{u(x)}{c_{1}+c_{2}x}\equiv\frac{1}{c_{1}+c_{2}x}\sum_{i=0}^{3}\tilde{u}_{1}x^{i}\,.\label{expression small u}
\end{equation}
We can then express it via its roots as
\begin{equation}
U(x)=\frac{\tilde{u}_{3}(x-x_{-})(x-x_{+})(x-x_{*})}{c_{1}+c_{2}x}\,,\label{expression in terms of roots}
\end{equation}
where we have denoted the third root by $x_{*}$. By equating the
expressions (\ref{expression small u}) and (\ref{expression in terms of roots})
we find that
\begin{eqnarray}
x_{-}+x_{+}+x_{*} & = & -\tilde{u}_{2}/\tilde{u}_{3}\,,\nonumber \\
x_{-}x_{+}+x_{-}x_{*}+x_{+}x_{*} & = & \tilde{u}_{1}/\tilde{u}_{3}\,,\label{eq reg}\\
x_{-}x_{+}x_{*} & = & -\tilde{u}_{0}/\tilde{u}_{3}\,.\nonumber 
\end{eqnarray}
Together with the regularity condition (\ref{regularity equation})
which can be expressed in terms of the roots using 
\be
U'(x_{\pm})=\frac{\tilde{u}_{3}(x_{\pm}-x_{\mp})(x_{\pm}-x_{*})}{c_{1}+c_{2}x_{\pm}}\,,
\ee
this gives four conditions for five variables. When $\kappa=\pm1,$
these can be solved using the substitutions
\be
a\rightarrow\frac{a'}{c_{1}^{2}}+1\,,\; \; x_{\pm}\rightarrow-2\frac{c_{1}}{c_2}\big(c_{1}-2\kappa+\sqrt{1-\kappa c_{1}+c_{1}^{2}}\big)x'_{\pm}\,,\; \; x_{*}\rightarrow-2\frac{c_{1}}{c_2}\big(c_{1}-2\kappa+\sqrt{1-\kappa c_{1}+c_{1}^{2}}\big)x'_{*}\,.
\ee
This way we can solve the equations (\ref{eq reg}) for $a',$ $x'_{+}$
and $x'_{-}$ in terms of $c_{1}$ and $x'_{*}$ . The regularity
condition (\ref{regularity equation}) needs then to be solved numerically
for $x'_{*}$. We found regular solutions for $\kappa=1$ when $c_{1,2}<0$
and $q>p>0.$ Note that the parameter regime for $p$ and $q$ is different from the field theory, which is only valid for $p>q>0$. We comment extensively on this issue in the main text. For $\kappa=-1$ regular solutions are obtained for values
$c_{1,2}>0$ and $p>q>0.$ The latter inequality is the same as in the field theory, thus this is evidence that we have found the dual to the field theory.

For $\kappa=0$ we can actually solve all
the equations explicitly and we use them to test the formulas we found
for the fluxes, central charge and R-charges. In this case we use
the transformations
\be
a\rightarrow\frac{a'}{c_{1}^{2}}+1\,,\qquad x_{\pm}\rightarrow3\frac{c_{1}}{c_2}x'_{\pm}\,,\qquad x_{*}\rightarrow3\frac{c_{1}}{c_2}x'_{*}\,,
\ee
to solve the equations (\ref{regularity equation}) and (\ref{eq reg})
for $a,$ $x'_{\pm}$ and $x'_{*}$ in terms of $p,$ $q,$ $c_{1}$
and $c_{2}$. For this we need that $p>0$ and $q>0.$ Transforming
back to the original variables, this results in 
\begin{eqnarray}
x_{-} &=&  -\frac{c_{1}}{6c_{2}}\frac{(p+3q)^{2}}{q(p+q)}\,,\qquad x_{+}=\frac{c_{1}}{6c_{2}}\frac{(p-3q)^{2}}{q(p-q)}\,,\qquad x_{*}=-\frac{4c_{1}}{3c_{2}}\frac{p^{2}}{p^{2}-q^{2}}\,,\nn \\
a & = & \frac{4c_{1}^{2}(p^{3}-9pq^{2})^{2}+81(q^{3}-qp^{2})^{2}}{4c_{1}^{2}p^{2}(p^{2}-9q^{2})^{2}}\,.
\end{eqnarray}
We find that to get regular solutions we need $c_{2}>0$ and $q>p>0$, i.e. to get $x_-<x_+$.
Previously we have already assumed that $c_{1}<0.$ We can then compute the
central charge and R-charges via formulas (\ref{central charge Ypq})
and (\ref{R-charges Ypq}) respectively, and find 
\be
c_{\text{geom}} = \frac{c_{1}N^{2}p(p^{4}-9q^{4})}{3q^{4}}\,
\ee
and
\begin{equation}
R_{x_{+}}  =  -\frac{N(p-q)(p+3q)}{3q^{2}}\,, \qquad
R_{x_{-}} =  -\frac{N(p+q)(p-3q)}{3q^{2}}\,, \qquad
R_{\theta_{0}}=R_{\theta_{\pi}}  =  \frac{Np^{2}}{3q^{2}}\,.
\end{equation}
We have used the flux
\be
N=-\frac{54c_{2}L^{4}}{\pi c_{1}g_{s}\ell_{s}^{4}}\frac{q^{4}}{p(p^{2}-9q^{2})^{2}}
\ee
to replace $g_{s}$ for $N$. Note that indeed the constant $c_{2}$
drops out of the expressions as it should. Also note that the R-charge $R_{x_{+}}$ becomes negative for $p>q$, which again implies that we need $q>p$.

The central charge and R-charges have been computed from the field theory and geometric
c-extremization side in \cite{Gauntlett:2018dpc} but a different basis for the toric data
was used. Repeating their computations for the toric data derived
from (\ref{jacobian Ypq}), i.e. 
\be
\vec{v}_{1}=(1,0,0)\,,\qquad\vec{v}_{2}=(1,1,0)\,,\qquad\vec{v}_{3}=(1,0,p)\,,\qquad\vec{v}_{4}=(1,-1,p-q)\,,
\ee
and taking $n_{2}=0,$ we find that the central charge is given by
\begin{equation}
c_{R}^{\text{geo}}=\frac{6N^{2}p(n_{3}-pm_{1})\big(qn_{3}^{2}+p(2p-q)m_{1}n_{3}+p^{2}(q-p)m_{1}^{2}\big)}{pq(p-q)m_{1}n_{3}+p^{2}q^{2}m_{1}^{2}+(p^{2}+pq+q^{2})n_{3}^{2}}\,.\label{central charge cextr}
\end{equation}
As before $M_{1}\equiv m_{1}N.$ The R-charges are equal to
\begin{eqnarray}
R_{1}=R_{x_+} & = & -\frac{N(p+q)\big((p-q)n_{3}^{2}+p(q-3p)m_{1}n_{3}+p^{2}(p-q)m_{1}^{2}\big)}{pq(p-q)m_{1}n_{3}+p^{2}q^{2}m_{1}^{2}+(p^{2}+pq+q^{2})n_{3}^{2}}\,,\nonumber \\
R_{3}=R_{x_-} & = & \frac{N(p+q)\big((p+q)n_{3}^{2}+p(p-q)m_{1}n_{3}+p^{2}(q-p)m_{1}^{2}\big)}{pq(p-q)m_{1}n_{3}+p^{2}q^{2}m_{1}^{2}+(p^{2}+pq+q^{2})n_{3}^{2}}\,,\label{rcharges cextr}\\
R_{2}=R_{\theta_0}=R_{4}=R_{\theta_\pi} & = & \frac{Np^{2}(n_{3}-pm_{1})^{2}}{pq(p-q)m_{1}n_{3}+p^{2}q^{2}m_{1}^{2}+(p^{2}+pq+q^{2})n_{3}^{2}}\,.\nonumber 
\end{eqnarray}
The central charge and R-charges indeed match the expressions in \cite{Gauntlett:2018dpc}
by taking $n_{3}\rightarrow n_{2}$ and $m_{1}\rightarrow m$ in our
expressions. We can match them to the expressions that we got from
the solution by computing $n_{3}$ and $m_{1}$ and substituting these values in (\ref{central charge cextr}) and
(\ref{rcharges cextr}). The values we obtain from the solution are
\be
n_{3}=-\frac{c_{1}(p^{2}+3q^{2})}{3q}\,,\qquad m_{1}=\frac{M_{x_{+}}}{N}=\frac{c_{1}(p-q)(p^{2}+3q^{2})}{6pq^{2}}\,.
\ee
With these values we can indeed match the solution to the geometric
c-extremization result. The latter has been matched to the field theory. As in the S$^2$ fibration case this is only a formal matching, as already mentioned the solution is only globally regular for $q>p>0$, while the field theory result is only valid for $p>q>0$. 

\bibliographystyle{utcaps}
\bibliography{ADSCFT}

\end{document}